\documentclass[aps,prc,twocolumn,showpacs,superscriptaddress,groupedaddress]{revtex4} 
\usepackage{graphicx}  % needed for figures
\usepackage{dcolumn}   % needed for some tables
\usepackage{bm}        % for math
\usepackage{amssymb}   % for math
\usepackage{amsfonts}
\usepackage{booktabs}
\usepackage{siunitx}
\usepackage{xcolor}
\usepackage{multirow}
\usepackage[normalem]{ulem}

\newcommand\tstrut{\rule{0pt}{3.0ex}}
\newcommand\bstrut{\rule[-1.5ex]{0pt}{0pt}}

\begin{document}

\title{Lifetime Measurements in the Even-Even $^{102-108}$Cd Isotopes}
\author{M.~Siciliano}
    \thanks{Present address: Physics Division, Argonne National Laboratory, Argonne (IL), United States.}
	\affiliation{Irfu/CEA, Universit\'e Paris-Saclay, Gif-sur-Yvette, France.}
	\affiliation{INFN, Laboratori Nazionali di Legnaro, Legnaro, Italy.}
	\affiliation{Dipartimento di Fisica e Astronomia, Universit\`a di Padova, Padua, Italy.}
\author{J.J.~Valiente-Dob\'on}
	\affiliation{INFN, Laboratori Nazionali di Legnaro, Legnaro, Italy.}
\author{A.~Goasduff}
	\affiliation{INFN, Laboratori Nazionali di Legnaro, Legnaro, Italy.}
	\affiliation{Dipartimento di Fisica e Astronomia, Universit\`a di Padova, Padua, Italy.}
	\affiliation{INFN, Sezione di Padova, Padua, Italy.}
\author{T.R.~Rodr\'{i}guez}
	\affiliation{Universidad Aut\'{o}noma de Madrid, Madrid, Spain.}
\author{D.~Bazzacco}
	\affiliation{INFN, Sezione di Padova, Padua, Italy.}
\author{G.~Benzoni}
	\affiliation{INFN, Sezione di Milano, Milan, Italy.}
\author{T.~Braunroth}
	\affiliation{Institut f\"{u}r Kernphysik, Universit\"{a}t zu K\"{o}ln, Cologne, Germany.}
\author{N.~Cieplicka-Ory\'nczak}
	\affiliation{INFN, Sezione di Milano, Milan, Italy.}
	\affiliation{Institute of Nuclear Physics Polish Academy of Sciences, PL-31342 Krakow, Poland}
\author{E.~Cl\'ement}
	\affiliation{Grand Acc\'el\'erateur National d’Ions Lourds, Irfu/CEA/DRF and CNRS/IN2P3, Caen, France.} 
\author{F.C.L.~Crespi}
	\affiliation{INFN, Sezione di Milano, Milan, Italy.}
	\affiliation{Dipartimento di Fisica, Universit\`a di Milano, Milan, Italy.}
\author{G.~de~France}
	\affiliation{Grand Acc\'el\'erateur National d’Ions Lourds, Irfu/CEA/DRF and CNRS/IN2P3, Caen, France.} 
\author{M.~Doncel}
	\affiliation{Universidad de Salamanca, Salamanca, Spain.}
\author{S.~Ert\"urk}
	\affiliation{\"Omer Halisdemir \"Universitesi, Ni\u{g}de, Turkey.}
\author{C.~Fransen}
	\affiliation{Institut f\"{u}r Kernphysik, Universit\"{a}t zu K\"{o}ln, Cologne, Germany.}
\author{A.~Gadea}
	\affiliation{Instituto de F\'isica Corpuscular, CSIC-Universidad de Valencia, Valencia, Spain.}
\author{G.~Georgiev}
	\affiliation{IJCLab, Universit\'e Paris-Saclay, Orsay, France.}
\author{A.~Goldkuhle}
	\affiliation{Institut f\"{u}r Kernphysik, Universit\"{a}t zu K\"{o}ln, Cologne, Germany.}
\author{U.~Jakobsson}
	\affiliation{Department of Physics, Royal Institute of Technology, Stockholm, Sweden.}
\author{G.~Jaworski}
	\affiliation{INFN, Laboratori Nazionali di Legnaro, Legnaro, Italy.}
	\affiliation{Heavy Ion Laboratory, University of Warsaw, Warsaw, Poland.}
\author{P.R.~John}
	\affiliation{Dipartimento di Fisica e Astronomia, Universit\`a di Padova, Padua, Italy.}
	\affiliation{INFN, Sezione di Padova, Padua, Italy.}
	\affiliation{Institut f\"ur Kernphysik, Technische Universit\"at Darmstadt, Darmstadt, Germany.}
\author{I.~Kuti}
	\affiliation{Institute of Nuclear Research ATOMKI, Debrecen, Hungary.}
\author{A.~Lemasson}
	\affiliation{Grand Acc\'el\'erateur National d’Ions Lourds, Irfu/CEA/DRF and CNRS/IN2P3, Caen, France.} 
\author{H.~Li}
	\affiliation{Department of Physics, Royal Institute of Technology, Stockholm, Sweden.}
\author{A.~Lopez-Martens}
	\affiliation{IJCLab, Universit\'e Paris-Saclay, Orsay, France.}
\author{T.~Marchi}
	\affiliation{INFN, Laboratori Nazionali di Legnaro, Legnaro, Italy.}
\author{D.~Mengoni}
	\affiliation{Dipartimento di Fisica e Astronomia, Universit\`a di Padova, Padua, Italy.}
	\affiliation{INFN, Sezione di Padova, Padua, Italy.}
\author{C.~Michelagnoli}
	\affiliation{Grand Acc\'el\'erateur National d’Ions Lourds, Irfu/CEA/DRF and CNRS/IN2P3, Caen, France.} 
	\affiliation{Institut Laue-Langevin, Grenoble, France.}
\author{T.~Mijatovi\'c}
	\affiliation{Ru{d\llap{\raise 1.22ex\hbox{\vrule height 0.09ex width 0.4em}}\rlap{\raise 1.22ex\hbox{\vrule height 0.09ex width 0.04em}}}er Bo\v{s}kovi\'{c} Institute and University of Zagreb, Zagreb, Croatia.}
\author{C.~M\"uller-Gatermann}
	\affiliation{Institut f\"{u}r Kernphysik, Universit\"{a}t zu K\"{o}ln, Cologne, Germany.}
	\affiliation{Physics Division, Argonne National Laboratory, Argonne (IL), United States.}
\author{D.R.~Napoli}
	\affiliation{INFN, Laboratori Nazionali di Legnaro, Legnaro, Italy.}
\author{J.~Nyberg}
	\affiliation{Department of Physics and Astronomy, Uppsala University, Uppsala, Sweden.}
\author{M.~Palacz}
	\affiliation{Heavy Ion Laboratory, University of Warsaw, Warsaw, Poland.}
\author{R.M.~P\'erez-Vidal}
	\affiliation{Instituto de F\'isica Corpuscular, CSIC-Universidad de Valencia, Valencia, Spain.}
	\affiliation{INFN, Laboratori Nazionali di Legnaro, Legnaro, Italy.}
\author{B.~Say\u{g}i}
	\affiliation{INFN, Laboratori Nazionali di Legnaro, Legnaro, Italy.}
	\affiliation{Ege \"Universitesi, \.Izmir, Turkey.}
	\affiliation{Department of Physics, Sakarya University, Sakarya, Turkey.}
\author{D.~Sohler}
	\affiliation{Institute of Nuclear Research ATOMKI, Debrecen, Hungary.}
\author{S.~Szilner}
	\affiliation{Ru{d\llap{\raise 1.22ex\hbox{\vrule height 0.09ex width 0.4em}}\rlap{\raise 1.22ex\hbox{\vrule height 0.09ex width 0.04em}}}er Bo\v{s}kovi\'{c} Institute and University of Zagreb, Zagreb, Croatia.}
\author{D.~Testov}
	\affiliation{Dipartimento di Fisica e Astronomia, Universit\`a di Padova, Padua, Italy.}
	\affiliation{INFN, Sezione di Padova, Padua, Italy.}
	\affiliation{Joint Institute for Nuclear Research, Dubna, Russia.}

\vskip 0.25cm    

\begin{abstract}
\noindent
\textbf{Background:} The heaviest $T_z=0$ doubly-magic nucleus, $^{100}$Sn, and the neighboring nuclei offer unique opportunities to investigate the properties of nuclear interaction. %in extreme conditions. 
For instance, the structure of light-Sn nuclei has been shown to be affected by the delicate balance between nuclear-interaction components, such as pairing and quadrupole correlations.
%\textcolor{red}{\sout{Studies of light Sn nuclei are hindered by their relatively high mass, proton-rich character and the presence of low-lying isomers.}}
From Cd to Te, many common features and phenomena have been observed experimentally along the isotopic chains, leading to theoretical studies devoted to a more general and comprehensive study of the region. 
In this context, having only two proton holes in the $Z = 50$ shell, the Cd isotopes are expected to present properties similar to those found in the Sn isotopic chain.

\noindent
\textbf{Purpose:} The aim of this work was to measure lifetimes of excited states in neutron-deficient nuclei in the vicinity of $^{100}$Sn. 

\noindent
\textbf{Methods:} The neutron-deficient nuclei in the $N \approx Z \approx 50$ region were populated using a multi-nucleon transfer reaction with a $^{106}$Cd beam and a $^{92}$Mo target. 
The beam-like products were identified by the VAMOS++ spectrometer, while the $\gamma$ rays were detected using the AGATA array. 
Lifetimes of excited states were determined using the Recoil Distance Doppler-Shift method, employing the Cologne differential plunger.

\noindent
\textbf{Results:} Lifetimes of low-lying states were measured in the even-mass $^{102-108}$Cd isotopes. In particular, multiple states with excitation energy up to $\approx 3$~MeV, belonging to various bands, were populated in $^{106}$Cd via inelastic scattering. 
The transition strengths corresponding to the measured lifetimes were compared with those resulting from state-of-the-art beyond-mean-field calculations using the symmetry-conserving configuration-mixing approach. 

\noindent
\textbf{Conclusions:} Despite the similarities in the electromagnetic properties of the low-lying states, there is a fundamental structural difference between the ground-state bands in the $Z=48$ and $Z=50$ isotopes. 
The comparison between experimental and theoretical results revealed a rotational character of the Cd nuclei, which have prolate-deformed ground states with $\beta_2 \approx 0.2$. 
At this deformation $Z=48$ becomes a closed-shell configuration, which is favored with respect to the spherical one. 
\end{abstract}

\maketitle

%\vspace{150mm}
In recent years, the interest in studies of nuclear structure around $Z=50$ has significantly increased. 
This region presents unique conditions to investigate observables, such as excitation energies, quadrupole moments and reduced transition probabilities, starting from neutron-deficient nuclei close to the proton drip line, up to neutron-rich isotopes towards and beyond the $N=82$ neutron shell closure.
Consequently, the longest isotopic chains between two experimentally accessible shell closures -- i.e. tellurium ($Z= 52$), tin ($Z=50$) and cadmium ($Z=48$) isotopes -- are being extensively studied in order to probe the evolution of nuclear properties in both stable and exotic nuclei. 
Various experimental works have pointed out similarities (e.g. transition-strength and excitation-energy systematics, neutron-transfer spectroscopic factors, shape coexistence) between the three isotopic chains~\cite{STRUWE1974605, FIELDING1977389, back2012BE, saxena2014rotational, garrett2018collective} and these common features have yielded theoretical investigations devoted to study the $Z\approx50$ region in a more general and comprehensive way~\cite{BF02769692, LIBERT200747, maheshwari2019seniority, fortune2019similarities, maheshwari2020unified}. 

Due to the rather constant excitation energies of the $2_1^+$ and $4_1^+$ states and the presence of low-lying isomers in the even-mass nuclei, the $Z=50$ semi-magic Sn isotopes have been considered for decades to be excellent examples of pairing dominance, showing the typical features of seniority schemes~\cite{talmi1971generalized, talmi1993simple, ressler2004transition, morales2011generalized, maheshwari2019seniority, maheshwari2020unified}. 
On the other hand, the $B(E2; 2_1^+ \to 0_{g.s.})$ reduced transition probabilities remain almost constant for the $106 \leq A \leq 114$ Sn nuclei, instead of following the parabolic trend expected for the pairing domination. 
This observation casts doubts on the validity of the generalized seniority interpretation. 
In particular, recent works~\cite{siciliano2020plb, zuker2020CdSn} highlighted the key role of the $4_1^+ \to 2_1^+$ transition strengths in revealing the delicate balance between pairing and quadrupole correlations in the light Sn isotopes. 
Furthermore, thanks to the precise determination of the $B(E2; 4_1^+ \to 2_1^+)$ value in $^{108}$Sn~\cite{siciliano2020plb}, Zuker~\cite{zuker2020CdSn} proved how the sole information on the $2_1^+$ states is not sufficient for an in-depth description of the nuclei in this mass region: any ``sufficiently good'' interaction is capable of reproducing the electromagnetic properties of the $2_1^+$ states. 

\begin{figure}[ht]
\includegraphics[width=0.48\textwidth]{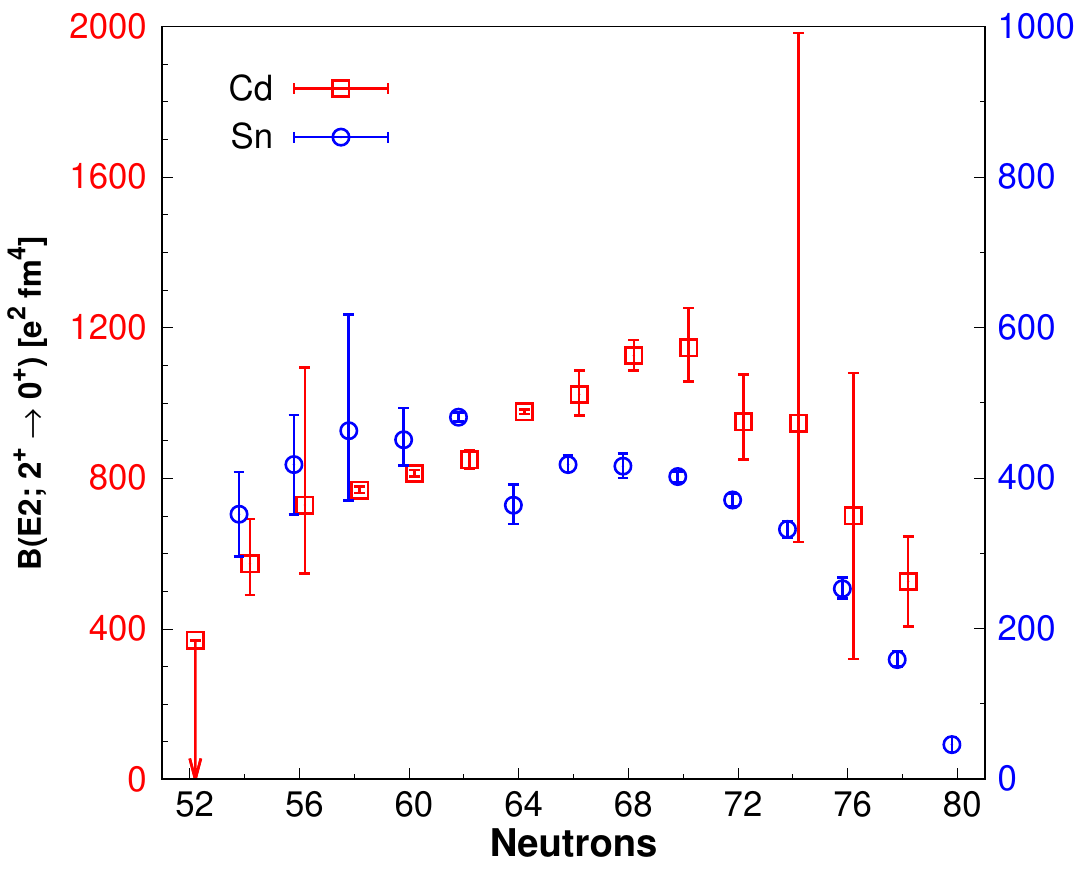}
\vspace{-7mm}
\caption{\label{fig:BE2_systematics}(Color online) Systematics of the experimental $B(E2; 2_1^+ \to 0_{g.s.}^+)$ reduced transition probabilities for the even-mass Cd (red squares) and Sn (blue circles) isotopic chains. Results are taken from Ref.~\cite{nudatDatabase}.}
\end{figure}

The Cd isotopes, which have only two proton holes in the $Z = 50$ shell, are expected to present features similar to those found for the Sn nuclei. 
For instance, Figure~\ref{fig:BE2_systematics} shows that the $B(E2; 2_1^+ \to 0_{g.s.}^+)$ values display similar trends in the Sn and Cd isotopic chains, except for the larger collectivity of the latter. 
%The competition between collective and non-collective degrees of freedom is an important issue in understanding the structure of the atomic nucleus.
%Depending on the polarization induced by the valence nucleons, the behaviour of a particular nucleus can evolve with spin and excitation energy between these two cases. 
In addition, the excitation energies of the $2_1^+$ and $4_1^+$ states in even-mass Cd nuclei are rather constant, similarly to the Sn isotopes.
Therefore, one can expect that the experimental information on the $Z=48$ nuclei may not only be important in itself, but it may also provide an insight into the structure of the corresponding $Z=50$ isotones. 

Based on the excitation energies of their low-lying states, the cadmium isotopes have been considered a textbook example of harmonic quadrupole-vibrational nuclei~\cite{arima1976interacting, iachello1987interacting, bohrMottelson1998nuclear, yates2005probing, rowe2010nuclear} with a two-phonon triplet and a three-phonon quintuplet of levels at approximately twice and three times the energy of the $2_1^+$ state, respectively. 
On the other hand, the electromagnetic properties of the Cd isotopes, i.e. quadrupole moments and transition strengths, put their vibrational character in doubt. 
In fact, recent multi-step Coulomb excitation and lifetime measurements have demonstrated a substantial disagreement with a vibrational structure and revealed a systematic trend of the $B(E2)$ values in the even-even $^{110-116}$Cd isotopes~\cite{fahlander1988hasselgrenet, lehmann1996nature, corminboeuf2000characterization, kadi2003vibrational, garrett2007properties, bandyopadhyay2007investigation, garrett2008breakdown, green2009degeneracy, garrett2019multiple, garrett2020multiple}. 
These experimental results have lead to a reinterpretation of these Cd nuclei, whose structure is seen as the coexistence of various rotational bands. 
However, the lack of precise experimental information makes it difficult to assess whether the vibrational picture still holds for the neutron-deficient species. 

The experiment described in this work aimed at the determination of the $2_1^+ \to 0_{g.s.}^+$ and $4_1^+ \to 2_1^+$ transition strengths in neutron-deficient $Z \leq 50$ nuclei by measuring lifetimes of the $2_1^+$ and $4_1^+$ states. 
The results concerning the light Sn isotopes were discussed in Ref.~\cite{siciliano2020plb}, while the present manuscript focuses on the lifetimes of low-lying states in even-mass $^{102-108}$Cd. 
The experimental values are compared with the predictions of new beyond-mean-field calculations using the symmetry-conserving configuration-mixing approach. 
The general features and the evolution of the ground-state structure are discussed for the whole Cd isotopic chain, with a particular focus on the variety of excited bands in $^{106}$Cd.

\section{Experiment}

%\textcolor{red}{\sout{Multi-nucleon transfer (MNT) is a reaction mechanism widely adopted to study neutron-rich nuclei~\cite{szilner2007multinucleon, corradi2009multinucleon, valiente2016gamma}. However, in the present experiment this reaction mechanism was applied, somehow unusually, to populate neutron-deficient nuclei in the vicinity of $^{100}$Sn.}}
Multi-nucleon transfer (MNT) is a reaction mechanism rarely applied to study neutron-deficient nuclei~\cite{szilner2007multinucleon, corradi2009multinucleon, valiente2016gamma}, but it was selected in the present work to study nuclei approaching $^{100}$Sn.
A $^{106}$Cd beam at 770~MeV energy, provided by the separated-sector cyclotron of the GANIL facility, impinged on a 0.8 mg/cm$^2$ $^{92}$Mo target. 
The lifetime measurement was performed with the Recoil Distance Doppler-Shift (RDDS) method~\cite{dewald1989differential, dewald2012developing, valiente2009lifetime}. 
The target was mounted on the differential Cologne plunger with a 1.6 mg/cm$^2$ thick $^{nat}$Mg degrader placed downstream. 
Eight different target-degrader distances in the 31-521~$\mu$m range were used to measure the lifetimes of interest. 
The complete identification of the beam-like reaction products was performed with the VAMOS++ magnetic spectrometer~\cite{pullanhiotan2008improvement, rejmund2011performance, vandebrouck2016dual}, placed at the grazing angle $\theta_{lab}$=25$^\circ$. 
The emitted $\gamma$ rays were detected by the $\gamma$-ray tracking detector array AGATA~\cite{akkoyun2012agata, clement2017conceptual}, consisting of 8 triple-cluster detectors placed in a compact configuration (18.5 cm from the target) at backward angles with respect to the beam direction. 
The combination of the pulse-shape analysis~\cite{bruyneel2013correction} and the Orsay Forward-Tracking (OFT) algorithm~\cite{lopez2004gamma} allowed reconstruction of the trajectories of the $\gamma$ rays emitted by the reaction products. 
More details can be found in Refs.~\cite{siciliano2017appb, siciliano2017ncc, Siciliano2019EPJ, siciliano2020plb}.

\section{Lifetime Analysis}

Combining the precise determination of the ion velocity vector given by VAMOS++ and the identification of the first interaction point of each $\gamma$ ray inside AGATA, Doppler correction was applied on an event-by-event basis. 
The magnetic spectrometer directly measured the ion velocity after the degrader ($\beta_{after} \approx 9\%$) and this velocity was used to correct the detected $\gamma$-ray energy. 
The velocity of the ions before the degrader ($\beta_{before} \approx 10\%$) was reconstructed by taking into account the direction and the energy loss of the ions inside the Mg foil. 
For each transition two peaks were observed, related to the emission of the $\gamma$-ray before the Mg foil (\textit{shifted component}) and after it (\textit{unshifted component}). 
The relative intensities  of the unshifted ($I^u$) and shifted ($I^s$) components depend on the ratio between the lifetime of the investigated state and the target-degrader time of flight, which depends on the $\beta_{before}$ velocity and the plunger distance~\cite{dewald2012developing}. 
Specifically, the ratio $R(x) \equiv \frac{I^u}{I^u +I^s}$, called \textit{decay curve}, is described by the Bateman equations. 

The lifetimes of the excited states were extracted using the NAPATAU software~\cite{napatau}, applying the Differential Decay Curve Method (DDCM)~\cite{dewald2012developing} by fitting the area of both the shifted and the unshifted components with a polynomial piecewise function. 
These intensities were scaled according to an external normalization, given by the number of ions identified in VAMOS++~\cite{RDDSions}. 
This normalization is not only proportional to the beam intensity and duration of the measurement, but it also provides a measure of possible degradation of the target during the experiment.
For each $i$-th target-stopper distance the lifetime $\tau_i$ is obtained as 

\begin{equation}
\tau_i = \frac{I^{u}_i -\Sigma_j \left( Br {\,} \alpha {\,} I^{u}_i \right)_j}{\frac{d}{dt}I^{s}_i} \, ,
\label{eq:ddcm}
\end{equation}
where the summation is extended over $j$ feeding transitions, each with a certain branching ratio ($Br$) and parameter $\alpha$, which includes the efficiency correction and the angular correlation between the transition of interest and the feeding one. 
The $\alpha$ parameters were extracted from a $\gamma$-ray energy spectrum obtained by summing the statistics collected for all target-degrader distances~\cite{napatau, dewald2012developing}. 
In the case of the $\gamma$-$\gamma$ coincidence procedure with a gate placed on the feeding transition, the contributions from feeding transitions are eliminated and this term is null.
The final result is given by a weighted average of the lifetimes within the sensitive region of the technique, i.e. where the derivative of the fitting function is largest. 

For the less intense channels, the Decay-Curve Method (DCM) was adopted. 
Since it relies on well-defined fitting functions, whose parameters can be deduced experimentally, this technique permits to measure lifetimes even if the number of experimental points is limited. 
A particular application of the method is the $R_{sum}$ approach~\cite{litzinger2015trans}. 
In this approach, if the statistics is not sufficient to determine the area of the $\gamma$-ray transition components for each target-degrader distance, the spectra obtained for the different target-degrader distances are summed. 
The lifetime is then calculated from the solution of the weighted average of the decay curves $R_j(x_j, \tau)$

\begin{equation}
R_{sum} \equiv \frac{ \Sigma_j I_j^u }{ \Sigma_j ( I_j^u + I_j^s )} = \Sigma_j n_j R(x_j, \tau)
\label{eq:sumBat}
\end{equation}
where $x_j$ denotes the plunger distance and $n_j$ is the normalization factor for each distance. 
The normalization factors $n_j$ were given by the total number of $\gamma$ rays detected by AGATA in time coincidence with the shifted component.

\section{Results}

\begin{figure*}[ht]
    \centering
    \includegraphics[width=0.95\textwidth]{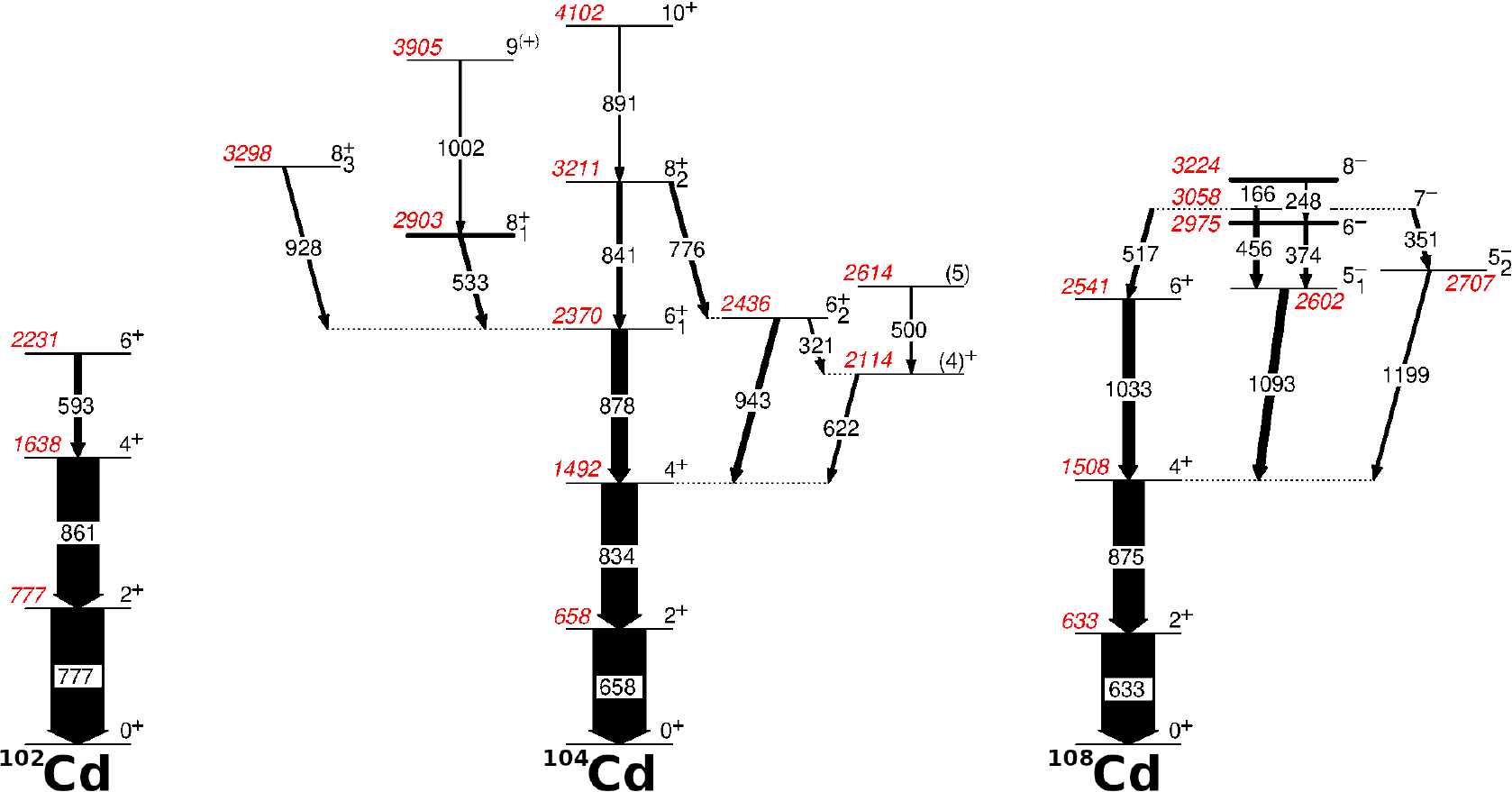}\vspace{8mm}
    \includegraphics[width=0.97\textwidth]{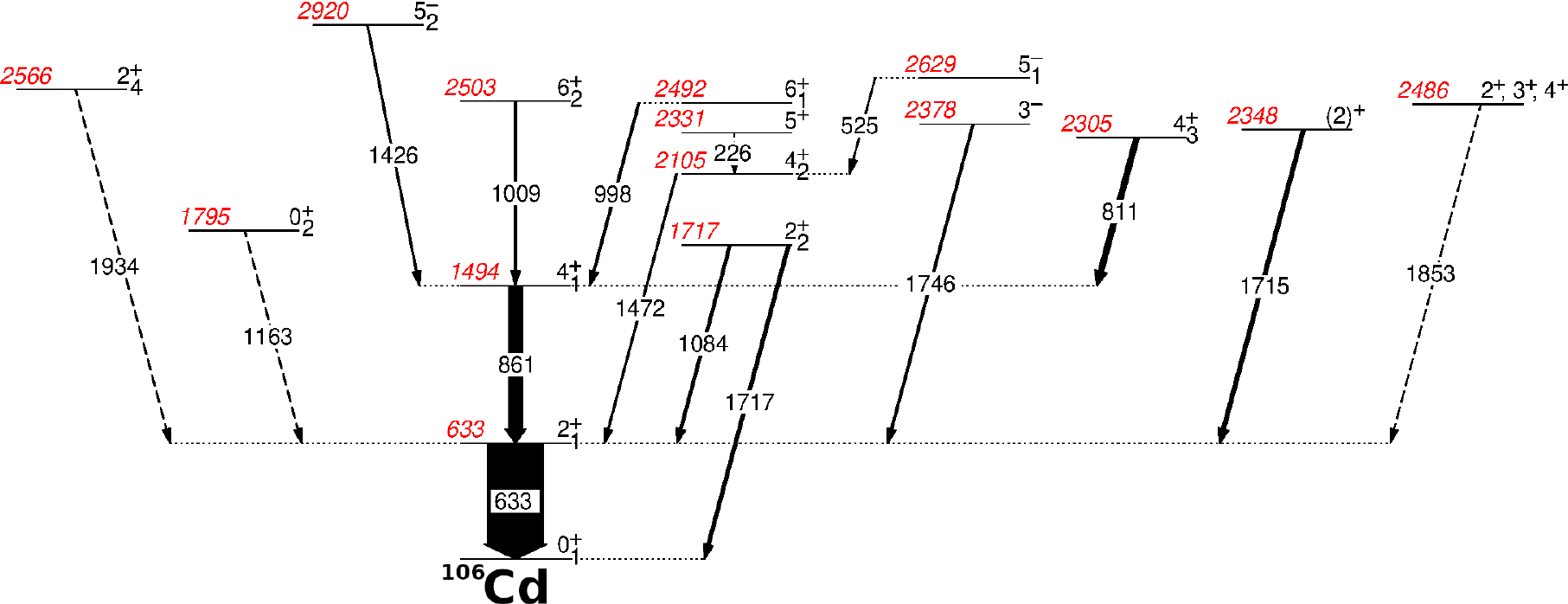}
    %\vspace{-3mm}
    \caption{\label{fig:lvl} (Color online) Partial level scheme of the even-mass $^{102-108}$Cd presenting the transitions observed in the current measurement. The arrow widths represent the efficiency-corrected transition yields normalized over the $2_1^+ \to 0_{g.s.}^+$ one: intensities below 1$\%$ are shown with dashed arrows, while transitions with yield lower than $0.1\%$ and not useful for the analysis have been omitted. The excitation energy of the states is highlighted in red. Spin, parity and excitation energy of the states are assigned according to the NNDC On-Line Data Service from the ENSDF database~\cite{ESNDF} (file revised as of August 2009 for $^{102}$Cd, September 2009 for $^{104}$Cd, June 2008 for $^{106}$Cd, and October 2008 for $^{108}$Cd).}
\end{figure*}

The coupling of the AGATA and VAMOS++ spectrometers represents a powerful tool for high precision spectroscopy. 
By requiring a time coincidence between the identified reaction products and the detected $\gamma$ rays, it is possible to clearly select the channels of interest. 
Additionally, the combination of a MNT reaction, which is a binary mechanism (secondary processes, such as particle evaporation, are negligible), with the complete recoil identification in VAMOS++ allowed us to reconstruct the Total Kinetic-Energy Loss (TKEL) on an event-by-event basis~\cite{brown1951excited}. 
This quantity is proportional to the total excitation energy of the investigated nucleus~\cite{valiente2009lifetime, mengoni2009lifetime}. 
While, in the Sn case, a TKEL gate was applied to control the direct feeding of the states~\cite{Siciliano2019EPJ, siciliano2020plb}, in the present analysis it was used to reduce the possible presence of the inelastically scattered $^{106}$Cd beam, which could contaminate other channels despite the extraordinary performance of the magnetic spectrometer. 

In the following, the lifetime measurements in the even-mass $^{102-108}$Cd isotopes are presented, with each case discussed in detail. 
Figure~\ref{fig:lvl} shows the partial level schemes of the investigated nuclei with the $\gamma$-ray transitions observed in the current measurement. 
Table~\ref{tab:Results} summarizes the measured lifetimes obtained with the different techniques.

%\newpage
\subsection{$^{102}$Cd}

\begin{figure}[ht]
    \includegraphics[width=0.46\textwidth]{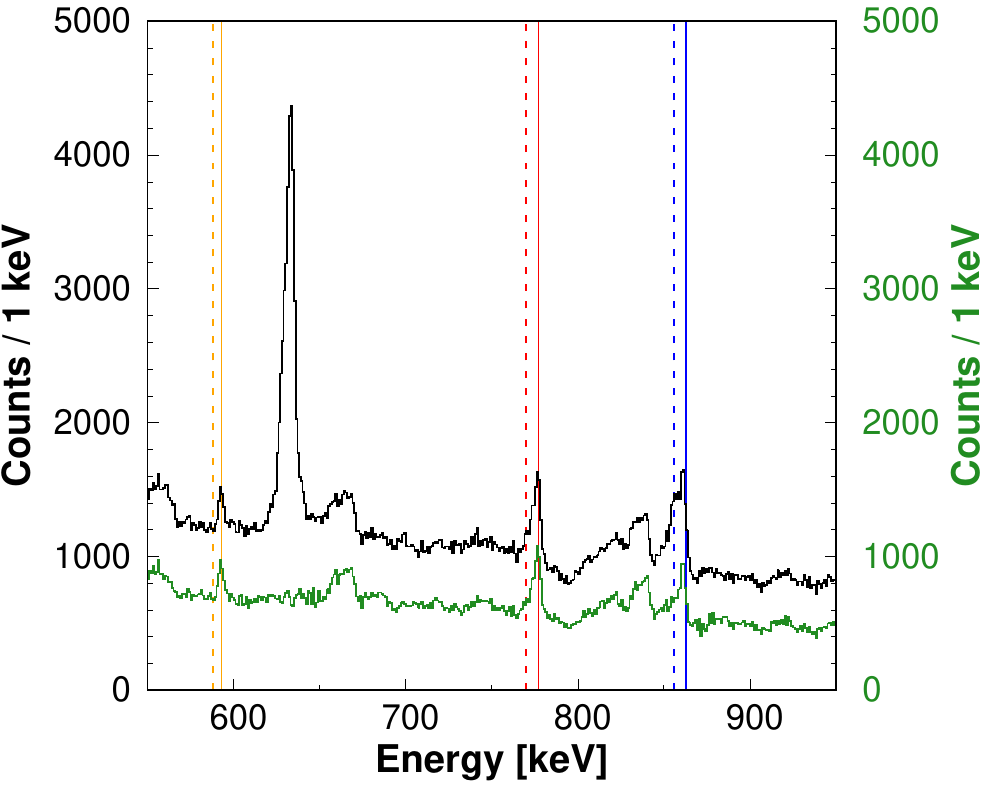}
    \vspace{-3mm}
    \caption{\label{fig:spectrum_102Cd}(Color online) Doppler-corrected $\gamma$-ray energy spectrum of $^{102}$Cd before (black) and after (green) the gate on the Total Kinetic-Energy Loss (TKEL), obtained by summing the statistics of all the target-degrader distances. The $2^+ \to 0_{g.s.}$ (red), $4^+ \to 2^+$ (blue) and $6^+ \to 4^+$ (orange) transitions are marked, indicating the unshifted and shifted centroids with a solid and a dashed line, respectively.}
\end{figure}

In order to reduce the contamination caused by the $^{106}$Cd beam, a TKEL$> 32$~MeV condition was imposed. 
Such threshold permitted to limit as much as possible the presence of $\gamma$-ray peaks related to the inelastically scattered beam, without decreasing the statistics in the transitions of interest. 
In these conditions, the $2_1^+ \to 0_{g.s.}^+$, $4_1^+ \to 2_1^+$ and $6_1^+ \to 4_1^+$ transitions in $^{102}$Cd were clearly identified in the $\gamma$-ray energy spectrum obtained by summing the statistics from all the distances, as shown in Figure~\ref{fig:spectrum_102Cd}.

\begin{figure}[ht]
    \includegraphics[width=0.46\textwidth]{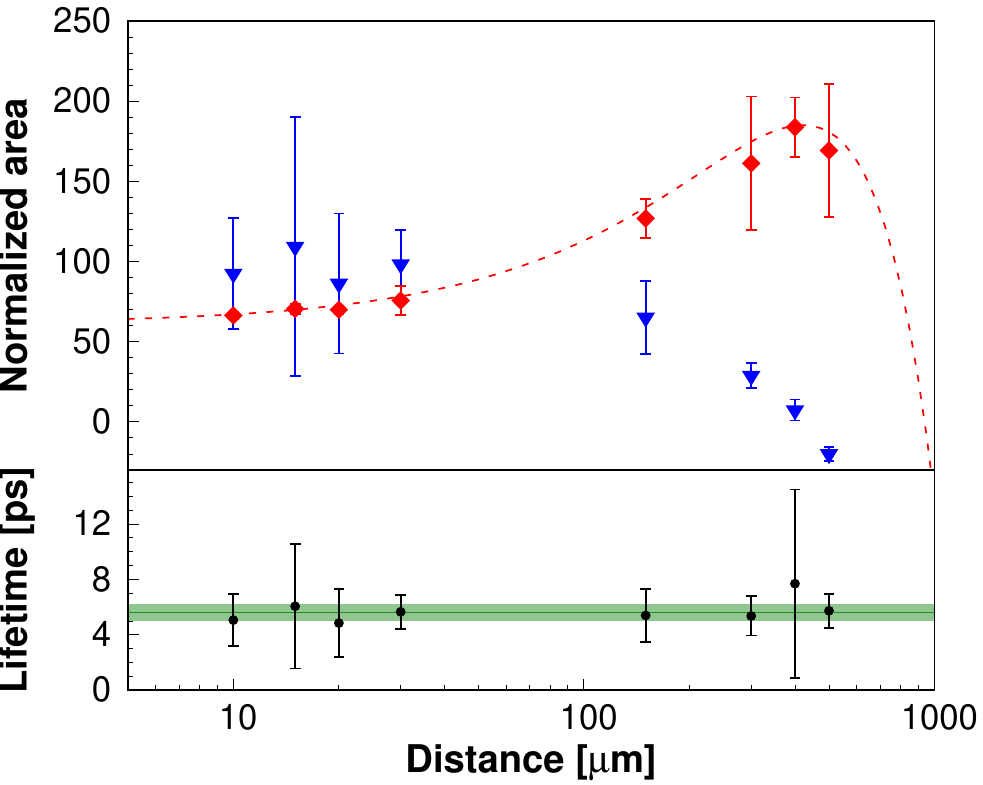}
    \vspace{-3mm}
    \caption{\label{fig:lt_102Cd}(Color online) DDCM analysis for the lifetime measurement of the $2_1^+$ excited state of $^{102}$Cd. (top) Area of the shifted (red diamonds) and feeding-corrected unshifted (blue triangles) components, normalized to the number of ions detected in VAMOS++. The dashed line represents a fit to the shifted-component points. (bottom) Corresponding lifetimes obtained for individual distances. The solid line denotes the weighted average of the lifetimes, while the filled area corresponds to 1$\sigma$ statistical uncertainty.}
\end{figure}

Since the statistics of the $4_1^+ \to 2_1^+$ shifted component were not sufficient for a coincidence measurement, the lifetime of the $2_1^+$ state was obtained via DDCM by subtracting the contribution of the unshifted component of the $4_1^+ \rightarrow 2_1^+$ transition. 
Figure~\ref{fig:lt_102Cd} presents the DDCM analysis, resulting in a lifetime $\tau(2_1^+) = 5.6$~(6)~ps, which is in agreement with literature~\cite{boelaert2007low, ekstrom2009cd}. 

Due to the limited statistics of the $6_1^+ \to 4_1^+$ transition it was not possible to determine the area of the shifted and unshifted components for the individual distances. 
%Therefore, since the contribution of the feeder cannot be evaluated, for the $4_1^+$ state only an upper limit was define for its lifetime, resulting $\tau(4_1^+) < 11$~ps, which is consistent with the threshold previously set in Ref.~\cite{lieb2001proton}.
The lifetime was, therefore, measured via DDCM adopting the so-called ``gate from below'' approach~\cite{PETKOV2001527}, resulting in $\tau(4_1^+) = 3.6$~(12)~ps. 
%However, gating on the unshifted component of the $2_1^+ \to 0_{g.s.}^+$ transition, only the shortest distances were in the sensitive region of the DDCM. 

The results are reported in Table~\ref{tab:Results}.


\subsection{$^{104}$Cd}

Figure~\ref{fig:lvl} presents the partial level scheme of $^{104}$Cd, populated via two-neutrons stripping, showing the transitions observed in the singles $\gamma$-ray energy spectrum and in the $\gamma-\gamma$ matrix obtained by summing the statistics of the different target-degrader distances. 
Due to the complexity of the decay pattern and the presence of $\gamma$-ray transitions with a similar energy, coincidence techniques were necessary to extract the lifetimes of several excited states. 

The lifetime of the $2_1^+$ state, fed by a single transition, was measured via DDCM by subtracting the contribution from the unshifted component of the $4_1^+ \to 2_1^+$ transition,  yielding $\tau(2_1^+) = 9.6$~(3)~ps. 
By gating on the shifted component of the same feeding transition, similar result was obtained via the $R_{sum}$ approach, resulting in $\tau(2_1^+) = 10.0^{+0.6}_{-0.4}$~ps. 
These results are in a perfect agreement with values reported in literature~\cite{muller2001high, boelaert2007low, ekstrom2009cd}.

Due to the presence of multiple feeding transitions, the lifetime of the $4^+_1$ state was measured via DDCM by gating on the unshifted component of the $2_1^+ \to 0_{g.s.}^+$ transition, and via the $R_{sum}$ approach by gating on the shifted component of the 878-keV $6_1^+ \to 4_1^+$ transition. 
The two techniques yielded $\tau(4_1^+) = 1.6$~(5)~ps and $\tau(4_1^+) = 1.44^{+0.33}_{-0.24}$~ps, respectively. 
Both results are compatible with the most accurate and recent measurement, reported in Ref.~\cite{boelaert2007low}. 

Due to the limited statistics and the presence of various feeding transitions, the $6_1^+$ excited state was studied only via the $R_{sum}$ method. 
Unfortunately, due to the close proximity of the unshifted component of the $4_1^+ \to 2_1^+$ transition, a very narrow gate had to be set on the shifted component of the 841-keV $8_2^+ \to 6_1^+$ transition.
The consequent limited statistics was not sufficient to determine the lifetime of the $6_1^+$ state, but an upper limit  $\tau(6_1^+) < 6$ ps could be obtained. %~\cite{muller2001high}.

%\begin{figure}[h]
%    \centering
%    \includegraphics[width=0.48\textwidth]{LevelScheme_104Cd.pdf}
%    \vspace{-5mm}
%    \caption{\label{fig:lvl_104Cd} Partial level scheme of $^{104}$Cd presenting the transitions observed in the current measurement.}
%\end{figure}

The lifetimes of the $2_1^+$, $4_1^+$ and $6_1^+$ states are presented in Table~\ref{tab:Results}.

\subsection{$^{106}$Cd}

%\begin{figure*}[h!]
%    \centering
%    \includegraphics{spectrum_106Cd.eps}
%    \caption{\label{fig:spectrum_106Cd} Doppler-corrected $\gamma$-ray energy spectrum of $^{106}$Cd, obtained by summing up the statistics of all the target-degrader distances. The unshifted components of the identified transitions are marked.}
%\end{figure*}

\begin{figure}[b]
    \includegraphics[width=0.48\textwidth]{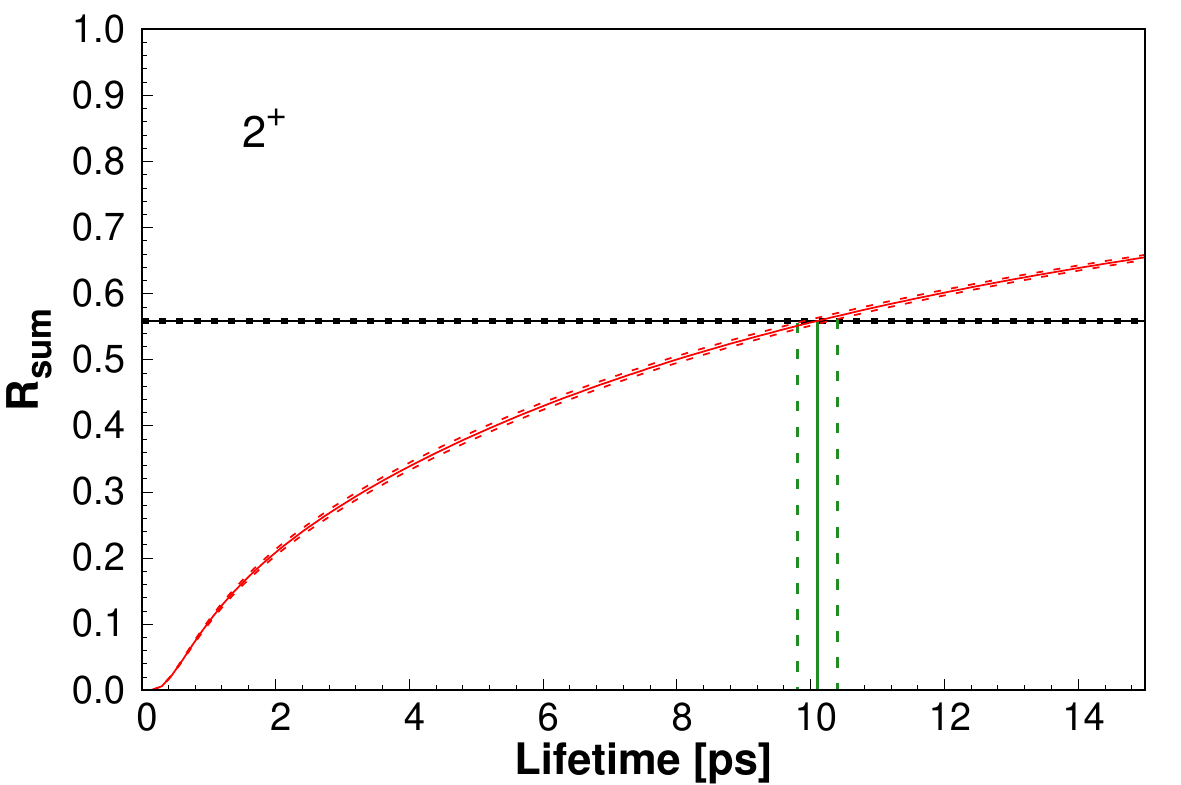}
    \vspace{-5mm}
    \caption{\label{fig:lt_106Cd}(Color online) Decay curve as a function of the lifetime of the $2_1^+$ excited state in $^{106}$Cd, obtained with the $R_{sum}$ approach. The black line represents the experimental value obtained by summing the statistics of all  target-degrader distances and gating on the shifted component of the $4_1^+ \to 2_1^+$ transition. The red curve is the expected value calculated with Equation~\ref{eq:sumBat}. The interception between the experimental and expected values (green line) represents the lifetime of the state. All the dashed curves denote the 1$\sigma$ uncertainty. }
\end{figure}

The lifetime of the $2_1^+$ excited state in $^{106}$Cd, equal to 10.4~(2)~ps, was obtained via DDCM from the present data~\cite{siciliano2017ncc}. 
The same result was obtained via DCM, see Table~\ref{tab:Results}, and the excellent agreement between the results of the two approaches validated the calibration of the plunger device. 
In the $R_{sum}$ approach, since it is based on the Bateman equations, the knowledge of the absolute target-degrader distances is crucial to properly measure lifetimes~\cite{siciliano2017ncc}.
In the spectrum obtained by summing the statistics of all the distances and then gating on the shifted component of the $4_1^+ \to 2_1^+$ transition, the intensity ratio resulted 0.56~(1). 
Considering this experimental value and exponential functions as $R_j(x_j,\tau)$ decay curves, Figure~\ref{fig:lt_106Cd} presents the $R_{sum}$ analysis for the $2_1^+$ state, yielding a lifetime of $\tau(2_1^+) = 10.1\,(3)$~ps, which confirms the validity of this  approach.

The lifetime of the $5_1^-$ excited state was obtained via the $R_{sum}$ approach by gating on the shifted component of the 691-keV transition de-exciting the $6^-$ state. 
The resulting lifetime is $\tau(5^-) = 8.2 (4)$~ps.
The obtained precision is over an order of magnitude better than that of the previous lifetime measurement of this state~\cite{gusinsky1983}.  

Since several $\gamma$-ray transitions were observed feeding the $4_1^+$ state and the statistics in their shifted components were not sufficient for a coincidence measurement, the lifetime of the $4_1^+$ state was extracted via DDCM by gating on the unshifted component of the $2_1^+ \to 0_{g.s.}^+$ transition. 
The analysis yielded $\tau(4_1^+) = 1.4 (2)$~ps and the result is in agreement with the measurements of Refs.~\cite{milner1969, rhodes2021high}.

For the following states: $0_2^+$ ($E_x = 1795$~keV), $2_2^+$ ($E_x = 1717$~keV), $4_3^+$ ($E_x = 2305$~keV), $(2)^+$ ($E_x = 2348$~keV), $6_2^+$ ($E_x = 2503$~keV) and $2^+,3^+,4^+$ ($E_x = 2486$~keV) no feeding transitions were observed in both singles $\gamma$-ray spectra and $\gamma-\gamma$ matrices. 
Therefore, their lifetimes were determined via DCM using an exponential function and, for the most intense channels, via DDCM as well. 

The $4_2^+$ excited state was investigated via DCM using second-order Bateman equations. 
This state was observed, in both singles $\gamma$-ray energy spectra and $\gamma$-$\gamma$ matrices, to be fed only by the $5^+ \to 4_2^+$ and $5_1^- \to 4_2^+$ transitions. 
The direct population of these states, which is a parameter of Bateman equations, was extracted from the $\gamma$-ray spectrum obtained by summing the statistics of all target-degrader distances~\cite{dewald2012developing}. 
%\textcolor{red}{\sout{The direct populations resulting from the efficiency and branching-ratio corrected intensities of the observed transitions  were $59 (4) \%$, $28 (3) \%$ and $14 (3) \%$ for the $4_2^+$, $5^+$ and $5^-_1$ states, respectively.}}
From the efficiency and branching-ratio corrected areas of the 1472-, 226- and 525-keV $\gamma$-rays transitions (see Figure~\ref{fig:lvl}), the direct feeding of the $4_2^+$, $5^+$ and $5^-_1$ states resulted $59 (4) \%$, $28 (3) \%$ and $14 (3) \%$, respectively. 
Taking into account the known lifetime $\tau(5^+) = 0.9 (3)$~ns~\cite{ANDREJTSCHEFF1985167} and the $\tau(5_1^-)$ measured in the present work, the lifetime of the $4_2^+$ level was determined to be $\tau(4_2^+) = 4.1 (7)$~ps. 

\begin{figure}[h]
    \centering
    \includegraphics[width=0.48\textwidth]{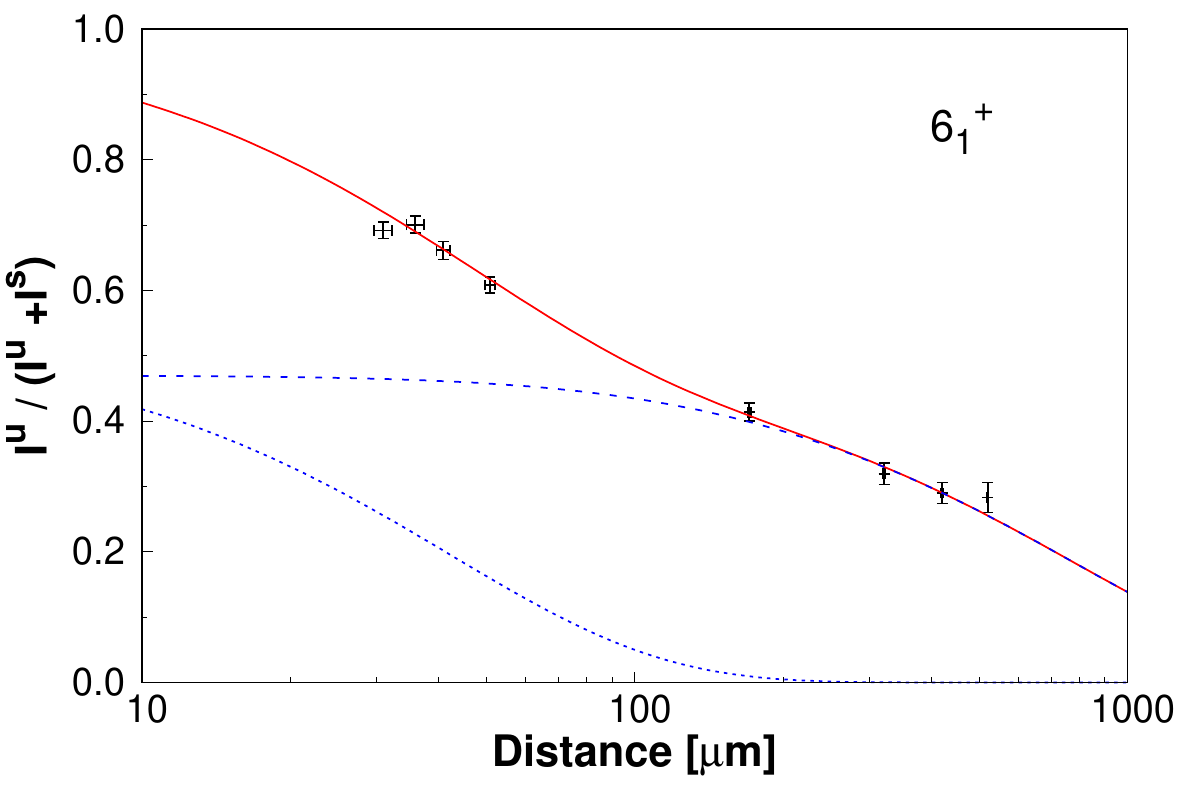}
    \vspace{-5mm}
    \caption{\label{fig:lt_106Cd-6}(Color online) Ratio of the  component intensities as a function of the target-degrader distance for the $6^+_1 \to 4_1^+$ transition in $^{106}$Cd. The solid red line represents the fitted decay curve obtained with second-order Bateman equations, whose components are shown with dashed and dotted blue lines for the feeder and the direct population, respectively.}
\end{figure}

For the $6_1^+$ excited state, no feeding transitions were  observed in singles $\gamma$-ray spectra and $\gamma-\gamma$ matrices. 
On the other hand, the decay curve of the $6_1^+ \to 4_1^+$ transition (see Figure~\ref{fig:lt_106Cd-6}) suggested feeding from a state with a lifetime longer than that of the $6_1^+$ state. 
Assuming a two-step decay cascade, the lifetime of the $6_1^+$ state was determined to be $\tau(6_1^+) = 1.3 (2)$~ps, while for the feeding state a value of 24~(5)~ps was obtained. 
As the nature of this feeding level is unknown, an upper limit of 2 ps for the $6_1^+$ lifetime was determined by fitting the decay curve with an exponential function. 
%Under such condition, the effective lifetime is $\tau_{eff}(6_1^+) = 2$~ps. 

An upper limit of 0.3~ps can be set for the lifetimes of the $2_4^+$, $3^-$ and $5_2^-$ excited states, since in the $\gamma$-ray spectra only the shifted components of the $2_4^+ \to 2_1^+$, $3^- \to 2_1^+$ and $5_2^- \to 4_1^+$ transitions were observed for the shortest plunger distances. 
The same limit can be determined for a state de-exciting via a $\approx 2080$-keV transition. 
Unfortunately, this decay may originate from either of the levels at the excitation energy of 2718~keV and 2721~keV, so it is not possible to unambiguously attribute this upper limit. 

The results obtained via both DCM and DDCM are summarized in Table~\ref{tab:Results}.

\subsection{$^{108}$Cd}

In both singles $\gamma$-ray energy spectra and $\gamma$-$\gamma$ matrices no contamination from the $^{106}$Cd beam was observed. 
This result is important for the study of $^{108}$Cd, since the two isotopes present similar structures with $\gamma$-ray transitions very close in energy. 
Figure~\ref{fig:lvl} shows the partial level scheme of $^{108}$Cd, indicating the observed transitions. 

%\begin{figure}[h]
%    \centering
%    \includegraphics[width=0.35\textwidth]{LevelScheme_108Cd.pdf}
%    \caption{\label{fig:lvl_108Cd} Partial level scheme of $^{108}$Cd presenting the transitions observed in the current measurement.}
%\end{figure}

The lifetime of the $2_1^+$ excited state was measured with the $R_{sum}$ method by gating on the shifted component of the $4_1^+ \to 2_1^+$ transition, resulting in $\tau(2_1^+) = 10^{+3}_{-2}$~ps. 
The large uncertainty of the lifetime is mostly due to the limited statistics resulting from the use of $\gamma-\gamma$ coincidences, even though the statistics of all target-degrader distances were summed together. 
Therefore a DDCM analysis  for this lifetime was performed by subtracting the intensity of the unshifted component of the $4_1^+ \to 2_1^+$ transition, yielding $\tau(2_1^+) = 10.8$~(9)~ps.

Due to the limited statistics and the presence of various feeding transitions, the lifetime of the $4_1^+$ state was obtained via DDCM by gating on the unshifted component of the $2_1^+ \to 0_{g.s.}^+$ transition. 
This approach yielded $\tau(4_1^+) = 1.4$~(5)~ps, which is in agreement with the result reported in Ref.~\cite{milner1969}.

Table~\ref{tab:Results} reports the lifetimes extracted for the $2_1^+$ and $4_1^+$ states.

\begin{table*}
\caption{\label{tab:Results} Measured lifetimes of the excited states $I_i^\pi$ in the even-mass $^{102-108}$Cd isotopes. 
The lifetimes are extracted from the $I_i^\pi \to I_f^\pi$ transitions and the results are compared with literature values. The results from Ref.~\cite{siciliano2017ncc} were presented in a former publication but they are from the current experiment. From the weighted average of the results belonging to the current work and literature, the reduced transition probabilities $B(E2; I_i^\pi \to I_f^\pi$) and $B(M1; I_i^\pi \to I_f^\pi$) are calculated and compared with the SCCM predictions. For $^{106}$Cd, branching and mixing ratios are taken from the ENSDF database of NNDC On-Line Data Service ~\cite{ESNDF} (file revised as of June 2008). }
\vspace*{3mm}
\begin{tabular}{l | c c c | c c c c | c c c | c c c}
\hline%\toprule
				&   \multirow{2}{*}{$I_i^\pi$}	
				&	\multirow{2}{*}{$I_f^\pi$}	
				&   \multirow{2}{*}{E$_\gamma$ [keV]}	%%
				&	\multicolumn{3}{c}{$\tau$ [ps]}	
				&   %%
				&   \multicolumn{2}{c}{$B(E2)$ [$e^2$fm$^4$]}
				&
				&   \multicolumn{2}{c}{$B(M1)$ [$\mu$N$^2$] $\times 10^3$}
				&   \tstrut\bstrut \\ 
\cmidrule(lr){5-7}
\cmidrule(lr){9-10}
				&
				&
				&
				&	DDCM	
				&   DCM 	
				&   Literature  
				&   
				&   Exper.
				&   SCCM
				&   
				&   Exper.
				&   SCCM
				&   \tstrut\bstrut \\ 
\hline%\toprule
\multirow{3}{*}{$^{102}$Cd}	&	\multirow{3}{*}{$2_1^+$}		
                            &	\multirow{3}{*}{$0_{g.s.}^+$}		
                            &	\multirow{3}{*}{777}      
                            &	\multirow{3}{*}{5.6 (6)}
                            &   \multirow{3}{*}{-}
                            &   $<8.1$~\cite{lieb2001proton}
                            &   
                            &   \multirow{3}{*}{513 (31)}
                            &   \multirow{3}{*}{895}
                            &   
                            &   \multirow{3}{*}{-}
                            &   \multirow{3}{*}{-}
                            &   \tstrut \\
            	            &
            	            &
                            &	      
                            &	
                            &   
                            &	5.9 (5)~\cite{boelaert2007low}
                            &
                            &
                            &
                            &   
                            &
                            &
                            &   \\    
                            &
            	            &
                            &	      
                            &	
                            &   
                            &	5.2 (7)~\cite{ekstrom2009cd}
                            &
                            &
                            &
                            &   
                            &
                            &
                            &   \bstrut \\    
                            &	$4_1^+$
                            &   $2_1^+$
                            &	861      
                            &	3.6 (12)
                            &   -
                            &	$<8.1$~\cite{lieb2001proton}	
                            &
                            &   479$^{+240}_{-120}$
                            &   1396
                            &
                            &   -
                            &   -
                            &   \tstrut\bstrut \\
\hline%\toprule
\multirow{6}{*}{$^{104}$Cd}	&	\multirow{3}{*}{$2_1^+$}		
                            &   \multirow{3}{*}{$0_{g.s.}^+$}		
                            &	\multirow{3}{*}{658} 
                            &   \multirow{3}{*}{9.6(3)}	
                            &	\multirow{3}{*}{$10.0^{+0.6}_{-0.4}$}			
                            &	9 (3)~\cite{muller2001high} 
                            &
                            &   \multirow{3}{*}{741 (13)}
                            &   \multirow{3}{*}{999}
                            &   
                            &   \multirow{3}{*}{-}
                            &   \multirow{3}{*}{-}
                            &   \tstrut \\
                        	&
                        	&
                            &	 
                            &   
                            &				
                            &	8.5 (12)~\cite{boelaert2007low}
                            &
                            &
                            &
                            &   \\
                            &
                            &			
                            &	 
                            &   
                            &				
                            &	8.5 (2)~\cite{ekstrom2009cd} 
                            &
                            &
                            &
                            & \bstrut \\
				            &	\multirow{2}{*}{$4_1^+$}	
				            &	\multirow{2}{*}{$2_1^+$}	
		    	    	    &	\multirow{2}{*}{834}	
		    	    	    &   \multirow{2}{*}{1.6 (5)}
			            	&	\multirow{2}{*}{$1.44^{+0.33}_{-0.24}$}
				            &  	$<6$~\cite{muller2001high} 
				            &
				            &   \multirow{2}{*}{1367 (202)}
				            &   \multirow{2}{*}{1535}
				            &
				            &   \multirow{2}{*}{-}
				            &   \multirow{2}{*}{-}
				            &   \tstrut \\ 
				            &	
				            &
                            &	 
                            &   
                            &				
                            &	1.5 (5)~\cite{boelaert2007low} 
                            &
                            &
                            &
                            & 
                            &
                            &
                            &   \bstrut \\
                            &	$6_1^+$		
                            &	$4_1^+$		
                            &	878
                            &   -
                            &	$<6$			
                            &	$<6$~\cite{muller2001high}  
                            &
                            &   $>261$
                            &   -
                            &   
                            &   -
                            &   -
                            &   \tstrut\bstrut \\
\hline
\multirow{19}{*}{$^{106}$Cd} &	$0_2^+$		
                            &	$2_1^+$		
                            &	1163 
                            &   1.7 (5)   
                            &	1.3 (1)			
                            &	-         
                            &
                            &   1463 (110)
                            &   4
                            &
                            &   -
                            &   -
                            &   \tstrut\bstrut\\
                            &   \multirow{5}{*}{$2_1^+$}
                            &	\multirow{5}{*}{$0_{g.s.}^+$}
                            &	\multirow{5}{*}{633}	
                            &	\multirow{5}{*}{10.4 (2)~\cite{siciliano2017ncc}}	
                            &   \multirow{4}{*}{10.7 (4)~\cite{siciliano2017ncc}}
                            &	9.4 (4)~\cite{milner1969}
                            &
                            &   \multirow{5}{*}{796 (6)}
                            &   \multirow{5}{*}{1056}
                            &
                            &   \multirow{5}{*}{-}
                            &   \multirow{5}{*}{-}
                            &   \tstrut \\
                            &			
                            &			
                            &	 
                            &   
                            &   \multirow{4}{*}{10.1 (3)}
                            &	10.1 (8)~\cite{KLEINFELD197081} 
                            &
                            &
                            &
                            &
                            &
                            &
                            &   \\
                            &			
                            &			
                            &	 
                            &   
                            &				
                            &	10.5 (1)~\cite{esat1976mass} 
                            &
                            &
                            &
                            &
                            &
                            &
                            &   \\
                            &			
                            &			
                            &	 
                            &   
                            &				
                            &	7.0 (3)~\cite{benczerkoller2016} 
                            &
                            &
                            &
                            &
                            &
                            &
                            &   \\
                            &			
                            &			
                            &	 
                            &   
                            &				
                            &	9.5 (3)~\cite{rhodes2021high} 
                            &
                            &
                            &
                            &
                            &
                            &
                            &   \bstrut \\                            
                            &	\multirow{3}{*}{$2_2^+$}		
                            &	\multirow{2}{*}{$0_{g.s.}^+$}		
                            &	\multirow{2}{*}{1717 }
                            &   \multirow{2}{*}{0.46 (10)}   
                            &	\multirow{2}{*}{0.51 (2)}			
                            &	0.45 (7)~\cite{milner1969}
                            &   
                            &   \multirow{2}{*}{70 (11)}
                            &   \multirow{2}{*}{104}
                            &
                            &   \multirow{2}{*}{-}
                            &   \multirow{2}{*}{-}
                            &   \tstrut\\   
                            &			
                            &	\multirow{2}{*}{$2_1^+$}		
                            &	\multirow{2}{*}{1084}
                            &   \multirow{2}{*}{0.41 (4)}  
                            &	\multirow{2}{*}{0.55 (3)}			
                            &	0.28 (2)~\cite{benczerkoller2016}
                            &
                            &   \multirow{2}{*}{375 (17)}
                            &   \multirow{2}{*}{346}
                            &
                            &   \multirow{2}{*}{15 (3)}
                            &   \multirow{2}{*}{1.2}
                            &   \\ 
                            &			
                            &			
                            &	 
                            &      
                            &				
                            &	0.50 (3)~\cite{rhodes2021high}
                            &
                            &   
                            &   
                            &
                            &
                            &
                            &   \bstrut\\ 
                            &	$2_4^+$		
                            &	$2_1^+$		
                            &	1934 
                            &   -   
                            &	$< 0.3$			
                            &	- 
                            &
                            &   $> 87$
                            &   -
                            &
                            &   $> 3.6$
                            &   -
                            &   \tstrut\bstrut\\ 
                            &	$(2)^+$		
                            &	$2_1^+$		
                            &	1715 
                            &   -   
                            &	0.59 (2)			
                            &	-
                            &
                            &   -
                            &   -
                            &
                            &   18.9 (6)
                            &   -
                            &   \tstrut\bstrut\\   
                            &	$2^+$,$3^+$,$4^+$ 		
                            &	$2_1^+$		
                            &	1853 
                            &   2.2 (3)   
                            &	2.4 (2)			
                            &	-          
                            &
                            &   $7.8^{+2.0}_{-1.3}$
                            &   -
                            &
                            &   -
                            &   -
                            &   \tstrut\bstrut\\   
                            &	$3^-$		
                            &	$2_1^+$		
                            &	1746 
                            &   -   
                            &	$< 0.3$			
                            &	0.16 (1)~\cite{benczerkoller2016}     
                            &
                            &   -
                            &   -
                            &
                            &   -
                            &   -
                            &   \tstrut\bstrut\\   
				            &	\multirow{3}{*}{$4_1^+$}		
				            &	\multirow{3}{*}{$2_1^+$}		
				            &	\multirow{3}{*}{861}	
				            &	\multirow{3}{*}{1.4 (2)}	
				            &   \multirow{3}{*}{-}
				            &	1.26 (16)~\cite{milner1969} 
				            &
				            &   \multirow{3}{*}{1159 (46)}
				            &   \multirow{3}{*}{1992}
				            &
				            &   \multirow{3}{*}{-}
				            &   \multirow{3}{*}{-}
				            &   \tstrut \\
				            &	
				            &
                            &	 
                            &   
                            &				
                            &	2.5 (2)~\cite{benczerkoller2016} 
                            &
                            &
                            &
                            &   \\
                            &	
				            &
                            &	 
                            &   
                            &				
                            &	1.42 (7)~\cite{rhodes2021high}
                            &
                            &
                            &
                            &   \bstrut\\
				            &	\multirow{2}{*}{$4_2^+$}
				            &	\multirow{2}{*}{$2_1^+$}		
				            &	\multirow{2}{*}{1472}
				            &	\multirow{2}{*}{-}	
				            &   \multirow{2}{*}{4.1 (7)}
				            &	$\leq 2.9$~\cite{gusinsky1983}
				            &
				            &   \multirow{2}{*}{$10.5_{-1.8}^{+2.5}$}
				            &   \multirow{2}{*}{43}
				            &   
				            &   \multirow{2}{*}{-}
				            &   \multirow{2}{*}{-}
				            &   \tstrut \\
				            &	
				            &
                            &	 
                            &   
                            &				
                            &	$>10$~\cite{benczerkoller2016} 
                            &
                            &
                            &
                            &   \bstrut\\
				            &	$4_3^+$		
				            &	$4_1^+$		
				            &	811	
				            &	-	
				            &   1.1 (1)
				            &	1.1 (1)~\cite{benczerkoller2016} 
				            &
				            &   6.8 (24)
				            &   -
				            &
				            &   11 (7)
				            &   -
				            &   \tstrut\bstrut \\
				            &	$5_1^-$		
				            &	$4_2^+$		
				            &	525	
				            &	-	
				            &   8.2 (4)
				            &	$7^{+6}_{-3}$~\cite{gusinsky1983}	 
				            &
				            &   -
				            &   -
				            &
				            &   -
				            &   -
				            &   \tstrut\bstrut \\
				            &	$5_2^-$		
				            &	$4_1^+$		
				            &	1426	
				            &	-	
				            &   $< 0.3$
				            &	-	 
				            &
				            &   -
				            &   -
				            &
				            &   -
				            &   -
				            &   \tstrut\bstrut \\
				            &	$6_1^+$		
				            &	$4_1^+$		
				            &	998	
				            &	-	
				            &   $< 2$
				            &	0.54 (8)~\cite{rhodes2021high}	 
				            &
				            &   145 (21)
				            &   21
				            &
				            &   -
				            &   -
				            &   \tstrut\bstrut \\
				            &	$6_2^+$		
				            &	$4_1^+$			
				            &	1009	
				            &	1.3 (6)	
				            &   1.21 (15)
				            &	-	 
				            &
				            &   $627^{+268}_{-144}$
				            &   1900
				            &
				            &   -
				            &   -
				            &   \tstrut\bstrut \\
\hline
\multirow{3}{*}{$^{108}$Cd} &	\multirow{3}{*}{$2_1^+$}	
                            &	\multirow{3}{*}{$0_{g.s.}^+$}	
                            &	\multirow{3}{*}{633}	
                            &	\multirow{3}{*}{10.8 (9)}		
                            &   \multirow{3}{*}{$10^{+3}_{-2}$}
                            &	9.9 (1)~\cite{esat1976mass} 
                            &
                            &   \multirow{3}{*}{815 (8)}
                            &   \multirow{3}{*}{1028}
                            &
                            &   \multirow{3}{*}{-}
                            &   \multirow{3}{*}{-}
                            &   \tstrut \\
                            &	
                            &
                            &	 
                            &   
                            &				
                            &	9.1 (4)~\cite{milner1969} 
                            &
                            &
                            &
                            &
                            &
                            &
                            &   \\
                            &	
                            &
                            &	 
                            &   
                            &				
                            &	10.1 (8)~\cite{THORSLUND1994306} 
                            &
                            &
                            &
                            &
                            &
                            &
                            &   \bstrut \\
				            &	\multirow{2}{*}{$4_1^+$}		
				            &	\multirow{2}{*}{$2_1^+$}		
				            &	\multirow{2}{*}{876	}
				            &	\multirow{2}{*}{1.4 (5)}	
				            &   \multirow{2}{*}{-}
                            &   1.28 (16)~\cite{milner1969} 
                            &
                            &   \multirow{2}{*}{1228 (145)}
                            &   \multirow{2}{*}{1589}
                            &
                            &   \multirow{2}{*}{-}
                            &   \multirow{2}{*}{-}
                            &   \tstrut \\
                            &	
                            &
                            &	 
                            &   
                            &				
                            &	$<5$~\cite{THORSLUND1994306} 
                            &
                            &
                            &
                            &
                            &
                            &
                            &   \bstrut \\
\hline%\bottomrule
\end{tabular}
\end{table*}

\section{Discussion}

In view of the measured lifetimes, the even-mass Cd nuclei were studied within a self-consistent beyond-mean-field framework~\cite{PS_91_073003_2016, JPG_46_013001_2019}, i.e. the symmetry-conserving configuration mixing (SCCM)~\cite{rodriguez2010triaxial, rodriguez2014structure} method, with the Gogny-D1S~\cite{PhysRevC.21.1568, BERGER1991365} interaction. %\textcolor{red}{\sout{and a model space consisting of $N_{h.o.}=9$ harmonic-oscillator orbitals.}} 
The calculations are based on the mixing of a set of intrinsic states with different quadrupole (axial and non axial) deformations. 
These states are Hartree-Fock-Bogoliubov (HFB) like wave functions obtained self-consistently through the particle-number variation-after-projection (PNVAP) method~\cite{anguiano2001particle}.
Since the HFB states break the rotational invariance of the system, this symmetry is consequently restored by projecting onto good angular momentum (particle-number and angular-momentum projection, PNAMP). 
The final spectrum and the nuclear wave functions are obtained by mixing such PNAMP states within the generator coordinate method.

\begin{figure*}[ht]
    \centering
    \includegraphics[width=0.88\textwidth]{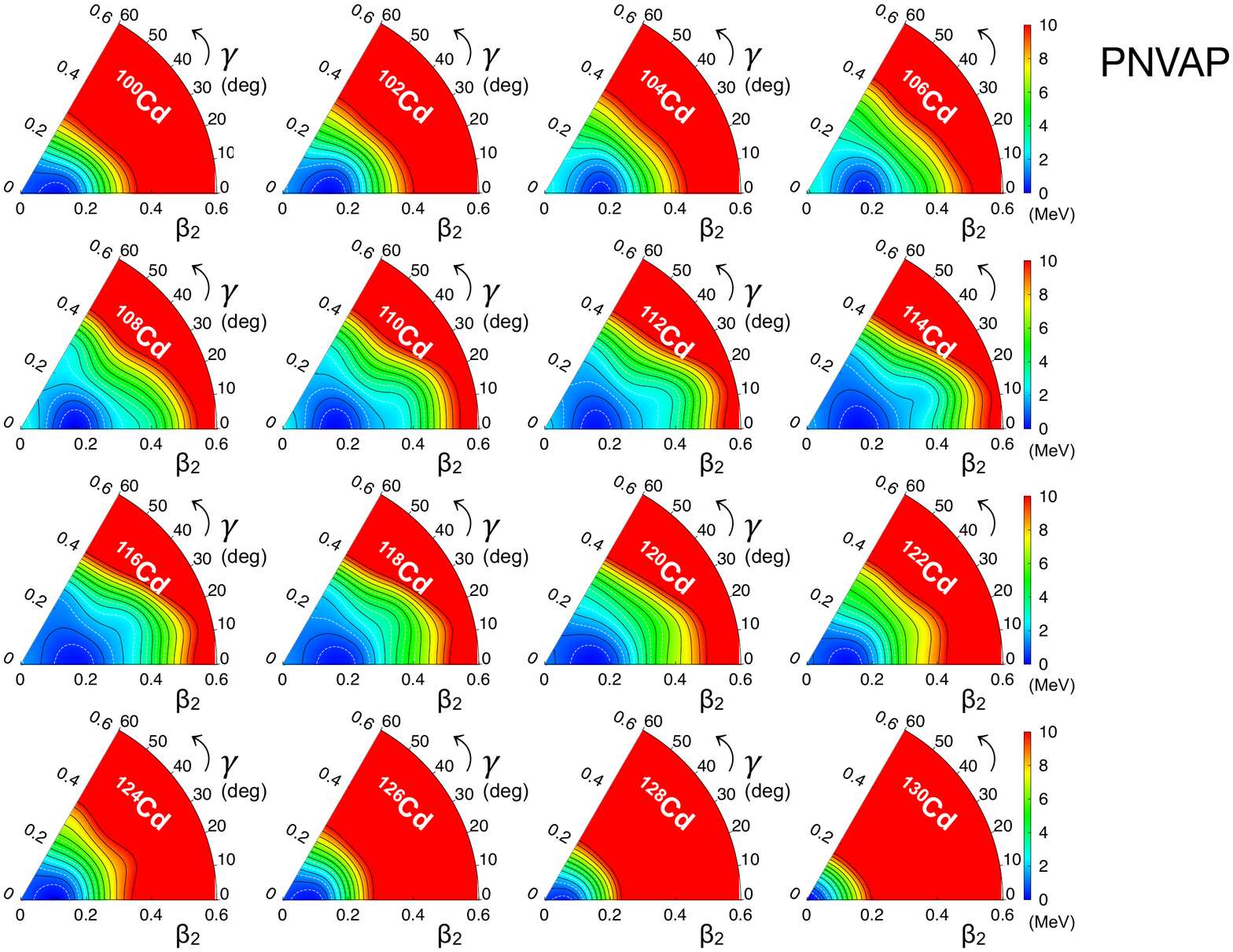}
    \vspace{-8mm}
    \caption{\label{fig:SCCM-PNVAP} (Color online) PNVAP potential energy surfaces as a function of the ($\beta_2$,~$\gamma$) deformation parameters for the even-mass $^{100-130}$Cd isotopes. The results are obtained with the Gogny-D1S interaction within the SCCM approach.}
\end{figure*}

A first estimation of the structure of the Cd isotopes can be obtained by analysing the calculated potential-energy surfaces (PES) as a function of deformation parameters. 
Figure~\ref{fig:SCCM-PNVAP} presents the PNVAP energies as a function of the ($\beta_{2}$, $\gamma$) deformation parameters for the even-mass $^{100-130}$Cd. 
For all studied isotopes a well defined prolate minimum with $\beta_2 = 0.1-0.2$ is present in the PES, except for $^{128-130}$Cd, which exhibit  practically spherical minima, due to the vicinity of the $N=82$ shell closure. 
Moreover, for the $^{110-118}$Cd isotopes a  shallow second triaxial-prolate  minimum with ($\beta_2$, $\gamma$)=(0.3, $20^\circ$) is obtained. %becomes more pronounced, competing with the ground-state prolate configuration. 

\begin{figure}[h]
    \centering
    \includegraphics[width=0.48\textwidth]{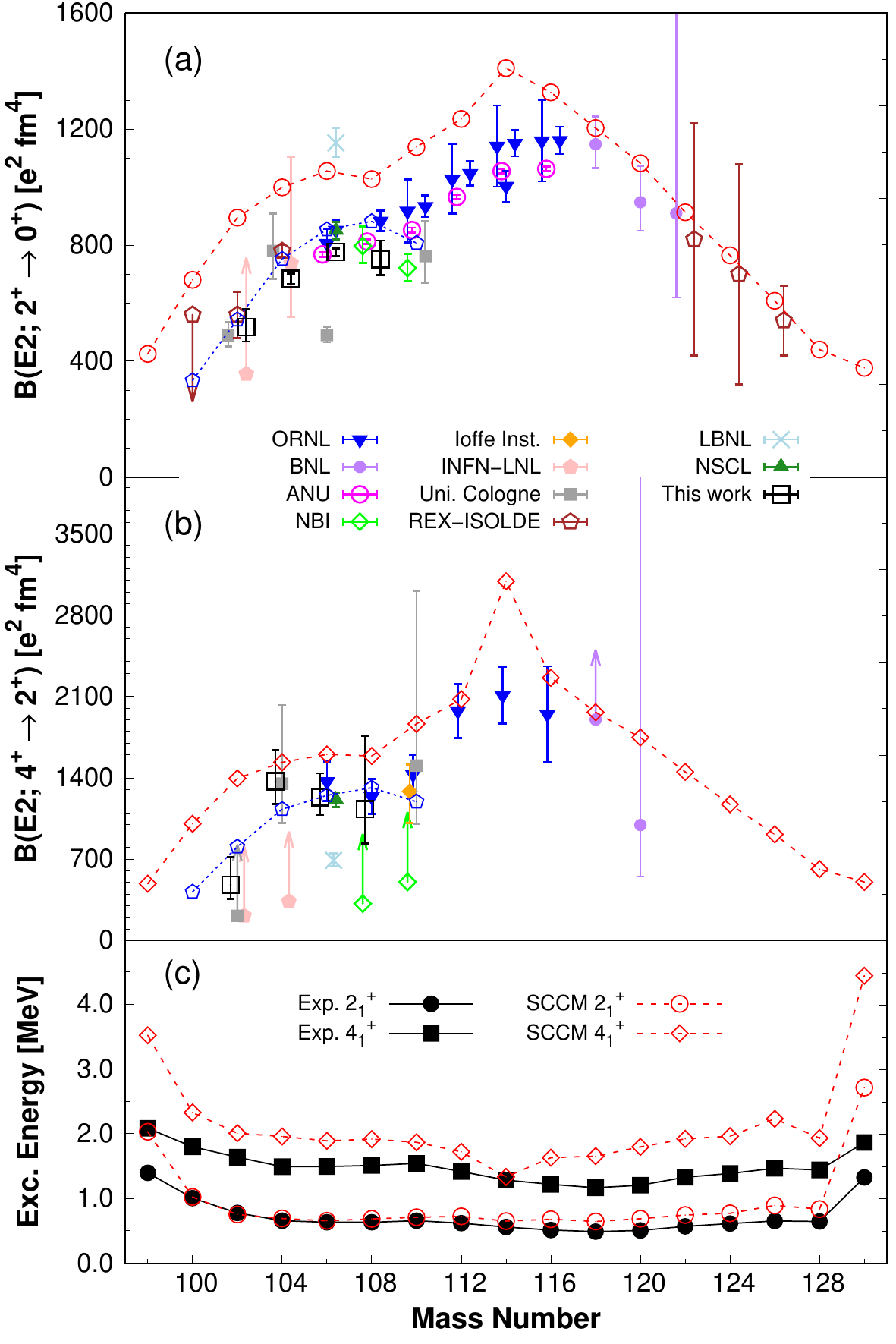}
    \vspace{-6mm}
    \caption{\label{fig:Syst} (Color online) Reduced transition probabilities (a) $B(E2; 2_1^+ \to 0_{g.s.}^+)$ and (b) $B(E2; 4_1^+ \to 2_1^+)$, and (c) 2$^+_1$ and 4$^+_1$ excitation energy systematics for even-mass Cd isotopes. The experimental results~\cite{MCGOWAN1965, milner1969, KLEINFELD197081, esat1976mass, mach1989, piiparinen1993, THORSLUND1994306, zamfir1995, lobach1999, muller2001high, lieb2001proton, Harissopulos2001, boelaert2007low, ashley2007, ekstrom2009cd, ilieva2014, benczerkoller2016, rhodes2021high} are compared with the recent Large-Scale Shell-Model (LSSM) calculations of Ref.~\cite{zuker2020CdSn} (blue open pentagons) and the present SCCM predictions (red open circles and squares). The results of this work (black open squares) are obtained by adopting the weighted average of the DCM and DDCM lifetimes, reported in Table~\ref{tab:Results}. }
\end{figure}

The final theoretical spectra, obtained by mixing intrinsic HFB-like states with different quadrupole deformations, are presented in Figure~\ref{fig:Syst}(c), which shows the systematics of the excitation energies for the $2_1^+$ and $4_1^+$ states. 
The theoretical predictions correctly reproduce the trends observed experimentally for the $2_1^+$ and $4_1^+$ states, although they overestimate the absolute values, especially for the $2_1^+$ energies above $N=64$ and for the $4_1^+$ energies in the whole range of neutron numbers.
This is a well-known effect in the present form of the SCCM method where only static intrinsic shapes are considered in the mixing. 
Thus, the ground state is variationally favored with respect to the excited states and, as a result, a stretched spectrum is obtained. 
A better approach would be an SCCM method that includes intrinsically rotating (cranking) states. 
Within such a framework it is possible to explore on an equal footing collective ground and excited states and the variational approach does not produce such a stretching~\cite{Borrajo2015cranking}. 
However, this improvement is very demanding from the computational point of view, especially for nuclei in this medium-mass region~\cite{Rodriguez2016cranking}. 
Nevertheless, the inclusion of the triaxial degree of freedom in the SCCM calculations improves significantly the agreement with the experimental data with respect to previous axial calculations~\cite{rodriguez2008cadmium}. 
Notably, the intriguing lowering of the $2_1^+$ energy from $^{126}$Cd to $^{128}$Cd is still reproduced by the present calculations. 
In the vicinity of a shell closure,  the excitation energy of the $2_1^+$ state is expected to increase and display a parabolic trend as a function of nucleon number. 
Not only such a parabolic increase has not been observed experimentally, but the excitation energy of the $2_1^+$ states slightly decreases. 
This pattern was reproduced, for the first time, by the previous axial calculations~\cite{rodriguez2008cadmium} and a better agreement is found with the present ones.

In Figure~\ref{fig:Syst} the experimental $B(E2; 2_1^+ \to 0_{g.s.}^+)$ and $B(E2; 4_1^+ \to 2_1^+)$ strengths are compared to the theoretical results of SCCM, together with the predictions of Ref.~\cite{zuker2020CdSn} for the neutron-deficient isotopes. 
An unusual behaviour is found for $^{114}$Cd, where prolate and triaxial-prolate configurations cross for $J^\pi=4^+$ and their corresponding $2^{+}$ states show a mild mixing between these shapes. 
The overall effect of this mixing is the decrease (increase) of the $4^{+}$ energy of the triaxial-prolate (prolate) configuration. 
Moreover, the overlap between the yrast $4^{+}$ and $2^{+}$ states is smaller but is found at larger $\beta_{2}$ values than their neighbors, producing a net increase of the $B(E2; 4^+_1 \to 2^+_1)$ reduced transition probability.  
%\textcolor{red}{\sout{, leading to a decrease of the $4^+_1$ excitation energy and a consequent increase of the $B(E2; 4^+_1 \to 2^+_1)$ reduced transition probability.}} 
Except for this single case, the calculated strengths of the $2_1^+ \to 0_{g.s.}^+$ and $4_1^+ \to 2_1^+$ transitions well reproduce the trend of the experimental results, slightly overestimating the $\beta_2$ deformation. 
This slight overestimation is a plausible explanation for the almost perfect reproduction of the $2_1^+$ excitation energies in $^{100-110}$Cd. 
The theoretical values, indeed, should be larger than the experimental ones for a SCCM method without cranking terms. 

%\begin{figure*}[p]
%    \centering
%    \includegraphics[width=0.82\textwidth]{SCCM_Cadmiums-0.pdf}
%    \caption{\label{fig:SCCM-CWF0} (Color online) Collective wave functions (CWF) as a  function  of  the ($\beta_2$, $\gamma$) deformation parameters for the $0_1^+$ states in the even-mass $^{100-130}$Cd isotopes. The results are obtained with the Gogny-D1S interaction within the SCCM approach.}
%\end{figure*}
\begin{figure*}[p]
    \centering
    \includegraphics[width=0.88\textwidth]{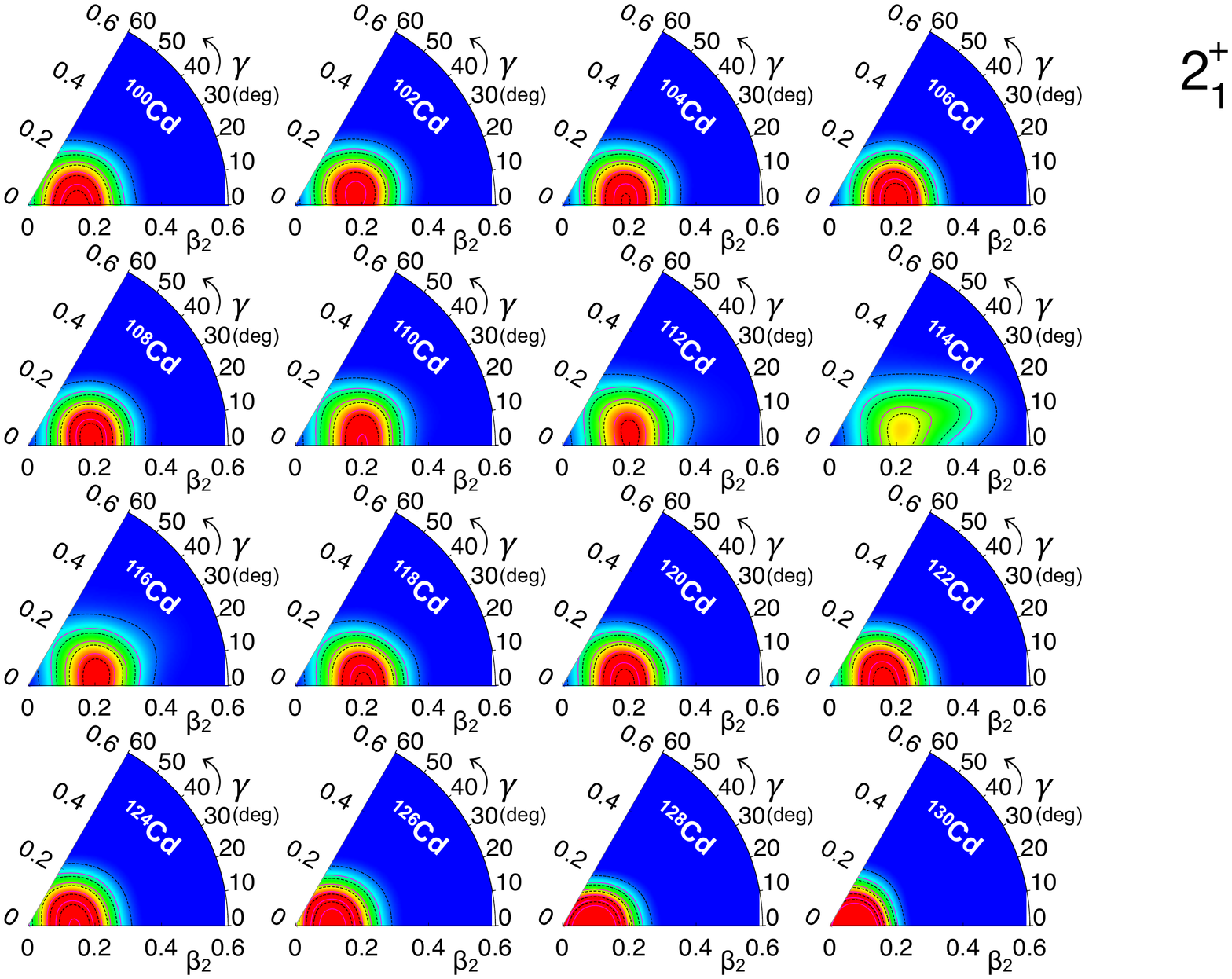}
    \vspace{-8mm}
    \caption{\label{fig:SCCM-CWF2} (Color online) Collective wave functions (CWF) as a  function  of  the ($\beta_2$, $\gamma$) deformation parameters for the $2_1^+$ states in the even-mass $^{100-130}$Cd isotopes. The results are obtained with the Gogny-D1S interaction within the SCCM approach.}
\end{figure*}
\begin{figure*}[p]
    \centering
    \includegraphics[width=0.88\textwidth]{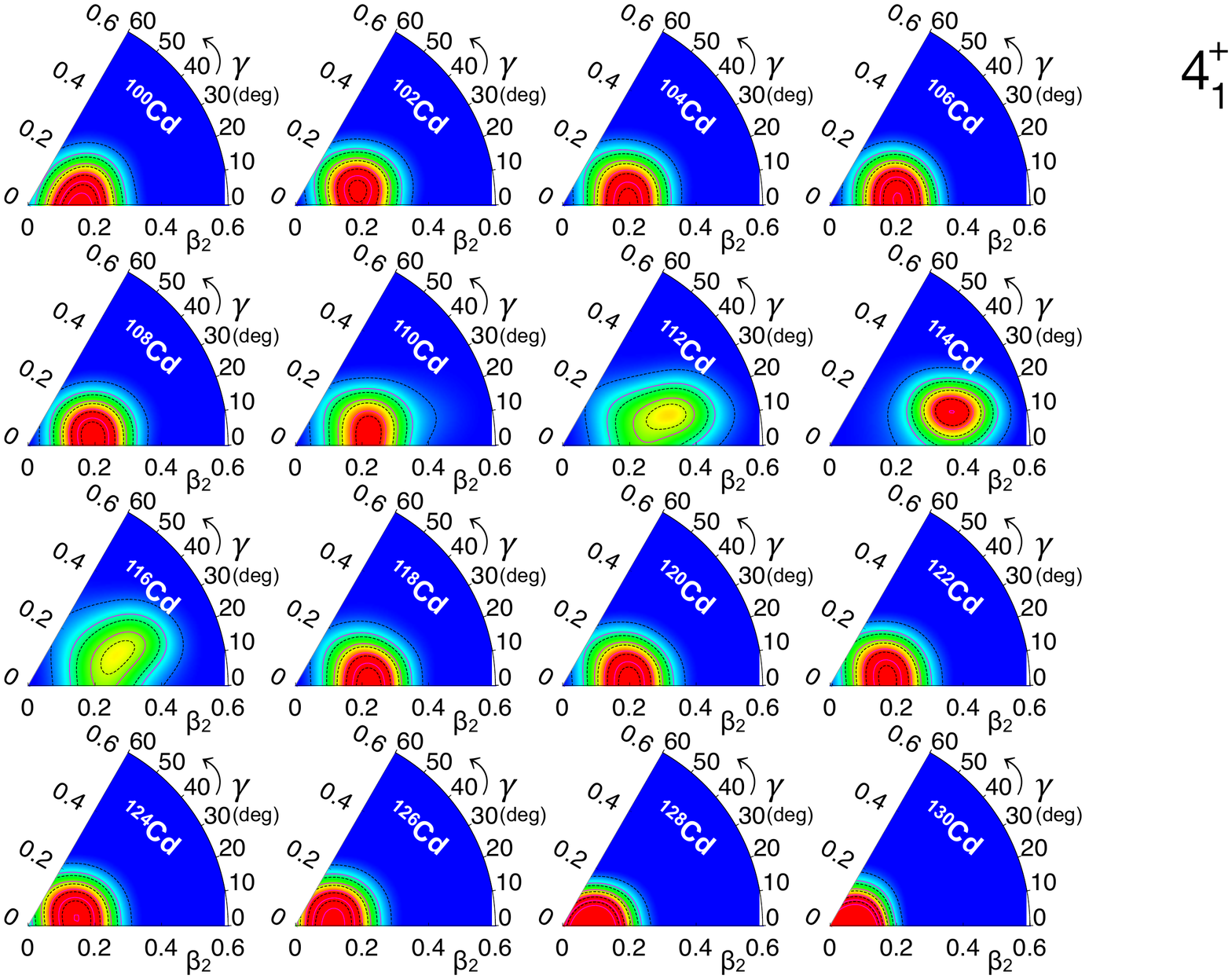}
    \vspace{-8mm}
    \caption{\label{fig:SCCM-CWF4} (Color online) Collective wave functions (CWF) as a  function  of  the ($\beta_2$, $\gamma$) deformation parameters for the $4_1^+$ states in the even-mass $^{100-130}$Cd isotopes. The results are obtained with the Gogny-D1S interaction within the SCCM approach.}
\end{figure*}

The collective wave functions (CWF), i.e. the weights of the intrinsic quadrupole deformations in each nuclear state, are presented for the $2_1^+$ and $4_1^+$ excited states in Figure~\ref{fig:SCCM-CWF2} and Figure~\ref{fig:SCCM-CWF4}, respectively. 
For all  $0_1^+$ states (not shown), the SCCM calculations predict a well-defined prolate minimum with deformation  $\beta_2 = 0.2$, which slightly decreases for $N \ge 76$ due to the proximity of the neutron shell closure. 
A non-zero  deformation of the ground states in the Cd nuclei was also deduced from the LSSM calculations of Zuker~\cite{zuker2020CdSn} and its origin was attributed to the pseudo-SU(3) symmetry, due to the evident quadrupole dominance in the nuclear interaction. 
These results are consistent also with former studies that interpreted these nuclei as deformed rotors~\cite{SAKAI1977441, Garrett_2010}.
Similar behavior is predicted for the $2_1^+$ and $4_1^+$ states, except for the $^{110-118}$Cd nuclei presenting a second triaxial-prolate minimum in the PES of Figure~\ref{fig:SCCM-PNVAP}. 
For those nuclei, the CWFs of the $2_1^+$ and $4_1^+$ states are  spread in both $\beta_2$ and $\gamma$. 
This suggests that they constitute perfect candidates for shape coexistence, as  investigated for $^{110,112}$Cd in the recent work of Garrett {\it et al.}~\cite{garrett2019multiple, garrett2020multiple}. 

As the ground-state bands are expected to present the features of prolate-deformed rotors, the intrinsic quadrupole moments and, consequently, the $\beta_2$ deformation parameters can be extracted from the measured lifetimes of the $2_1^+$ and $4_1^+$ states, as discussed in detail in Ref.~\cite[Sec.IV]{siciliano2020coexistence}. 
Assuming an axially symmetric rotational model and adopting the weighted average of the values reported in Table~\ref{tab:Results}, the deduced average $\beta_2$ parameters are 0.14 and 0.17 for $^{102}$Cd and for the even-mass $^{104-108}$Cd, respectively. 
These results are in agreement with the constant deformation predicted by the SCCM calculations, even though the theoretical predictions slightly overestimate its magnitude. 
%On the other hand, the quadrupole deformation deduced from the experimental results assuming the axial rotor model slightly decreases with spin in the ground-state bands. 
%However, due to the paucity of experimental information on the lifetimes of higher-spin states, this trend cannot be investigated further. 
For $^{106}$Cd another estimation of the ($\beta_2$, $\gamma$) quadrupole-deformation parameters of the $0_{g.s.}^+$ state can be obtained by adopting the so-called ``quadrupole sum rules'' method~\cite{poves2020shape, rocchini2021Zn66}.
As discussed in Ref.~\cite[Sec.IV-A]{rocchini2021Zn66}, this model-independent approach permits to extract the value of $\langle \beta_2^2 \rangle$, by calculating the lowest-order shape invariant $\langle Q^2 \rangle$ from the $B(E2; 2_i^+ \to 0_{g.s.}^+)$ reduced transition probabilities. 
The lower-order invariant results $\langle Q^2 \rangle = 0.419 ~(8)~e^2$b$^2$ and, assuming $\beta_2 \approx \sqrt{\langle \beta_2^2 \rangle}$, the quadrupole deformation strength is $\beta_2 = 0.175~(2)$. 
The $2_1^+$ and $2_2^+$ excited states were considered in the sum rules, while higher-lying states are expected to contribute to the value of $\beta_2$ by less than 1\%~\cite{kasiaw2012Mo100, kasiah2018Ca42, morrison2020Xe130, rocchini2021Zn66}. 
With the same procedure, the next-order shape invariant $\langle Q^3 cos(3 \delta) \rangle$ can be calculated, leading to an estimation of the triaxiality. 
Assuming the diagonal E2 matrix elements of Ref.~\cite[Tab. I]{rhodes2021high} and considering the lifetimes measured in this work (the sign of the non-diagonal matrix elements has been chosen consistently with the results of Ref.~\cite{rhodes2021high}), the next-order invariant results $\langle Q^3 cos(3 \delta) \rangle = -0.025~(18)~e^3$b$^3$, yelding the triaxial parameter $\gamma = 32~(1)^\circ$. 

Contrary to what is observed in the neutron-deficient Sn isotopes~\cite{siciliano2020plb}, no unusual trends are present for the reduced transition probabilities between the low-lying states in the light Cd nuclei. 
The $Z=48$ nuclei behave, instead, as prolate-deformed rotors, as suggested also in Refs.~\cite{garrett2019multiple, garrett2020multiple, zuker2020CdSn}. 
Even though these two isotopic chains differ by only two protons, the Sn and Cd nuclei present completely different structures whose origin can be attributed to a rearrangement of the nuclear orbitals. 
While for the Gogny-D1S interaction the spherical $Z=50$ gap remains rather constant along the Cd isotopic chain and has the same size as for the Sn nuclei, in the Nilsson plots a gap is produced at a prolate-deformed configuration, due to the lowering of the $d_{5/2}$ and $g_{7/2}$, and the rise of the $g_{9/2}$ proton orbitals.
At this deformation, $Z=48$ is a closed-shell configuration and is favored with respect to the spherical one~\cite{rodriguez2008cadmium}. 
Thus, the structure of the ground-state band changes completely between the Cd and the Sn isotopes: the former are dominated by rotational structures, while the latter have seniority spectra associated to particle-pair breaking. 
This fundamental structural change is obscured by the observed similarities between the $Z=48$ and $Z=50$ nuclei in terms of several experimental observables.
The present  study demonstrates that, contrary to what one could naively imagine, it is not possible to infer details of the structure of $Z=50$ nuclei from the properties of the $Z=48$ ones and vice versa.
This does not preclude, however, using the experimental data on one chain in order to refine the model description of the neighbouring one. 
For instance, Zuker~\cite{zuker2020CdSn} tuned the adopted nuclear interaction to the experimental information on the Cd isotopes, where the quadrupole dominance is evident, and subsequently used this interaction to investigate the Sn nuclei. 
Finally, the lowering of the $d_{5/2}$, $g_{7/2}$ orbits and the rise of the $g_{9/2}$ neutron orbitals could also favor the proton-neutron coupling that would eventually produce the $5/2^+$ ground states found for the odd-mass cadmium isotopes $^{101-109}$Cd~\cite{PhysRevC.98.011303}. 
Nevertheless, a detailed SCCM calculation that could confirm such ground state properties of the odd-even Cd isotopes is beyond the scope of the present work.

\subsection{$^{106}$Cd}

\begin{figure*}[p]%[ht]
    \centering
    \includegraphics[width=\textwidth]{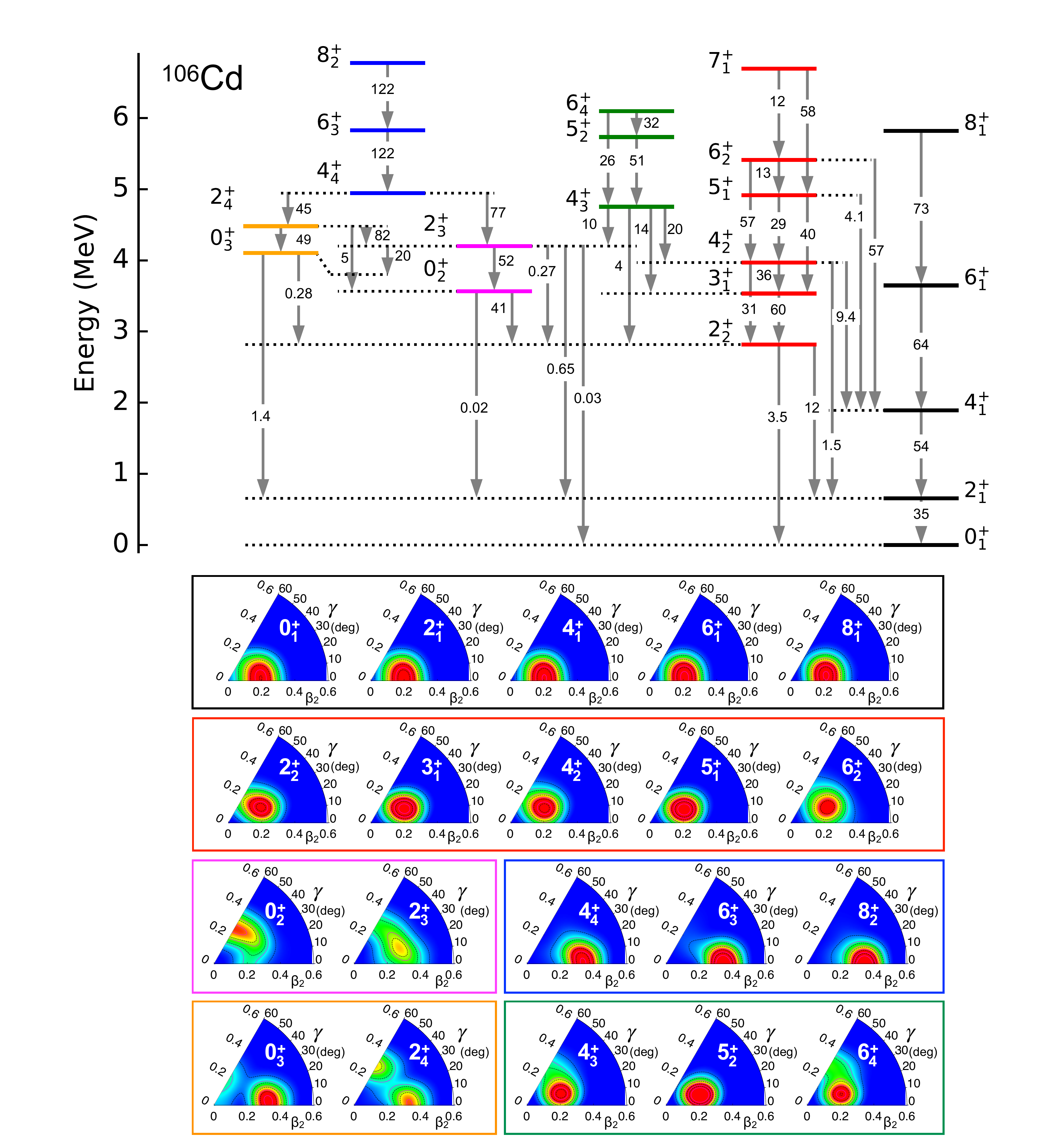}
    \vspace{-5mm}
    \caption{\label{fig:SCCM-106Cd-rotor} (Color online) Top: Level scheme of $^{106}$Cd predicted by the present SCCM calculations. The numbers on the arrows represent the $B(E2)$ reduced transition probabilities in Weisskopf units. Bottom: The CWFs are shown for all the investigated states and grouped in bands with different $(\beta_{2},\gamma)$:  prolate-deformed ground-state band (black),  predominantly $K=2$ pseudo-gamma band (red),  predominantly $K=4$ band (green), oblate-deformed shape-mixed band (magenta), prolate-deformed shape-mixed band (orange) and the continuation of the latter at larger angular momenta (blue).}
    %Excitation energy of low-lying states in $^{106}$Cd as a function of their angular momentum, calculated  with the Gogny-D1S interaction within the SCCM approach. \textcolor{red}{The CWFs are shown for all the investigated states and grouped in bands with different $(\beta_{2},\gamma)$}:  prolate-deformed ground-state band (black),  \textcolor{red}{predominantly $K=2$} pseudo-gamma band (red),  oblate-deformed shape-mixed band (blue), \textcolor{red}{predominantly $K=4$} band (green), and  prolate-deformed shape-mixed band (magenta). }
\end{figure*}

%In view of the new experimental results that concern both its ground-state and excited bands, the structure of $^{106}$Cd was investigated in detail with the SCCM approach. 
%Despite the Cd nuclei have always been considered as the textbook example of harmonic quadrupole vibrators, 
Contrary to the common view of the Cd nuclei as harmonic quadrupole vibrators, the theoretical calculations predict the low-lying bands to be due to rotation of deformed structures (see Figure~\ref{fig:SCCM-106Cd-rotor}).
This result is in agreement with the conclusions of a recent study on $^{110,112}$Cd~\cite{garrett2019multiple,garrett2020multiple}, which suggested that an interpretation of these nuclei as exhibiting coexistence of multiple distinct structures is more appropriate than a vibrational picture. 

As discussed previously, the calculations predict a prolate-deformed ground-state band with $(\beta_{2},\gamma)$~$\approx$~$(0.2,0^\circ)$. 
This prediction of the quadrupole-deformation strength is slightly larger than the values deduced by assuming the quadrupole sum rules method, yielding $\beta_2$~=~$0.175\,(2)$, or the axial-rotor model, resulting $\beta_2 = 0.169\,(3)$. 
%\textcolor{red}{\sout{and the measured lifetimes of the $2_1^+$, $4_1^+$ and $6_2^+$ states}}
%However, the lifetime of the $6_2^+$ state leads to a transitional quadrupole moment that is smaller than for the other states, leading to $\beta_2 = 0.12 (2)$.  
However, with the latter approach, the lifetime of the $6_2^+$ state leads to $\beta_2$~=~$0.12\,(2)$, which is not compatible with the quadrupole-deformation strengths estimated for the $2_1^+$ and $4_1^+$ excited states. 
Such a difference may be attributed to the mixing between the $6_1^+$ and $6_2^+$ states, since these levels are very close in excitation energy and they both decay to the $4_1^+$ state. 
A (mostly $K=2$) pseudo-gamma band built on top of the $2_2^+$ state, having $(\beta_2 , \gamma) \approx (0.2 , 25^\circ)$, and a mostly $K=4$ band are associated to the ground-state band. 
Additionally, another prolate-deformed band with $(\beta_{2}, \gamma) \approx (0.35,0^\circ)$ develops above the $4_4^+$ state. 
Below the $4^{+}_{4}$ state this band splits into two branches corresponding to strongly mixed configurations. 
One of them includes the $0^{+}_{2}$ (oblate shape-mixing) and $2^+_3$ (triaxial shape-mixing) states, coupled by a strong E2 transition of 52 W.u. 
The second branch is formed by the $0^+_3$ (prolate-deformed) and the $2^+_4$ (oblate-prolate shape-mixing) states. 

%\begin{figure*}
%    \centering
%    \includegraphics[width=0.88\textwidth]{LevelScheme_106Cd-comparison.pdf}
%    \caption{\label{fig:SCCM-106Cd} (Color online) Partial level scheme of $^{106}$Cd, reporting low-lying excited states. The numbers on the arrows represent the $B(E2)$ reduced transition probabilities in Weisskopf units. The experimental results are compared with the energy spectrum predicted by the SCCM approach, for which the band-head CWFs are shown.}
%\end{figure*}

%\textcolor{red}{\sout{Figure~\ref{fig:SCCM-106Cd} shows the comparison of the theoretical predictions with experimental data, including  excitation energies and reduced transition probabilities deduced from the lifetimes measured in the present work. This comparison allowed us to identify four of these configurations. }}
From the comparison between experimental and theoretical level schemes (see Figure~\ref{fig:lvl} and Figure~\ref{fig:SCCM-106Cd-rotor}, respectively) and reduced transition probabilities of Table~\ref{tab:Results}, it has been possible to identify four of these configurations.  
In particular, thanks to a good agreement between the theoretical and experimental transition strengths it is possible to firmly identify the pseudo-gamma band. 
%The underestimation of the $4_2^+ \to 4_1^+$ transition strength may suggest a strong mixing between the states belonging to the two different bands. 
However, the experimental information on the structures built on top of the $0_2^+$ and $0_3^+$ states is too limited to draw conclusions regarding the strongly mixed configurations. 

Consistently with the overestimation of the quadrupole deformation strengths and reduced transition probabilities, the SCCM calculations predict the spectroscopic quadrupole moment $Q_{sp}(2_1^+) = -0.62~e$b, whose absolute value is larger than what measured experimentally, i.e. $-0.28~(8)~e$b~\cite{esat1976mass} and  $-0.19~(4)~e$b~\cite{rhodes2021high}. 
In the recent work of Rhodes {\it et al.}~\cite{rhodes2021high} the spectroscopic quadrupole moments are obtained also for the $4_1^+$ and $6_1^+$ excited states, yielding $-0.39~(18)~e$b and $-0.8~(5)~e$b respectively. 
While the experimental results seem to rapidly increase with the spin, the theoretical predictions are rather constant, i.e. $Q_{sp} \approx -0.76~e$b, for the same states. 
Such an interruption of the increasing quadrupole moment would be in agreement with the trend of the $\beta_2$ values estimated with the axial-rotor model, supporting the hypothesis of mixing between $6_1^+$ and $6_2^+$ excited states. 
On the other hand, it is worth nothing that both theoretical predictions and experiment give negative values of the quadrupole moments, confirming the prolate-deformed structure of the ground-state band. 
In addition to the ground-state band, Rhodes {\it et al.} measure also $Q_{sp}(2_2^+) = 1.01~(5)~e$b. 
This result is in agreement with the SCCM calculations, which predicts $0.66~e$b for the pseudo-gamma band-head.

\section{Conclusions}

The structure of even-mass $^{102-108}$Cd isotopes was investigated via lifetime measurements at GANIL. 
These neutron-deficient nuclei were populated via an unconventional use of a multi-nucleon transfer reaction and, thanks to the powerful capabilities of the AGATA and VAMOS++ spectrometers, an unambiguous identification of the channels of interest was possible. 
Moreover, the combination of the magnetic spectrometer with the adopted binary reaction mechanism permitted the reconstruction of the TKEL on an event-by-event basis, which was used in the present work to reduce the contamination of the $\gamma$-ray spectra by the scattered $^{106}$Cd beam. 

Using the RDDS technique, the lifetimes of the $2_1^+$ and $4_1^+$ states in even-mass $^{102-108}$Cd were obtained. 
Additionally, lifetimes of other 8 states in $^{106}$Cd were determined, providing a deep insight into the structure of excited bands in this nucleus.

In view of these experimental results, state-of-the-art beyond-mean-field calculations were performed for the even-mass $^{100-130}$Cd nuclei using the symmetry-conserving configuration-mixing approach. 
Except for the nuclei in proximity of the neutron shell closures, these calculations predict prolate-deformed ground-state bands in the whole Cd isotopic chain. 
For $^{106}$Cd the comparison between theoretical results and recent measurements of spectroscopic quadrupole moments confirms the structure of the ground-state band. 
The quadrupole deformation $\beta_2$ is in fair agreement with the estimation obtained from the measured lifetimes by adopting the sum-rules method and by assuming an axially-symmetric rotor model. 
According to the LSSM calculations of Zuker~\cite{zuker2020CdSn}, the presence of deformation along the Cd isotopic chain can be attributed to the quadrupole dominance observed for the $Z=48$ nuclei. 
The calculations within the SCCM approach show that, due to a rearrangement of the $d_{5/2}$ and $g_{9/2}$ orbitals, a deformed closed-shell configuration is obtained for $Z=48$. 
As discussed in details in Ref.~\cite{rodriguez2008cadmium}, the semi-magic character of Cd nuclei impacts the $N=82$ shell quenching problem: all the observables that were attributed to a possible reduction of the $N=82$ shell closure in proximity of $^{132}$Sn can simply be described by invoking the structure of the Cd nuclei.

Despite the similarities between the $Z=48$ and $Z=50$ nuclei, in particular with regard to the electromagnetic properties of the $2_1^+$ states, the structures of the two isotopic chains are completely different. 
This result supports the conclusions of our previous work~\cite{siciliano2020plb} concerning the limited role of the observables related to the $2_1^+$ states in investigations of the structure of the $Z \approx 50$ nuclei. 
%Along the Cd and Sn isotopic chains, the $2_1^+$ states present similar electromagnetic properties, besides the completely different nature of these nuclei.
%In the same way, for the light-Sn isotopes the experimental information on $2_1^+$ was not sufficient to disentangle between nuclear interactions with very different features.
%Only the precise knowledge of the $4_1^+$ states allows us to shed light on the nuclear structure in the $Z \approx 50$ region.
For both Cd and Sn, only a precise knowledge of the properties of the $4_1^+$ states allows us to shed light on the structure of the nuclei in question. 
Experimental information on these states has been shown to be crucial to disentangle between different models and interpretations. 

Further experimental and theoretical studies are necessary to fully understand the structure of  neutron-deficient Cd nuclei. 
Theoretical and experimental results suggest that the multiple shape-coexistence interpretation, proposed by Garrett {\it et al.} for $^{110-112}$Cd, can be extended to the neutron-deficient region.
In this context,  future multi-step Coulomb-excitation measurements, benefiting from the lifetimes measured in the present work, will permit to directly determine the ($\beta_2$, $\gamma$) deformation parameters of the individual nuclear states. 
Moreover, in view of the predictions presented in this manuscript, the investigation of the structures built on top of the two excited $0^+$ states is of great interest. 
The identification of these bands together with a precise determination of $E0$ transition strengths for the decay of their band-heads will be crucial to verify the shape-coexistence scenario in these neutron-deficient nuclei. 

\section*{Acknowledgement}
The authors would like to thank the AGATA and VAMOS collaborations. 
Special thanks go to the GANIL technical staff for their help in setting up the apparatuses and the good quality beam. 
This research was partially supported by the European Union’s Seventh Framework Programme for Research and Technological Development (grant no. 262010). 
The authors are also grateful to CloudVeneto~\cite{CloudVeneto} for the use of computing and storage facilities. 
A.G. acknowledges the support of the Fondazione Cassa di Risparmio Padova e Rovigo under the project CONPHYT, starting grant in 2017. 
The work of T.R.R. was supported by the Spanish MICINN under Grant No. PGC2018-094583-B-I00.
The research was also supported 
(H.L., J.N. and U.J.) by Swedish Research Council under the grant agreements nos. 822-2005-3332, 821-2010-6024, 821-2013-2304, 621-2014-5558 and 2017-0065, and by the Knut and Alice Wallenberg Foundation grant no. 2005.0184, 
(B.S.) by the Scientific and Technological Council of Turkey (TUBITAK) under the project no. 114F473, 
(A.G. and R.P.) by the Ministerio de Ciencia e Innovac\'{i}on under the contracts nos. SEV-2014-0398, FPA2017-84756-C4 and EEBB-I-15-09671, by the Generalitat Valenciana under the grant agreement no. PROMETEO/2019/005 and by the EU-FEDER funds, 
(T.M. and S.S.) by the Croatian Science Foundation under the project no. 7194, 
(I.K. and D.S.) by the Hungarian National Research and Innovation Office (NKFIH) under the project nos. K128947, PD124717 and GINOP-2.3.3-15-2016-00034, 
(M.P. and G.J.) by the Polish National Science Centre with the grants nos. 2014-14-M-ST2-00738, 2016-22-M-ST2-00269 and 2017-25-B-ST2-01569, and the COPIN-IN2P3, COPIGAL and POLITA projects,
(C.M.-G.) by the U.S. Department of Energy, Office of Science, Office of Nuclear Physics, under contract number DE-AC02-06CH11357. 
This manuscript owes much to the collaboration with M.~Zieli\'nska, M.~Rocchini and D.~Kalaydjieva.  

\bibliography{Siciliano_Cd}

\begin{thebibliography}{95}
\expandafter\ifx\csname natexlab\endcsname\relax\def\natexlab#1{#1}\fi
\expandafter\ifx\csname bibnamefont\endcsname\relax
  \def\bibnamefont#1{#1}\fi
\expandafter\ifx\csname bibfnamefont\endcsname\relax
  \def\bibfnamefont#1{#1}\fi
\expandafter\ifx\csname citenamefont\endcsname\relax
  \def\citenamefont#1{#1}\fi
\expandafter\ifx\csname url\endcsname\relax
  \def\url#1{\texttt{#1}}\fi
\expandafter\ifx\csname urlprefix\endcsname\relax\def\urlprefix{URL }\fi
\providecommand{\bibinfo}[2]{#2}
\providecommand{\eprint}[2][]{\url{#2}}

\bibitem[{\citenamefont{Struwe and Winkler}(1974)}]{STRUWE1974605}
\bibinfo{author}{\bibfnamefont{W.}~\bibnamefont{Struwe}} \bibnamefont{and}
  \bibinfo{author}{\bibfnamefont{G.}~\bibnamefont{Winkler}},
  \bibinfo{journal}{Nucl. Phys. A} \textbf{\bibinfo{volume}{222}},
  \bibinfo{pages}{605} (\bibinfo{year}{1974}).

\bibitem[{\citenamefont{Fielding et~al.}(1977)\citenamefont{Fielding, Anderson,
  Zafiratos, Lind, Cecil, Wieman, and Alford}}]{FIELDING1977389}
\bibinfo{author}{\bibfnamefont{H.}~\bibnamefont{Fielding}},
  \bibinfo{author}{\bibfnamefont{R.}~\bibnamefont{Anderson}},
  \bibinfo{author}{\bibfnamefont{C.}~\bibnamefont{Zafiratos}},
  \bibinfo{author}{\bibfnamefont{D.}~\bibnamefont{Lind}},
  \bibinfo{author}{\bibfnamefont{F.}~\bibnamefont{Cecil}},
  \bibinfo{author}{\bibfnamefont{H.}~\bibnamefont{Wieman}}, \bibnamefont{and}
  \bibinfo{author}{\bibfnamefont{W.}~\bibnamefont{Alford}},
  \bibinfo{journal}{Nucl. Phys. A} \textbf{\bibinfo{volume}{281}},
  \bibinfo{pages}{389} (\bibinfo{year}{1977}).

\bibitem[{\citenamefont{B{\"a}ck et~al.}(2012)\citenamefont{B{\"a}ck, Qi,
  Cederwall, Liotta, {Ghazi Moradi}, Johnson, Wyss, and
  Wadsworth}}]{back2012BE}
\bibinfo{author}{\bibfnamefont{T.}~\bibnamefont{B{\"a}ck}},
  \bibinfo{author}{\bibfnamefont{C.}~\bibnamefont{Qi}},
  \bibinfo{author}{\bibfnamefont{B.}~\bibnamefont{Cederwall}},
  \bibinfo{author}{\bibfnamefont{R.}~\bibnamefont{Liotta}},
  \bibinfo{author}{\bibfnamefont{F.}~\bibnamefont{{Ghazi Moradi}}},
  \bibinfo{author}{\bibfnamefont{A.}~\bibnamefont{Johnson}},
  \bibinfo{author}{\bibfnamefont{R.}~\bibnamefont{Wyss}}, \bibnamefont{and}
  \bibinfo{author}{\bibfnamefont{R.}~\bibnamefont{Wadsworth}},
  \bibinfo{journal}{Phys. Scripta} \textbf{\bibinfo{volume}{T150}},
  \bibinfo{pages}{014003} (\bibinfo{year}{2012}).

\bibitem[{\citenamefont{Saxena et~al.}(2014)\citenamefont{Saxena, Kumar,
  Jhingan, Mandal, Stolarz, Banerjee, Bhowmik, Dutt, Kaur, Kumar
  et~al.}}]{saxena2014rotational}
\bibinfo{author}{\bibfnamefont{M.}~\bibnamefont{Saxena}},
  \bibinfo{author}{\bibfnamefont{R.}~\bibnamefont{Kumar}},
  \bibinfo{author}{\bibfnamefont{A.}~\bibnamefont{Jhingan}},
  \bibinfo{author}{\bibfnamefont{S.}~\bibnamefont{Mandal}},
  \bibinfo{author}{\bibfnamefont{A.}~\bibnamefont{Stolarz}},
  \bibinfo{author}{\bibfnamefont{A.}~\bibnamefont{Banerjee}},
  \bibinfo{author}{\bibfnamefont{R.~K.} \bibnamefont{Bhowmik}},
  \bibinfo{author}{\bibfnamefont{S.}~\bibnamefont{Dutt}},
  \bibinfo{author}{\bibfnamefont{J.}~\bibnamefont{Kaur}},
  \bibinfo{author}{\bibfnamefont{V.}~\bibnamefont{Kumar}},
  \bibnamefont{et~al.}, \bibinfo{journal}{Phys. Rev. C}
  \textbf{\bibinfo{volume}{90}}, \bibinfo{pages}{024316}
  (\bibinfo{year}{2014}).

\bibitem[{\citenamefont{Garrett}(2018)}]{garrett2018collective}
\bibinfo{author}{\bibfnamefont{P.~E.} \bibnamefont{Garrett}},
  \bibinfo{journal}{Eur. Phys. J.: Conf.} \textbf{\bibinfo{volume}{178}},
  \bibinfo{pages}{02011} (\bibinfo{year}{2018}).

\bibitem[{\citenamefont{Yao-Song and Zhongzhou}(1996)}]{BF02769692}
\bibinfo{author}{\bibfnamefont{S.}~\bibnamefont{Yao-Song}} \bibnamefont{and}
  \bibinfo{author}{\bibfnamefont{R.}~\bibnamefont{Zhongzhou}},
  \bibinfo{journal}{Z. Phys. A} \textbf{\bibinfo{volume}{355}},
  \bibinfo{pages}{247} (\bibinfo{year}{1996}).

\bibitem[{\citenamefont{Libert et~al.}(2007)\citenamefont{Libert,
  Roussi{\`e}re, and Sauvage}}]{LIBERT200747}
\bibinfo{author}{\bibfnamefont{J.}~\bibnamefont{Libert}},
  \bibinfo{author}{\bibfnamefont{B.}~\bibnamefont{Roussi{\`e}re}},
  \bibnamefont{and} \bibinfo{author}{\bibfnamefont{J.}~\bibnamefont{Sauvage}},
  \bibinfo{journal}{Nucl. Phys. A} \textbf{\bibinfo{volume}{786}},
  \bibinfo{pages}{47} (\bibinfo{year}{2007}).

\bibitem[{\citenamefont{Maheshwari et~al.}(2019)\citenamefont{Maheshwari,
  Abu~Kassim, Yusof, and Kumar~Jain}}]{maheshwari2019seniority}
\bibinfo{author}{\bibfnamefont{B.}~\bibnamefont{Maheshwari}},
  \bibinfo{author}{\bibfnamefont{H.}~\bibnamefont{Abu~Kassim}},
  \bibinfo{author}{\bibfnamefont{N.}~\bibnamefont{Yusof}}, \bibnamefont{and}
  \bibinfo{author}{\bibfnamefont{A.}~\bibnamefont{Kumar~Jain}},
  \bibinfo{journal}{Nucl. Phys. A} \textbf{\bibinfo{volume}{992}},
  \bibinfo{pages}{121619} (\bibinfo{year}{2019}).

\bibitem[{\citenamefont{Fortune}(2019)}]{fortune2019similarities}
\bibinfo{author}{\bibfnamefont{H.~T.} \bibnamefont{Fortune}},
  \bibinfo{journal}{Phys. Rev. C} \textbf{\bibinfo{volume}{100}},
  \bibinfo{pages}{054328} (\bibinfo{year}{2019}).

\bibitem[{\citenamefont{Maheshwari}(2020)}]{maheshwari2020unified}
\bibinfo{author}{\bibfnamefont{B.}~\bibnamefont{Maheshwari}},
  \bibinfo{journal}{Eur. Phys. J.: Special Topics}
  \textbf{\bibinfo{volume}{229}}, \bibinfo{pages}{2485} (\bibinfo{year}{2020}).

\bibitem[{\citenamefont{Talmi}(1971)}]{talmi1971generalized}
\bibinfo{author}{\bibfnamefont{I.}~\bibnamefont{Talmi}},
  \bibinfo{journal}{Nuclear Physics A} \textbf{\bibinfo{volume}{172}},
  \bibinfo{pages}{1} (\bibinfo{year}{1971}).

\bibitem[{\citenamefont{Talmi}(1993)}]{talmi1993simple}
\bibinfo{author}{\bibfnamefont{I.}~\bibnamefont{Talmi}},
  \emph{\bibinfo{title}{{Simple models of complex nuclei}}},
  vol.~\bibinfo{volume}{7} (\bibinfo{publisher}{CRC Press},
  \bibinfo{year}{1993}).

\bibitem[{\citenamefont{Ressler et~al.}(2004)\citenamefont{Ressler, Casten,
  Zamfir, Beausang, Cakirli, Ai, Amro, Caprio, Hecht, Heinz
  et~al.}}]{ressler2004transition}
\bibinfo{author}{\bibfnamefont{J.~J.} \bibnamefont{Ressler}},
  \bibinfo{author}{\bibfnamefont{R.~F.} \bibnamefont{Casten}},
  \bibinfo{author}{\bibfnamefont{N.~V.} \bibnamefont{Zamfir}},
  \bibinfo{author}{\bibfnamefont{C.~W.} \bibnamefont{Beausang}},
  \bibinfo{author}{\bibfnamefont{R.~B.} \bibnamefont{Cakirli}},
  \bibinfo{author}{\bibfnamefont{H.}~\bibnamefont{Ai}},
  \bibinfo{author}{\bibfnamefont{H.}~\bibnamefont{Amro}},
  \bibinfo{author}{\bibfnamefont{M.~A.} \bibnamefont{Caprio}},
  \bibinfo{author}{\bibfnamefont{A.~A.} \bibnamefont{Hecht}},
  \bibinfo{author}{\bibfnamefont{A.}~\bibnamefont{Heinz}},
  \bibnamefont{et~al.}, \bibinfo{journal}{Phys. Rev. C}
  \textbf{\bibinfo{volume}{69}}, \bibinfo{pages}{034317}
  (\bibinfo{year}{2004}).

\bibitem[{\citenamefont{Morales et~al.}(2011)\citenamefont{Morales,
  {Van~Isacker}, and Talmi}}]{morales2011generalized}
\bibinfo{author}{\bibfnamefont{I.~O.} \bibnamefont{Morales}},
  \bibinfo{author}{\bibfnamefont{P.}~\bibnamefont{{Van~Isacker}}},
  \bibnamefont{and} \bibinfo{author}{\bibfnamefont{I.}~\bibnamefont{Talmi}},
  \bibinfo{journal}{Phys. Lett. B} \textbf{\bibinfo{volume}{703}},
  \bibinfo{pages}{606} (\bibinfo{year}{2011}).

\bibitem[{\citenamefont{Siciliano
  et~al.}(2020{\natexlab{a}})\citenamefont{Siciliano, Valiente-Dob\'{o}n,
  Goasduff, Nowacki, Zuker, Bazzacco, Lopez-Martens, Cl\'{e}ment, Benzoni,
  Braunroth et~al.}}]{siciliano2020plb}
\bibinfo{author}{\bibfnamefont{M.}~\bibnamefont{Siciliano}},
  \bibinfo{author}{\bibfnamefont{J.}~\bibnamefont{Valiente-Dob\'{o}n}},
  \bibinfo{author}{\bibfnamefont{A.}~\bibnamefont{Goasduff}},
  \bibinfo{author}{\bibfnamefont{F.}~\bibnamefont{Nowacki}},
  \bibinfo{author}{\bibfnamefont{A.}~\bibnamefont{Zuker}},
  \bibinfo{author}{\bibfnamefont{D.}~\bibnamefont{Bazzacco}},
  \bibinfo{author}{\bibfnamefont{A.}~\bibnamefont{Lopez-Martens}},
  \bibinfo{author}{\bibfnamefont{E.}~\bibnamefont{Cl\'{e}ment}},
  \bibinfo{author}{\bibfnamefont{G.}~\bibnamefont{Benzoni}},
  \bibinfo{author}{\bibfnamefont{T.}~\bibnamefont{Braunroth}},
  \bibnamefont{et~al.}, \bibinfo{journal}{Phys. Lett. B}
  \textbf{\bibinfo{volume}{806}}, \bibinfo{pages}{135474}
  (\bibinfo{year}{2020}{\natexlab{a}}).

\bibitem[{\citenamefont{Zuker}(2021)}]{zuker2020CdSn}
\bibinfo{author}{\bibfnamefont{A.}~\bibnamefont{Zuker}},
  \bibinfo{journal}{Phys. Rev. C} \textbf{\bibinfo{volume}{103}},
  \bibinfo{pages}{024322} (\bibinfo{year}{2021}).

\bibitem[{\citenamefont{Sonzogni et~al.}(2016)}]{nudatDatabase}
\bibinfo{author}{\bibfnamefont{A.}~\bibnamefont{Sonzogni}}
  \bibnamefont{et~al.},
  \emph{\bibinfo{title}{\href{http://www.nndc.bnl.gov/nudat2}{Nudat 2}}}
  (\bibinfo{year}{2016}).

\bibitem[{\citenamefont{Arima and Iachello}(1976)}]{arima1976interacting}
\bibinfo{author}{\bibfnamefont{A.}~\bibnamefont{Arima}} \bibnamefont{and}
  \bibinfo{author}{\bibfnamefont{F.}~\bibnamefont{Iachello}},
  \bibinfo{journal}{Ann. Phys.} \textbf{\bibinfo{volume}{99}},
  \bibinfo{pages}{253} (\bibinfo{year}{1976}).

\bibitem[{\citenamefont{Iachello and Arima}(1987)}]{iachello1987interacting}
\bibinfo{author}{\bibfnamefont{F.}~\bibnamefont{Iachello}} \bibnamefont{and}
  \bibinfo{author}{\bibfnamefont{A.}~\bibnamefont{Arima}},
  \emph{\bibinfo{title}{The interacting boson model}}
  (\bibinfo{publisher}{Cambridge University Press}, \bibinfo{year}{1987}).

\bibitem[{\citenamefont{Bohr and Mottelson}(1998)}]{bohrMottelson1998nuclear}
\bibinfo{author}{\bibfnamefont{A.}~\bibnamefont{Bohr}} \bibnamefont{and}
  \bibinfo{author}{\bibfnamefont{B.}~\bibnamefont{Mottelson}},
  \emph{\bibinfo{title}{{Nuclear structure: nuclear deformations}}},
  vol.~\bibinfo{volume}{2} (\bibinfo{publisher}{World Scientific},
  \bibinfo{year}{1998}).

\bibitem[{\citenamefont{Yates}(2005)}]{yates2005probing}
\bibinfo{author}{\bibfnamefont{S.~W.} \bibnamefont{Yates}},
  \bibinfo{journal}{J. Phys. G} \textbf{\bibinfo{volume}{31}},
  \bibinfo{pages}{S1393} (\bibinfo{year}{2005}).

\bibitem[{\citenamefont{Rowe}(2010)}]{rowe2010nuclear}
\bibinfo{author}{\bibfnamefont{D.~J.} \bibnamefont{Rowe}},
  \emph{\bibinfo{title}{Nuclear collective motion: models and theory}}
  (\bibinfo{publisher}{World Scientific}, \bibinfo{year}{2010}).

\bibitem[{\citenamefont{Fahlander et~al.}(1988)\citenamefont{Fahlander,
  B{\"a}cklin, Hasselgren, Kavka, Mittal, Svensson, Varnestig, Cline,
  Kotlinski, Grein et~al.}}]{fahlander1988hasselgrenet}
\bibinfo{author}{\bibfnamefont{C.}~\bibnamefont{Fahlander}},
  \bibinfo{author}{\bibfnamefont{A.}~\bibnamefont{B{\"a}cklin}},
  \bibinfo{author}{\bibfnamefont{L.}~\bibnamefont{Hasselgren}},
  \bibinfo{author}{\bibfnamefont{A.}~\bibnamefont{Kavka}},
  \bibinfo{author}{\bibfnamefont{V.}~\bibnamefont{Mittal}},
  \bibinfo{author}{\bibfnamefont{L.}~\bibnamefont{Svensson}},
  \bibinfo{author}{\bibfnamefont{B.}~\bibnamefont{Varnestig}},
  \bibinfo{author}{\bibfnamefont{D.}~\bibnamefont{Cline}},
  \bibinfo{author}{\bibfnamefont{B.}~\bibnamefont{Kotlinski}},
  \bibinfo{author}{\bibfnamefont{H.}~\bibnamefont{Grein}},
  \bibnamefont{et~al.}, \bibinfo{journal}{Nucl. Phys. A}
  \textbf{\bibinfo{volume}{485}}, \bibinfo{pages}{327} (\bibinfo{year}{1988}).

\bibitem[{\citenamefont{Lehmann et~al.}(1996)\citenamefont{Lehmann, Garrett,
  Jolie, McGrath, Yeh, and Yates}}]{lehmann1996nature}
\bibinfo{author}{\bibfnamefont{H.}~\bibnamefont{Lehmann}},
  \bibinfo{author}{\bibfnamefont{P.~E.} \bibnamefont{Garrett}},
  \bibinfo{author}{\bibfnamefont{J.}~\bibnamefont{Jolie}},
  \bibinfo{author}{\bibfnamefont{C.~A.} \bibnamefont{McGrath}},
  \bibinfo{author}{\bibfnamefont{M.}~\bibnamefont{Yeh}}, \bibnamefont{and}
  \bibinfo{author}{\bibfnamefont{S.~W.} \bibnamefont{Yates}},
  \bibinfo{journal}{Phys. Lett. B} \textbf{\bibinfo{volume}{387}},
  \bibinfo{pages}{259} (\bibinfo{year}{1996}).

\bibitem[{\citenamefont{Corminboeuf et~al.}(2000)\citenamefont{Corminboeuf,
  Brown, Genilloud, Hannant, Jolie, Kern, Warr, and
  Yates}}]{corminboeuf2000characterization}
\bibinfo{author}{\bibfnamefont{F.}~\bibnamefont{Corminboeuf}},
  \bibinfo{author}{\bibfnamefont{T.~B.} \bibnamefont{Brown}},
  \bibinfo{author}{\bibfnamefont{L.}~\bibnamefont{Genilloud}},
  \bibinfo{author}{\bibfnamefont{C.~D.} \bibnamefont{Hannant}},
  \bibinfo{author}{\bibfnamefont{J.}~\bibnamefont{Jolie}},
  \bibinfo{author}{\bibfnamefont{J.}~\bibnamefont{Kern}},
  \bibinfo{author}{\bibfnamefont{N.}~\bibnamefont{Warr}}, \bibnamefont{and}
  \bibinfo{author}{\bibfnamefont{S.~W.} \bibnamefont{Yates}},
  \bibinfo{journal}{Phys. Rev. Lett.} \textbf{\bibinfo{volume}{84}},
  \bibinfo{pages}{4060} (\bibinfo{year}{2000}).

\bibitem[{\citenamefont{Kadi et~al.}(2003)\citenamefont{Kadi, Warr, Garrett,
  Jolie, and Yates}}]{kadi2003vibrational}
\bibinfo{author}{\bibfnamefont{M.}~\bibnamefont{Kadi}},
  \bibinfo{author}{\bibfnamefont{N.}~\bibnamefont{Warr}},
  \bibinfo{author}{\bibfnamefont{P.~E.} \bibnamefont{Garrett}},
  \bibinfo{author}{\bibfnamefont{J.}~\bibnamefont{Jolie}}, \bibnamefont{and}
  \bibinfo{author}{\bibfnamefont{S.~W.} \bibnamefont{Yates}},
  \bibinfo{journal}{Phys. Rev. C} \textbf{\bibinfo{volume}{68}},
  \bibinfo{pages}{031306(R)} (\bibinfo{year}{2003}).

\bibitem[{\citenamefont{Garrett et~al.}(2007)\citenamefont{Garrett, Green,
  Lehmann, Jolie, McGrath, Yeh, and Yates}}]{garrett2007properties}
\bibinfo{author}{\bibfnamefont{P.~E.} \bibnamefont{Garrett}},
  \bibinfo{author}{\bibfnamefont{K.~L.} \bibnamefont{Green}},
  \bibinfo{author}{\bibfnamefont{H.}~\bibnamefont{Lehmann}},
  \bibinfo{author}{\bibfnamefont{J.}~\bibnamefont{Jolie}},
  \bibinfo{author}{\bibfnamefont{C.~A.} \bibnamefont{McGrath}},
  \bibinfo{author}{\bibfnamefont{M.}~\bibnamefont{Yeh}}, \bibnamefont{and}
  \bibinfo{author}{\bibfnamefont{S.~W.} \bibnamefont{Yates}},
  \bibinfo{journal}{Phys. Rev. C} \textbf{\bibinfo{volume}{75}},
  \bibinfo{pages}{054310} (\bibinfo{year}{2007}).

\bibitem[{\citenamefont{Bandyopadhyay et~al.}(2007)\citenamefont{Bandyopadhyay,
  Lesher, Fransen, Boukharouba, Garrett, Green, McEllistrem, and
  Yates}}]{bandyopadhyay2007investigation}
\bibinfo{author}{\bibfnamefont{D.}~\bibnamefont{Bandyopadhyay}},
  \bibinfo{author}{\bibfnamefont{S.~R.} \bibnamefont{Lesher}},
  \bibinfo{author}{\bibfnamefont{C.}~\bibnamefont{Fransen}},
  \bibinfo{author}{\bibfnamefont{N.}~\bibnamefont{Boukharouba}},
  \bibinfo{author}{\bibfnamefont{P.~E.} \bibnamefont{Garrett}},
  \bibinfo{author}{\bibfnamefont{K.~L.} \bibnamefont{Green}},
  \bibinfo{author}{\bibfnamefont{M.~T.} \bibnamefont{McEllistrem}},
  \bibnamefont{and} \bibinfo{author}{\bibfnamefont{S.~W.} \bibnamefont{Yates}},
  \bibinfo{journal}{Phys. Rev. C} \textbf{\bibinfo{volume}{76}},
  \bibinfo{pages}{054308} (\bibinfo{year}{2007}).

\bibitem[{\citenamefont{Garrett et~al.}(2008)\citenamefont{Garrett, Green, and
  Wood}}]{garrett2008breakdown}
\bibinfo{author}{\bibfnamefont{P.~E.} \bibnamefont{Garrett}},
  \bibinfo{author}{\bibfnamefont{K.~L.} \bibnamefont{Green}}, \bibnamefont{and}
  \bibinfo{author}{\bibfnamefont{J.~L.} \bibnamefont{Wood}},
  \bibinfo{journal}{Phys. Rev. C} \textbf{\bibinfo{volume}{78}},
  \bibinfo{pages}{044307} (\bibinfo{year}{2008}).

\bibitem[{\citenamefont{Green et~al.}(2009)\citenamefont{Green, Garrett,
  Austin, Ball, Bandyopadhyay, Colosimo, Cross, Demand, Grinyer, Hackman
  et~al.}}]{green2009degeneracy}
\bibinfo{author}{\bibfnamefont{K.~L.} \bibnamefont{Green}},
  \bibinfo{author}{\bibfnamefont{P.~E.} \bibnamefont{Garrett}},
  \bibinfo{author}{\bibfnamefont{R.~A.~E.} \bibnamefont{Austin}},
  \bibinfo{author}{\bibfnamefont{G.~C.} \bibnamefont{Ball}},
  \bibinfo{author}{\bibfnamefont{D.~S.} \bibnamefont{Bandyopadhyay}},
  \bibinfo{author}{\bibfnamefont{S.}~\bibnamefont{Colosimo}},
  \bibinfo{author}{\bibfnamefont{D.}~\bibnamefont{Cross}},
  \bibinfo{author}{\bibfnamefont{G.~A.} \bibnamefont{Demand}},
  \bibinfo{author}{\bibfnamefont{G.~F.} \bibnamefont{Grinyer}},
  \bibinfo{author}{\bibfnamefont{G.}~\bibnamefont{Hackman}},
  \bibnamefont{et~al.}, \bibinfo{journal}{Phys. Rev. C}
  \textbf{\bibinfo{volume}{80}}, \bibinfo{pages}{032502(R)}
  (\bibinfo{year}{2009}).

\bibitem[{\citenamefont{Garrett et~al.}(2019)\citenamefont{Garrett,
  Rodr\'{\i}guez, Varela, Green, Bangay, Finlay, Austin, Ball, Bandyopadhyay,
  Bildstein et~al.}}]{garrett2019multiple}
\bibinfo{author}{\bibfnamefont{P.~E.} \bibnamefont{Garrett}},
  \bibinfo{author}{\bibfnamefont{T.~R.} \bibnamefont{Rodr\'{\i}guez}},
  \bibinfo{author}{\bibfnamefont{A.~D.} \bibnamefont{Varela}},
  \bibinfo{author}{\bibfnamefont{K.~L.} \bibnamefont{Green}},
  \bibinfo{author}{\bibfnamefont{J.}~\bibnamefont{Bangay}},
  \bibinfo{author}{\bibfnamefont{A.}~\bibnamefont{Finlay}},
  \bibinfo{author}{\bibfnamefont{R.~A.~E.} \bibnamefont{Austin}},
  \bibinfo{author}{\bibfnamefont{G.~C.} \bibnamefont{Ball}},
  \bibinfo{author}{\bibfnamefont{D.~S.} \bibnamefont{Bandyopadhyay}},
  \bibinfo{author}{\bibfnamefont{V.}~\bibnamefont{Bildstein}},
  \bibnamefont{et~al.}, \bibinfo{journal}{Phys. Rev. Lett.}
  \textbf{\bibinfo{volume}{123}}, \bibinfo{pages}{142502}
  (\bibinfo{year}{2019}).

\bibitem[{\citenamefont{Garrett et~al.}(2020)\citenamefont{Garrett,
  Rodr\'{\i}guez, Diaz~Varela, Green, Bangay, Finlay, Austin, Ball,
  Bandyopadhyay, Bildstein et~al.}}]{garrett2020multiple}
\bibinfo{author}{\bibfnamefont{P.~E.} \bibnamefont{Garrett}},
  \bibinfo{author}{\bibfnamefont{T.~R.} \bibnamefont{Rodr\'{\i}guez}},
  \bibinfo{author}{\bibfnamefont{A.}~\bibnamefont{Diaz~Varela}},
  \bibinfo{author}{\bibfnamefont{K.~L.} \bibnamefont{Green}},
  \bibinfo{author}{\bibfnamefont{J.}~\bibnamefont{Bangay}},
  \bibinfo{author}{\bibfnamefont{A.}~\bibnamefont{Finlay}},
  \bibinfo{author}{\bibfnamefont{R.~A.~E.} \bibnamefont{Austin}},
  \bibinfo{author}{\bibfnamefont{G.~C.} \bibnamefont{Ball}},
  \bibinfo{author}{\bibfnamefont{D.~S.} \bibnamefont{Bandyopadhyay}},
  \bibinfo{author}{\bibfnamefont{V.}~\bibnamefont{Bildstein}},
  \bibnamefont{et~al.}, \bibinfo{journal}{Phys. Rev. C}
  \textbf{\bibinfo{volume}{101}}, \bibinfo{pages}{044302}
  (\bibinfo{year}{2020}).

\bibitem[{\citenamefont{Szilner et~al.}(2007)\citenamefont{Szilner, Ur,
  Corradi, M\u{a}rginean, Pollarolo, Stefanini, Beghini, Behera, Fioretto,
  Gadea et~al.}}]{szilner2007multinucleon}
\bibinfo{author}{\bibfnamefont{S.}~\bibnamefont{Szilner}},
  \bibinfo{author}{\bibfnamefont{C.~A.} \bibnamefont{Ur}},
  \bibinfo{author}{\bibfnamefont{L.}~\bibnamefont{Corradi}},
  \bibinfo{author}{\bibfnamefont{N.}~\bibnamefont{M\u{a}rginean}},
  \bibinfo{author}{\bibfnamefont{G.}~\bibnamefont{Pollarolo}},
  \bibinfo{author}{\bibfnamefont{A.~M.} \bibnamefont{Stefanini}},
  \bibinfo{author}{\bibfnamefont{S.}~\bibnamefont{Beghini}},
  \bibinfo{author}{\bibfnamefont{B.~R.} \bibnamefont{Behera}},
  \bibinfo{author}{\bibfnamefont{E.}~\bibnamefont{Fioretto}},
  \bibinfo{author}{\bibfnamefont{A.}~\bibnamefont{Gadea}},
  \bibnamefont{et~al.}, \bibinfo{journal}{Phys. Rev. C}
  \textbf{\bibinfo{volume}{76}}, \bibinfo{pages}{024604}
  (\bibinfo{year}{2007}).

\bibitem[{\citenamefont{Corradi et~al.}(2009)\citenamefont{Corradi, Pollarolo,
  and Szilner}}]{corradi2009multinucleon}
\bibinfo{author}{\bibfnamefont{L.}~\bibnamefont{Corradi}},
  \bibinfo{author}{\bibfnamefont{G.}~\bibnamefont{Pollarolo}},
  \bibnamefont{and} \bibinfo{author}{\bibfnamefont{S.}~\bibnamefont{Szilner}},
  \bibinfo{journal}{J. Phys. G} \textbf{\bibinfo{volume}{36}},
  \bibinfo{pages}{113101} (\bibinfo{year}{2009}).

\bibitem[{\citenamefont{Valiente-Dob{\'o}n}(2016)}]{valiente2016gamma}
\bibinfo{author}{\bibfnamefont{J.~J.} \bibnamefont{Valiente-Dob{\'o}n}},
  \bibinfo{journal}{{Basic Concepts in Nuclear Physics: Theory, Experiments and
  Applications}} \textbf{\bibinfo{volume}{182}}, \bibinfo{pages}{87}
  (\bibinfo{year}{2016}).

\bibitem[{\citenamefont{Dewald et~al.}(1989)\citenamefont{Dewald, Harissopulos,
  and {Von Brentano}}}]{dewald1989differential}
\bibinfo{author}{\bibfnamefont{A.}~\bibnamefont{Dewald}},
  \bibinfo{author}{\bibfnamefont{S.}~\bibnamefont{Harissopulos}},
  \bibnamefont{and} \bibinfo{author}{\bibfnamefont{P.}~\bibnamefont{{Von
  Brentano}}}, \bibinfo{journal}{Z. Phys. A} \textbf{\bibinfo{volume}{334}},
  \bibinfo{pages}{163} (\bibinfo{year}{1989}).

\bibitem[{\citenamefont{Dewald et~al.}(2012)\citenamefont{Dewald, M{\"o}ller,
  and Petkov}}]{dewald2012developing}
\bibinfo{author}{\bibfnamefont{A.}~\bibnamefont{Dewald}},
  \bibinfo{author}{\bibfnamefont{O.}~\bibnamefont{M{\"o}ller}},
  \bibnamefont{and} \bibinfo{author}{\bibfnamefont{P.}~\bibnamefont{Petkov}},
  \bibinfo{journal}{Prog. Part. and Nucl. Phys.} \textbf{\bibinfo{volume}{67}},
  \bibinfo{pages}{786} (\bibinfo{year}{2012}).

\bibitem[{\citenamefont{Valiente-Dob\'on
  et~al.}(2009)\citenamefont{Valiente-Dob\'on, Mengoni, Gadea, Farnea, Lenzi,
  Lunardi, Dewald, Pissulla, Szilner, Broda et~al.}}]{valiente2009lifetime}
\bibinfo{author}{\bibfnamefont{J.~J.} \bibnamefont{Valiente-Dob\'on}},
  \bibinfo{author}{\bibfnamefont{D.}~\bibnamefont{Mengoni}},
  \bibinfo{author}{\bibfnamefont{A.}~\bibnamefont{Gadea}},
  \bibinfo{author}{\bibfnamefont{E.}~\bibnamefont{Farnea}},
  \bibinfo{author}{\bibfnamefont{S.~M.} \bibnamefont{Lenzi}},
  \bibinfo{author}{\bibfnamefont{S.}~\bibnamefont{Lunardi}},
  \bibinfo{author}{\bibfnamefont{A.}~\bibnamefont{Dewald}},
  \bibinfo{author}{\bibfnamefont{T.}~\bibnamefont{Pissulla}},
  \bibinfo{author}{\bibfnamefont{S.}~\bibnamefont{Szilner}},
  \bibinfo{author}{\bibfnamefont{R.}~\bibnamefont{Broda}},
  \bibnamefont{et~al.}, \bibinfo{journal}{Phys. Rev. Lett.}
  \textbf{\bibinfo{volume}{102}}, \bibinfo{pages}{242502}
  (\bibinfo{year}{2009}).

\bibitem[{\citenamefont{Pullanhiotan et~al.}(2008)\citenamefont{Pullanhiotan,
  Chatterjee, Jacquot, Navin, and Rejmund}}]{pullanhiotan2008improvement}
\bibinfo{author}{\bibfnamefont{S.}~\bibnamefont{Pullanhiotan}},
  \bibinfo{author}{\bibfnamefont{A.}~\bibnamefont{Chatterjee}},
  \bibinfo{author}{\bibfnamefont{B.}~\bibnamefont{Jacquot}},
  \bibinfo{author}{\bibfnamefont{A.}~\bibnamefont{Navin}}, \bibnamefont{and}
  \bibinfo{author}{\bibfnamefont{M.}~\bibnamefont{Rejmund}},
  \bibinfo{journal}{Nucl. Instr. and Meth. in Phys. B}
  \textbf{\bibinfo{volume}{266}}, \bibinfo{pages}{4148} (\bibinfo{year}{2008}).

\bibitem[{\citenamefont{Rejmund et~al.}(2011)\citenamefont{Rejmund, Lecornu,
  Navin, Schmitt et~al.}}]{rejmund2011performance}
\bibinfo{author}{\bibfnamefont{M.}~\bibnamefont{Rejmund}},
  \bibinfo{author}{\bibfnamefont{B.}~\bibnamefont{Lecornu}},
  \bibinfo{author}{\bibfnamefont{A.}~\bibnamefont{Navin}},
  \bibinfo{author}{\bibfnamefont{C.}~\bibnamefont{Schmitt}},
  \bibnamefont{et~al.}, \bibinfo{journal}{Nucl. Instr. and Meth. in Phys. A}
  \textbf{\bibinfo{volume}{646}}, \bibinfo{pages}{184} (\bibinfo{year}{2011}).

\bibitem[{\citenamefont{Vandebrouck et~al.}(2016)\citenamefont{Vandebrouck,
  Lemasson, Rejmund, Fremont, Pancin, Navin, Michelagnoli, Goupil, Spitaels,
  and Jacquot}}]{vandebrouck2016dual}
\bibinfo{author}{\bibfnamefont{M.}~\bibnamefont{Vandebrouck}},
  \bibinfo{author}{\bibfnamefont{A.}~\bibnamefont{Lemasson}},
  \bibinfo{author}{\bibfnamefont{M.}~\bibnamefont{Rejmund}},
  \bibinfo{author}{\bibfnamefont{G.}~\bibnamefont{Fremont}},
  \bibinfo{author}{\bibfnamefont{J.}~\bibnamefont{Pancin}},
  \bibinfo{author}{\bibfnamefont{A.}~\bibnamefont{Navin}},
  \bibinfo{author}{\bibfnamefont{C.}~\bibnamefont{Michelagnoli}},
  \bibinfo{author}{\bibfnamefont{J.}~\bibnamefont{Goupil}},
  \bibinfo{author}{\bibfnamefont{C.}~\bibnamefont{Spitaels}}, \bibnamefont{and}
  \bibinfo{author}{\bibfnamefont{B.}~\bibnamefont{Jacquot}},
  \bibinfo{journal}{Nucl. Instr. and Meth. in Phys. A}
  \textbf{\bibinfo{volume}{812}}, \bibinfo{pages}{112} (\bibinfo{year}{2016}).

\bibitem[{\citenamefont{Akkoyun et~al.}(2012)}]{akkoyun2012agata}
\bibinfo{author}{\bibfnamefont{S.}~\bibnamefont{Akkoyun}} \bibnamefont{et~al.},
  \bibinfo{journal}{Nucl. Instr. and Meth. in Phys. A}
  \textbf{\bibinfo{volume}{668}}, \bibinfo{pages}{26} (\bibinfo{year}{2012}).

\bibitem[{\citenamefont{Cl{\'e}ment et~al.}(2017)\citenamefont{Cl{\'e}ment,
  Michelagnoli, {de France}, Li, Lemasson, Dejean
  et~al.}}]{clement2017conceptual}
\bibinfo{author}{\bibfnamefont{E.}~\bibnamefont{Cl{\'e}ment}},
  \bibinfo{author}{\bibfnamefont{C.}~\bibnamefont{Michelagnoli}},
  \bibinfo{author}{\bibfnamefont{G.}~\bibnamefont{{de France}}},
  \bibinfo{author}{\bibfnamefont{H.~J.} \bibnamefont{Li}},
  \bibinfo{author}{\bibfnamefont{A.}~\bibnamefont{Lemasson}},
  \bibinfo{author}{\bibfnamefont{C.~B.} \bibnamefont{Dejean}},
  \bibnamefont{et~al.}, \bibinfo{journal}{Nucl. Instr. and Meth. in Phys. A}
  \textbf{\bibinfo{volume}{855}}, \bibinfo{pages}{1} (\bibinfo{year}{2017}).

\bibitem[{\citenamefont{Bruyneel et~al.}(2013)\citenamefont{Bruyneel,
  Birkenbach, Eberth, Hess, Pascovici, Reiter, Wiens
  et~al.}}]{bruyneel2013correction}
\bibinfo{author}{\bibfnamefont{B.}~\bibnamefont{Bruyneel}},
  \bibinfo{author}{\bibfnamefont{B.}~\bibnamefont{Birkenbach}},
  \bibinfo{author}{\bibfnamefont{J.}~\bibnamefont{Eberth}},
  \bibinfo{author}{\bibfnamefont{H.}~\bibnamefont{Hess}},
  \bibinfo{author}{\bibfnamefont{G.}~\bibnamefont{Pascovici}},
  \bibinfo{author}{\bibfnamefont{P.}~\bibnamefont{Reiter}},
  \bibinfo{author}{\bibfnamefont{A.}~\bibnamefont{Wiens}},
  \bibnamefont{et~al.}, \bibinfo{journal}{Eur. Phys. J. A}
  \textbf{\bibinfo{volume}{49}}, \bibinfo{pages}{1} (\bibinfo{year}{2013}).

\bibitem[{\citenamefont{Lopez-Martens et~al.}(2004)\citenamefont{Lopez-Martens,
  Hauschild, Korichi, Roccaz, and Thibaud}}]{lopez2004gamma}
\bibinfo{author}{\bibfnamefont{A.}~\bibnamefont{Lopez-Martens}},
  \bibinfo{author}{\bibfnamefont{K.}~\bibnamefont{Hauschild}},
  \bibinfo{author}{\bibfnamefont{A.}~\bibnamefont{Korichi}},
  \bibinfo{author}{\bibfnamefont{J.}~\bibnamefont{Roccaz}}, \bibnamefont{and}
  \bibinfo{author}{\bibfnamefont{J.~P.} \bibnamefont{Thibaud}},
  \bibinfo{journal}{Nucl. Instr. and Meth. in Phys. A}
  \textbf{\bibinfo{volume}{533}}, \bibinfo{pages}{454} (\bibinfo{year}{2004}).

\bibitem[{\citenamefont{Siciliano et~al.}(2017)\citenamefont{Siciliano,
  Valiente-Dob{\'o}n, Goasduff, Bazzacco et~al.}}]{siciliano2017appb}
\bibinfo{author}{\bibfnamefont{M.}~\bibnamefont{Siciliano}},
  \bibinfo{author}{\bibfnamefont{J.~J.} \bibnamefont{Valiente-Dob{\'o}n}},
  \bibinfo{author}{\bibfnamefont{A.}~\bibnamefont{Goasduff}},
  \bibinfo{author}{\bibfnamefont{D.}~\bibnamefont{Bazzacco}},
  \bibnamefont{et~al.}, \bibinfo{journal}{Acta Phys. Pol. B}
  \textbf{\bibinfo{volume}{48}}, \bibinfo{pages}{331} (\bibinfo{year}{2017}).

\bibitem[{\citenamefont{Siciliano}(2017)}]{siciliano2017ncc}
\bibinfo{author}{\bibfnamefont{M.}~\bibnamefont{Siciliano}},
  \bibinfo{journal}{Nuovo Cimento C} \textbf{\bibinfo{volume}{40}},
  \bibinfo{pages}{84} (\bibinfo{year}{2017}).

\bibitem[{\citenamefont{Siciliano et~al.}(2019)\citenamefont{Siciliano,
  Valiente-Dob\'{o}n, and Goasduff}}]{Siciliano2019EPJ}
\bibinfo{author}{\bibfnamefont{M.}~\bibnamefont{Siciliano}},
  \bibinfo{author}{\bibfnamefont{J.}~\bibnamefont{Valiente-Dob\'{o}n}},
  \bibnamefont{and} \bibinfo{author}{\bibfnamefont{A.}~\bibnamefont{Goasduff}},
  \bibinfo{journal}{Eur. Phys. J.: Web of Conference}
  \textbf{\bibinfo{volume}{223}}, \bibinfo{pages}{01060}
  (\bibinfo{year}{2019}).

\bibitem[{\citenamefont{Saha}(2004)}]{napatau}
\bibinfo{author}{\bibfnamefont{B.}~\bibnamefont{Saha}}, Ph.D. thesis,
  \bibinfo{school}{Universit{\"a}t zu K{\"o}ln} (\bibinfo{year}{2004}).

\bibitem[{\citenamefont{Doncel et~al.}(2017)\citenamefont{Doncel, Gadea,
  Valiente-Dob{\'o}n, Quintana, Modamio, Mengoni, M{\"o}ller, Dewald, and
  Pietralla}}]{RDDSions}
\bibinfo{author}{\bibfnamefont{M.}~\bibnamefont{Doncel}},
  \bibinfo{author}{\bibfnamefont{A.}~\bibnamefont{Gadea}},
  \bibinfo{author}{\bibfnamefont{J.~J.} \bibnamefont{Valiente-Dob{\'o}n}},
  \bibinfo{author}{\bibfnamefont{B.}~\bibnamefont{Quintana}},
  \bibinfo{author}{\bibfnamefont{V.}~\bibnamefont{Modamio}},
  \bibinfo{author}{\bibfnamefont{D.}~\bibnamefont{Mengoni}},
  \bibinfo{author}{\bibfnamefont{O.}~\bibnamefont{M{\"o}ller}},
  \bibinfo{author}{\bibfnamefont{A.}~\bibnamefont{Dewald}}, \bibnamefont{and}
  \bibinfo{author}{\bibfnamefont{N.}~\bibnamefont{Pietralla}},
  \bibinfo{journal}{Eur. Phys. J. A} \textbf{\bibinfo{volume}{53}},
  \bibinfo{pages}{211} (\bibinfo{year}{2017}).

\bibitem[{\citenamefont{Litzinger et~al.}(2015)\citenamefont{Litzinger,
  Blazhev, Dewald, Didierjean, Duch\^ene, Fransen, Lozeva, Sieja, Verney
  et~al.}}]{litzinger2015trans}
\bibinfo{author}{\bibfnamefont{J.}~\bibnamefont{Litzinger}},
  \bibinfo{author}{\bibfnamefont{A.}~\bibnamefont{Blazhev}},
  \bibinfo{author}{\bibfnamefont{A.}~\bibnamefont{Dewald}},
  \bibinfo{author}{\bibfnamefont{F.}~\bibnamefont{Didierjean}},
  \bibinfo{author}{\bibfnamefont{G.}~\bibnamefont{Duch\^ene}},
  \bibinfo{author}{\bibfnamefont{C.}~\bibnamefont{Fransen}},
  \bibinfo{author}{\bibfnamefont{R.}~\bibnamefont{Lozeva}},
  \bibinfo{author}{\bibfnamefont{K.}~\bibnamefont{Sieja}},
  \bibinfo{author}{\bibfnamefont{D.}~\bibnamefont{Verney}},
  \bibnamefont{et~al.}, \bibinfo{journal}{Phys. Rev. C}
  \textbf{\bibinfo{volume}{92}}, \bibinfo{pages}{064322}
  (\bibinfo{year}{2015}).

\bibitem[{\citenamefont{Bhat}(1992)}]{ESNDF}
\bibinfo{author}{\bibfnamefont{M.~R.} \bibnamefont{Bhat}}, in
  \emph{\bibinfo{booktitle}{Nuclear Data for Science and Technology}}, edited
  by \bibinfo{editor}{\bibfnamefont{S.~M.} \bibnamefont{Qaim}}
  (\bibinfo{publisher}{Springer Berlin Heidelberg}, \bibinfo{year}{1992}), p.
  \bibinfo{pages}{817}.

\bibitem[{\citenamefont{Brown et~al.}(1951)\citenamefont{Brown, Snyder, Fowler,
  and Lauritsen}}]{brown1951excited}
\bibinfo{author}{\bibfnamefont{A.~B.} \bibnamefont{Brown}},
  \bibinfo{author}{\bibfnamefont{C.~W.} \bibnamefont{Snyder}},
  \bibinfo{author}{\bibfnamefont{W.~A.} \bibnamefont{Fowler}},
  \bibnamefont{and} \bibinfo{author}{\bibfnamefont{C.~C.}
  \bibnamefont{Lauritsen}}, \bibinfo{journal}{Phys. Rev.}
  \textbf{\bibinfo{volume}{82}}, \bibinfo{pages}{159} (\bibinfo{year}{1951}).

\bibitem[{\citenamefont{Mengoni et~al.}(2009)\citenamefont{Mengoni,
  Valiente-Dob{\'o}n, Farnea, Gadea, Dewald, and Latina}}]{mengoni2009lifetime}
\bibinfo{author}{\bibfnamefont{D.}~\bibnamefont{Mengoni}},
  \bibinfo{author}{\bibfnamefont{J.}~\bibnamefont{Valiente-Dob{\'o}n}},
  \bibinfo{author}{\bibfnamefont{E.}~\bibnamefont{Farnea}},
  \bibinfo{author}{\bibfnamefont{A.}~\bibnamefont{Gadea}},
  \bibinfo{author}{\bibfnamefont{A.}~\bibnamefont{Dewald}}, \bibnamefont{and}
  \bibinfo{author}{\bibfnamefont{A.}~\bibnamefont{Latina}},
  \bibinfo{journal}{Eur. Phys. J. A} \textbf{\bibinfo{volume}{42}},
  \bibinfo{pages}{387} (\bibinfo{year}{2009}).

\bibitem[{\citenamefont{Boelaert et~al.}(2007)\citenamefont{Boelaert, Dewald,
  Fransen, Jolie, Linnemann, Melon, M{\"o}ller, Smirnova, and
  Heyde}}]{boelaert2007low}
\bibinfo{author}{\bibfnamefont{N.}~\bibnamefont{Boelaert}},
  \bibinfo{author}{\bibfnamefont{A.}~\bibnamefont{Dewald}},
  \bibinfo{author}{\bibfnamefont{C.}~\bibnamefont{Fransen}},
  \bibinfo{author}{\bibfnamefont{J.}~\bibnamefont{Jolie}},
  \bibinfo{author}{\bibfnamefont{A.}~\bibnamefont{Linnemann}},
  \bibinfo{author}{\bibfnamefont{B.}~\bibnamefont{Melon}},
  \bibinfo{author}{\bibfnamefont{O.}~\bibnamefont{M{\"o}ller}},
  \bibinfo{author}{\bibfnamefont{N.}~\bibnamefont{Smirnova}}, \bibnamefont{and}
  \bibinfo{author}{\bibfnamefont{K.}~\bibnamefont{Heyde}},
  \bibinfo{journal}{Phys. Rev. C} \textbf{\bibinfo{volume}{75}},
  \bibinfo{pages}{054311} (\bibinfo{year}{2007}).

\bibitem[{\citenamefont{Ekstr\"om et~al.}(2009)\citenamefont{Ekstr\"om,
  Cederk\"all, DiJulio, Fahlander, Hjorth-Jensen, Blazhev, Bruyneel, Butler,
  Davinson, Eberth et~al.}}]{ekstrom2009cd}
\bibinfo{author}{\bibfnamefont{A.}~\bibnamefont{Ekstr\"om}},
  \bibinfo{author}{\bibfnamefont{J.}~\bibnamefont{Cederk\"all}},
  \bibinfo{author}{\bibfnamefont{D.~D.} \bibnamefont{DiJulio}},
  \bibinfo{author}{\bibfnamefont{C.}~\bibnamefont{Fahlander}},
  \bibinfo{author}{\bibfnamefont{M.}~\bibnamefont{Hjorth-Jensen}},
  \bibinfo{author}{\bibfnamefont{A.}~\bibnamefont{Blazhev}},
  \bibinfo{author}{\bibfnamefont{B.}~\bibnamefont{Bruyneel}},
  \bibinfo{author}{\bibfnamefont{P.~A.} \bibnamefont{Butler}},
  \bibinfo{author}{\bibfnamefont{T.}~\bibnamefont{Davinson}},
  \bibinfo{author}{\bibfnamefont{J.}~\bibnamefont{Eberth}},
  \bibnamefont{et~al.}, \bibinfo{journal}{Phys. Rev. C}
  \textbf{\bibinfo{volume}{80}}, \bibinfo{pages}{054302}
  (\bibinfo{year}{2009}).

\bibitem[{\citenamefont{Petkov et~al.}(2001)\citenamefont{Petkov, Dewald, and
  {von Brentano}}}]{PETKOV2001527}
\bibinfo{author}{\bibfnamefont{P.}~\bibnamefont{Petkov}},
  \bibinfo{author}{\bibfnamefont{A.}~\bibnamefont{Dewald}}, \bibnamefont{and}
  \bibinfo{author}{\bibfnamefont{P.}~\bibnamefont{{von Brentano}}},
  \bibinfo{journal}{Nucl. Instr. and Meth. in Phys. A}
  \textbf{\bibinfo{volume}{457}}, \bibinfo{pages}{527} (\bibinfo{year}{2001}).

\bibitem[{\citenamefont{M{\"u}ller et~al.}(2001)\citenamefont{M{\"u}ller,
  Jungclaus, Yordanov, Galindo, Hausmann, Kast, Lieb, Brant, Krsti\'{c},
  Vrentenar et~al.}}]{muller2001high}
\bibinfo{author}{\bibfnamefont{G.~A.} \bibnamefont{M{\"u}ller}},
  \bibinfo{author}{\bibfnamefont{A.}~\bibnamefont{Jungclaus}},
  \bibinfo{author}{\bibfnamefont{O.}~\bibnamefont{Yordanov}},
  \bibinfo{author}{\bibfnamefont{E.}~\bibnamefont{Galindo}},
  \bibinfo{author}{\bibfnamefont{M.}~\bibnamefont{Hausmann}},
  \bibinfo{author}{\bibfnamefont{D.}~\bibnamefont{Kast}},
  \bibinfo{author}{\bibfnamefont{K.~P.} \bibnamefont{Lieb}},
  \bibinfo{author}{\bibfnamefont{S.}~\bibnamefont{Brant}},
  \bibinfo{author}{\bibfnamefont{V.}~\bibnamefont{Krsti\'{c}}},
  \bibinfo{author}{\bibfnamefont{D.}~\bibnamefont{Vrentenar}},
  \bibnamefont{et~al.}, \bibinfo{journal}{Phys. Rev. C}
  \textbf{\bibinfo{volume}{64}}, \bibinfo{pages}{014305}
  (\bibinfo{year}{2001}).

\bibitem[{\citenamefont{Gusinsky and Zvonov}(1983)}]{gusinsky1983}
\bibinfo{author}{\bibfnamefont{G.}~\bibnamefont{Gusinsky}} \bibnamefont{and}
  \bibinfo{author}{\bibfnamefont{V.}~\bibnamefont{Zvonov}},
  \bibinfo{journal}{Izv. Akad. Nauk SSSR, Ser.Fiz.}
  \textbf{\bibinfo{volume}{47}}, \bibinfo{pages}{862} (\bibinfo{year}{1983}).

\bibitem[{\citenamefont{Milner et~al.}(1969)\citenamefont{Milner, MvGowan,
  Stelson, Robinson, and Sayer}}]{milner1969}
\bibinfo{author}{\bibfnamefont{W.~T.} \bibnamefont{Milner}},
  \bibinfo{author}{\bibfnamefont{F.~K.} \bibnamefont{MvGowan}},
  \bibinfo{author}{\bibfnamefont{P.~H.} \bibnamefont{Stelson}},
  \bibinfo{author}{\bibfnamefont{R.~L.} \bibnamefont{Robinson}},
  \bibnamefont{and} \bibinfo{author}{\bibfnamefont{R.~O.} \bibnamefont{Sayer}},
  \bibinfo{journal}{Nucl. Phys. A} \textbf{\bibinfo{volume}{129}},
  \bibinfo{pages}{687} (\bibinfo{year}{1969}).

\bibitem[{\citenamefont{Rh{\"o}des et~al.}(2021)\citenamefont{Rh{\"o}des,
  Brown, Henderson, Gade, Ash, Bender, Elder, Elman, Grinder, Hjorth-Jensen
  et~al.}}]{rhodes2021high}
\bibinfo{author}{\bibfnamefont{D.}~\bibnamefont{Rh{\"o}des}},
  \bibinfo{author}{\bibfnamefont{B.~A.} \bibnamefont{Brown}},
  \bibinfo{author}{\bibfnamefont{J.}~\bibnamefont{Henderson}},
  \bibinfo{author}{\bibfnamefont{A.}~\bibnamefont{Gade}},
  \bibinfo{author}{\bibfnamefont{J.}~\bibnamefont{Ash}},
  \bibinfo{author}{\bibfnamefont{P.~C.} \bibnamefont{Bender}},
  \bibinfo{author}{\bibfnamefont{R.}~\bibnamefont{Elder}},
  \bibinfo{author}{\bibfnamefont{B.}~\bibnamefont{Elman}},
  \bibinfo{author}{\bibfnamefont{M.}~\bibnamefont{Grinder}},
  \bibinfo{author}{\bibfnamefont{M.}~\bibnamefont{Hjorth-Jensen}},
  \bibnamefont{et~al.}, \bibinfo{journal}{Phys. Rev. C}
  \textbf{\bibinfo{volume}{103}}, \bibinfo{pages}{L051301}
  (\bibinfo{year}{2021}).

\bibitem[{\citenamefont{Andrejtscheff et~al.}(1985)\citenamefont{Andrejtscheff,
  Kostov, Rotter, Prade, Stary, Senba, Tsoupas, Ding, and
  Raghavan}}]{ANDREJTSCHEFF1985167}
\bibinfo{author}{\bibfnamefont{W.}~\bibnamefont{Andrejtscheff}},
  \bibinfo{author}{\bibfnamefont{L.~K.} \bibnamefont{Kostov}},
  \bibinfo{author}{\bibfnamefont{H.}~\bibnamefont{Rotter}},
  \bibinfo{author}{\bibfnamefont{H.}~\bibnamefont{Prade}},
  \bibinfo{author}{\bibfnamefont{F.}~\bibnamefont{Stary}},
  \bibinfo{author}{\bibfnamefont{M.}~\bibnamefont{Senba}},
  \bibinfo{author}{\bibfnamefont{N.}~\bibnamefont{Tsoupas}},
  \bibinfo{author}{\bibfnamefont{Z.~Z.} \bibnamefont{Ding}}, \bibnamefont{and}
  \bibinfo{author}{\bibfnamefont{P.}~\bibnamefont{Raghavan}},
  \bibinfo{journal}{Nuclear Physics A} \textbf{\bibinfo{volume}{437}},
  \bibinfo{pages}{167} (\bibinfo{year}{1985}).

\bibitem[{\citenamefont{Lieb et~al.}(2001)\citenamefont{Lieb, Kast, Jungclaus,
  Johnstone, de~Angelis, Fahlander, de~Poli, Bizzeti, Dewald, Peusquens
  et~al.}}]{lieb2001proton}
\bibinfo{author}{\bibfnamefont{K.~P.} \bibnamefont{Lieb}},
  \bibinfo{author}{\bibfnamefont{D.}~\bibnamefont{Kast}},
  \bibinfo{author}{\bibfnamefont{A.}~\bibnamefont{Jungclaus}},
  \bibinfo{author}{\bibfnamefont{I.~P.} \bibnamefont{Johnstone}},
  \bibinfo{author}{\bibfnamefont{G.}~\bibnamefont{de~Angelis}},
  \bibinfo{author}{\bibfnamefont{C.}~\bibnamefont{Fahlander}},
  \bibinfo{author}{\bibfnamefont{M.}~\bibnamefont{de~Poli}},
  \bibinfo{author}{\bibfnamefont{P.~G.} \bibnamefont{Bizzeti}},
  \bibinfo{author}{\bibfnamefont{A.}~\bibnamefont{Dewald}},
  \bibinfo{author}{\bibfnamefont{R.}~\bibnamefont{Peusquens}},
  \bibnamefont{et~al.}, \bibinfo{journal}{Phys. Rev. C}
  \textbf{\bibinfo{volume}{63}}, \bibinfo{pages}{054304}
  (\bibinfo{year}{2001}).

\bibitem[{\citenamefont{Kleinfeld et~al.}(1970)\citenamefont{Kleinfeld, Rogers,
  Gastebois, Steadman, and {de Boer}}}]{KLEINFELD197081}
\bibinfo{author}{\bibfnamefont{A.}~\bibnamefont{Kleinfeld}},
  \bibinfo{author}{\bibfnamefont{J.}~\bibnamefont{Rogers}},
  \bibinfo{author}{\bibfnamefont{J.}~\bibnamefont{Gastebois}},
  \bibinfo{author}{\bibfnamefont{S.}~\bibnamefont{Steadman}}, \bibnamefont{and}
  \bibinfo{author}{\bibfnamefont{J.}~\bibnamefont{{de Boer}}},
  \bibinfo{journal}{Nucl. Phys. A} \textbf{\bibinfo{volume}{158}},
  \bibinfo{pages}{81} (\bibinfo{year}{1970}).

\bibitem[{\citenamefont{Esat et~al.}(1976)\citenamefont{Esat, Kean, Spear, and
  Baxter}}]{esat1976mass}
\bibinfo{author}{\bibfnamefont{M.~T.} \bibnamefont{Esat}},
  \bibinfo{author}{\bibfnamefont{D.~C.} \bibnamefont{Kean}},
  \bibinfo{author}{\bibfnamefont{R.~H.} \bibnamefont{Spear}}, \bibnamefont{and}
  \bibinfo{author}{\bibfnamefont{A.~M.} \bibnamefont{Baxter}},
  \bibinfo{journal}{Nucl. Phys. A} \textbf{\bibinfo{volume}{274}},
  \bibinfo{pages}{237} (\bibinfo{year}{1976}).

\bibitem[{\citenamefont{Benczer-Koller
  et~al.}(2016)\citenamefont{Benczer-Koller, Kumbartzki, Speidel, Torres,
  Robinson, Sharon, Allmond, Fallon, Abramovic, Bernstein
  et~al.}}]{benczerkoller2016}
\bibinfo{author}{\bibfnamefont{N.}~\bibnamefont{Benczer-Koller}},
  \bibinfo{author}{\bibfnamefont{G.~J.} \bibnamefont{Kumbartzki}},
  \bibinfo{author}{\bibfnamefont{K.-H.} \bibnamefont{Speidel}},
  \bibinfo{author}{\bibfnamefont{D.~A.} \bibnamefont{Torres}},
  \bibinfo{author}{\bibfnamefont{S.~J.~Q.} \bibnamefont{Robinson}},
  \bibinfo{author}{\bibfnamefont{Y.~Y.} \bibnamefont{Sharon}},
  \bibinfo{author}{\bibfnamefont{J.~M.} \bibnamefont{Allmond}},
  \bibinfo{author}{\bibfnamefont{P.}~\bibnamefont{Fallon}},
  \bibinfo{author}{\bibfnamefont{I.}~\bibnamefont{Abramovic}},
  \bibinfo{author}{\bibfnamefont{L.~A.} \bibnamefont{Bernstein}},
  \bibnamefont{et~al.}, \bibinfo{journal}{Phys. Rev. C}
  \textbf{\bibinfo{volume}{94}}, \bibinfo{pages}{034303}
  (\bibinfo{year}{2016}).

\bibitem[{\citenamefont{Thorslund et~al.}(1994)\citenamefont{Thorslund,
  Fahlander, Nyberg, Piiparinen, Julin, Juutinen, Virtanen, M{\"u}ller, Jensen,
  and Sugawara}}]{THORSLUND1994306}
\bibinfo{author}{\bibfnamefont{I.}~\bibnamefont{Thorslund}},
  \bibinfo{author}{\bibfnamefont{C.}~\bibnamefont{Fahlander}},
  \bibinfo{author}{\bibfnamefont{J.}~\bibnamefont{Nyberg}},
  \bibinfo{author}{\bibfnamefont{M.}~\bibnamefont{Piiparinen}},
  \bibinfo{author}{\bibfnamefont{R.}~\bibnamefont{Julin}},
  \bibinfo{author}{\bibfnamefont{S.}~\bibnamefont{Juutinen}},
  \bibinfo{author}{\bibfnamefont{A.}~\bibnamefont{Virtanen}},
  \bibinfo{author}{\bibfnamefont{D.}~\bibnamefont{M{\"u}ller}},
  \bibinfo{author}{\bibfnamefont{H.}~\bibnamefont{Jensen}}, \bibnamefont{and}
  \bibinfo{author}{\bibfnamefont{M.}~\bibnamefont{Sugawara}},
  \bibinfo{journal}{Nucl. Phys. A} \textbf{\bibinfo{volume}{568}},
  \bibinfo{pages}{306} (\bibinfo{year}{1994}).

\bibitem[{\citenamefont{Egido}(2016)}]{PS_91_073003_2016}
\bibinfo{author}{\bibfnamefont{J.}~\bibnamefont{Egido}},
  \bibinfo{journal}{Phys. Scripta} \textbf{\bibinfo{volume}{91}},
  \bibinfo{pages}{073003} (\bibinfo{year}{2016}).

\bibitem[{\citenamefont{Robledo et~al.}(2018)\citenamefont{Robledo,
  Rodr\'{i}guez, and Rodr\'{i}guez-Guzm\'{a}n}}]{JPG_46_013001_2019}
\bibinfo{author}{\bibfnamefont{L.}~\bibnamefont{Robledo}},
  \bibinfo{author}{\bibfnamefont{T.~R.} \bibnamefont{Rodr\'{i}guez}},
  \bibnamefont{and}
  \bibinfo{author}{\bibfnamefont{R.}~\bibnamefont{Rodr\'{i}guez-Guzm\'{a}n}},
  \bibinfo{journal}{J. Phys. G} \textbf{\bibinfo{volume}{46}},
  \bibinfo{pages}{013001} (\bibinfo{year}{2018}).

\bibitem[{\citenamefont{Rodr\'{\i}guez and
  Egido}(2010)}]{rodriguez2010triaxial}
\bibinfo{author}{\bibfnamefont{T.~R.} \bibnamefont{Rodr\'{\i}guez}}
  \bibnamefont{and} \bibinfo{author}{\bibfnamefont{J.~L.} \bibnamefont{Egido}},
  \bibinfo{journal}{Phys. Rev. C} \textbf{\bibinfo{volume}{81}},
  \bibinfo{pages}{064323} (\bibinfo{year}{2010}).

\bibitem[{\citenamefont{Rodr{\'i}guez}(2014)}]{rodriguez2014structure}
\bibinfo{author}{\bibfnamefont{T.~R.} \bibnamefont{Rodr{\'i}guez}},
  \bibinfo{journal}{Phys. Rev. C} \textbf{\bibinfo{volume}{90}},
  \bibinfo{pages}{034306} (\bibinfo{year}{2014}).

\bibitem[{\citenamefont{Decharg\'e and Gogny}(1980)}]{PhysRevC.21.1568}
\bibinfo{author}{\bibfnamefont{J.}~\bibnamefont{Decharg\'e}} \bibnamefont{and}
  \bibinfo{author}{\bibfnamefont{D.}~\bibnamefont{Gogny}},
  \bibinfo{journal}{Phys. Rev. C} \textbf{\bibinfo{volume}{21}},
  \bibinfo{pages}{1568} (\bibinfo{year}{1980}).

\bibitem[{\citenamefont{Berger et~al.}(1991)\citenamefont{Berger, Girod, and
  Gogny}}]{BERGER1991365}
\bibinfo{author}{\bibfnamefont{J.}~\bibnamefont{Berger}},
  \bibinfo{author}{\bibfnamefont{M.}~\bibnamefont{Girod}}, \bibnamefont{and}
  \bibinfo{author}{\bibfnamefont{D.}~\bibnamefont{Gogny}},
  \bibinfo{journal}{Comp. Phys. Comm.} \textbf{\bibinfo{volume}{63}},
  \bibinfo{pages}{365} (\bibinfo{year}{1991}).

\bibitem[{\citenamefont{Anguiano et~al.}(2001)\citenamefont{Anguiano, Egido,
  and Robledo}}]{anguiano2001particle}
\bibinfo{author}{\bibfnamefont{M.}~\bibnamefont{Anguiano}},
  \bibinfo{author}{\bibfnamefont{J.}~\bibnamefont{Egido}}, \bibnamefont{and}
  \bibinfo{author}{\bibfnamefont{L.}~\bibnamefont{Robledo}},
  \bibinfo{journal}{Nucl. Phys. A} \textbf{\bibinfo{volume}{696}},
  \bibinfo{pages}{467} (\bibinfo{year}{2001}).

\bibitem[{\citenamefont{McGowan et~al.}(1965)\citenamefont{McGowan, Robinson,
  Stelson, and Ford}}]{MCGOWAN1965}
\bibinfo{author}{\bibfnamefont{F.}~\bibnamefont{McGowan}},
  \bibinfo{author}{\bibfnamefont{R.}~\bibnamefont{Robinson}},
  \bibinfo{author}{\bibfnamefont{P.}~\bibnamefont{Stelson}}, \bibnamefont{and}
  \bibinfo{author}{\bibfnamefont{J.}~\bibnamefont{Ford}},
  \bibinfo{journal}{Nuclear Physics} \textbf{\bibinfo{volume}{66}},
  \bibinfo{pages}{97} (\bibinfo{year}{1965}).

\bibitem[{\citenamefont{Mach et~al.}(1989)\citenamefont{Mach, Moszy\'{n}ski,
  Casten, Gill, Brenner, Winger, Krips, Wesselborg, B{\"u}scher, Wohn
  et~al.}}]{mach1989}
\bibinfo{author}{\bibfnamefont{H.}~\bibnamefont{Mach}},
  \bibinfo{author}{\bibfnamefont{M.}~\bibnamefont{Moszy\'{n}ski}},
  \bibinfo{author}{\bibfnamefont{R.~F.} \bibnamefont{Casten}},
  \bibinfo{author}{\bibfnamefont{R.~L.} \bibnamefont{Gill}},
  \bibinfo{author}{\bibfnamefont{D.~S.} \bibnamefont{Brenner}},
  \bibinfo{author}{\bibfnamefont{J.~A.} \bibnamefont{Winger}},
  \bibinfo{author}{\bibfnamefont{W.}~\bibnamefont{Krips}},
  \bibinfo{author}{\bibfnamefont{C.}~\bibnamefont{Wesselborg}},
  \bibinfo{author}{\bibfnamefont{M.}~\bibnamefont{B{\"u}scher}},
  \bibinfo{author}{\bibfnamefont{F.~K.} \bibnamefont{Wohn}},
  \bibnamefont{et~al.}, \bibinfo{journal}{Phys. Rev. Lett.}
  \textbf{\bibinfo{volume}{63}}, \bibinfo{pages}{143} (\bibinfo{year}{1989}).

\bibitem[{\citenamefont{Piiparinen et~al.}(1993)\citenamefont{Piiparinen,
  Julin, Juutinen, Virtanen, Ahonen, Fahlander, Hattula, Lampinen,
  L{\"o}nnroth, Maj et~al.}}]{piiparinen1993}
\bibinfo{author}{\bibfnamefont{M.}~\bibnamefont{Piiparinen}},
  \bibinfo{author}{\bibfnamefont{R.}~\bibnamefont{Julin}},
  \bibinfo{author}{\bibfnamefont{S.}~\bibnamefont{Juutinen}},
  \bibinfo{author}{\bibfnamefont{A.}~\bibnamefont{Virtanen}},
  \bibinfo{author}{\bibfnamefont{P.}~\bibnamefont{Ahonen}},
  \bibinfo{author}{\bibfnamefont{C.}~\bibnamefont{Fahlander}},
  \bibinfo{author}{\bibfnamefont{J.}~\bibnamefont{Hattula}},
  \bibinfo{author}{\bibfnamefont{A.}~\bibnamefont{Lampinen}},
  \bibinfo{author}{\bibfnamefont{T.}~\bibnamefont{L{\"o}nnroth}},
  \bibinfo{author}{\bibfnamefont{A.}~\bibnamefont{Maj}}, \bibnamefont{et~al.},
  \bibinfo{journal}{Nucl. Phys. A} \textbf{\bibinfo{volume}{565}},
  \bibinfo{pages}{671} (\bibinfo{year}{1993}).

\bibitem[{\citenamefont{Zamfir et~al.}(1995)\citenamefont{Zamfir, Gill,
  Brenner, Casten, and Wolf}}]{zamfir1995}
\bibinfo{author}{\bibfnamefont{N.~V.} \bibnamefont{Zamfir}},
  \bibinfo{author}{\bibfnamefont{R.~L.} \bibnamefont{Gill}},
  \bibinfo{author}{\bibfnamefont{D.~S.} \bibnamefont{Brenner}},
  \bibinfo{author}{\bibfnamefont{R.~F.} \bibnamefont{Casten}},
  \bibnamefont{and} \bibinfo{author}{\bibfnamefont{A.}~\bibnamefont{Wolf}},
  \bibinfo{journal}{Phys. Rev. C} \textbf{\bibinfo{volume}{51}},
  \bibinfo{pages}{98} (\bibinfo{year}{1995}).

\bibitem[{\citenamefont{Lobach et~al.}(1999)\citenamefont{Lobach, Efimov, and
  Pasternak}}]{lobach1999}
\bibinfo{author}{\bibfnamefont{Y.}~\bibnamefont{Lobach}},
  \bibinfo{author}{\bibfnamefont{A.}~\bibnamefont{Efimov}}, \bibnamefont{and}
  \bibinfo{author}{\bibfnamefont{A.}~\bibnamefont{Pasternak}},
  \bibinfo{journal}{Eur. Phys. J. A} \textbf{\bibinfo{volume}{6}},
  \bibinfo{pages}{131} (\bibinfo{year}{1999}).

\bibitem[{\citenamefont{Harissopulos et~al.}(2001)\citenamefont{Harissopulos,
  Dewald, Gelberg, Zell, {von Bretano}, and Kern}}]{Harissopulos2001}
\bibinfo{author}{\bibfnamefont{S.}~\bibnamefont{Harissopulos}},
  \bibinfo{author}{\bibfnamefont{A.}~\bibnamefont{Dewald}},
  \bibinfo{author}{\bibfnamefont{A.}~\bibnamefont{Gelberg}},
  \bibinfo{author}{\bibfnamefont{K.}~\bibnamefont{Zell}},
  \bibinfo{author}{\bibfnamefont{P.}~\bibnamefont{{von Bretano}}},
  \bibnamefont{and} \bibinfo{author}{\bibfnamefont{J.}~\bibnamefont{Kern}},
  \bibinfo{journal}{Nucl. Phys. A} \textbf{\bibinfo{volume}{683}},
  \bibinfo{pages}{157} (\bibinfo{year}{2001}).

\bibitem[{\citenamefont{Ashley et~al.}(2007)\citenamefont{Ashley, Linnemann,
  Jolie, Regan, Andgren, Dewald, McCutchan, Melon, M{\"o}ller, Zamfir
  et~al.}}]{ashley2007}
\bibinfo{author}{\bibfnamefont{S.}~\bibnamefont{Ashley}},
  \bibinfo{author}{\bibfnamefont{A.}~\bibnamefont{Linnemann}},
  \bibinfo{author}{\bibfnamefont{J.}~\bibnamefont{Jolie}},
  \bibinfo{author}{\bibfnamefont{P.}~\bibnamefont{Regan}},
  \bibinfo{author}{\bibfnamefont{K.}~\bibnamefont{Andgren}},
  \bibinfo{author}{\bibfnamefont{A.}~\bibnamefont{Dewald}},
  \bibinfo{author}{\bibfnamefont{E.}~\bibnamefont{McCutchan}},
  \bibinfo{author}{\bibfnamefont{B.}~\bibnamefont{Melon}},
  \bibinfo{author}{\bibfnamefont{O.}~\bibnamefont{M{\"o}ller}},
  \bibinfo{author}{\bibfnamefont{N.}~\bibnamefont{Zamfir}},
  \bibnamefont{et~al.}, \bibinfo{journal}{Acta Phys. Pol. B}
  \textbf{\bibinfo{volume}{38}}, \bibinfo{pages}{1385} (\bibinfo{year}{2007}).

\bibitem[{\citenamefont{Ilieva et~al.}(2014)\citenamefont{Ilieva, Th{\"u}rauf,
  Kr{\"o}ll, Kr{\"u}cken, Behrens, Bildstein, Blazhev, B{\"o}nig, Butler,
  Cederk{\"a}ll et~al.}}]{ilieva2014}
\bibinfo{author}{\bibfnamefont{S.}~\bibnamefont{Ilieva}},
  \bibinfo{author}{\bibfnamefont{M.}~\bibnamefont{Th{\"u}rauf}},
  \bibinfo{author}{\bibfnamefont{T.}~\bibnamefont{Kr{\"o}ll}},
  \bibinfo{author}{\bibfnamefont{R.}~\bibnamefont{Kr{\"u}cken}},
  \bibinfo{author}{\bibfnamefont{T.}~\bibnamefont{Behrens}},
  \bibinfo{author}{\bibfnamefont{V.}~\bibnamefont{Bildstein}},
  \bibinfo{author}{\bibfnamefont{A.}~\bibnamefont{Blazhev}},
  \bibinfo{author}{\bibfnamefont{S.}~\bibnamefont{B{\"o}nig}},
  \bibinfo{author}{\bibfnamefont{P.~A.} \bibnamefont{Butler}},
  \bibinfo{author}{\bibfnamefont{J.}~\bibnamefont{Cederk{\"a}ll}},
  \bibnamefont{et~al.}, \bibinfo{journal}{Phys. Rev. C}
  \textbf{\bibinfo{volume}{89}}, \bibinfo{pages}{014313}
  (\bibinfo{year}{2014}).

\bibitem[{\citenamefont{Borrajo et~al.}(2015)\citenamefont{Borrajo,
  Rodr\'{i}guez, and {Luis Egido}}}]{Borrajo2015cranking}
\bibinfo{author}{\bibfnamefont{M.}~\bibnamefont{Borrajo}},
  \bibinfo{author}{\bibfnamefont{T.~R.} \bibnamefont{Rodr\'{i}guez}},
  \bibnamefont{and} \bibinfo{author}{\bibfnamefont{J.}~\bibnamefont{{Luis
  Egido}}}, \bibinfo{journal}{Phys. Lett. B} \textbf{\bibinfo{volume}{746}},
  \bibinfo{pages}{341} (\bibinfo{year}{2015}).

\bibitem[{\citenamefont{Rodr\'{i}guez}(2016)}]{Rodriguez2016cranking}
\bibinfo{author}{\bibfnamefont{T.~R.} \bibnamefont{Rodr\'{i}guez}},
  \bibinfo{journal}{Eur. Phys. J. A} \textbf{\bibinfo{volume}{52}},
  \bibinfo{pages}{190} (\bibinfo{year}{2016}).

\bibitem[{\citenamefont{Rodr{\'i}guez et~al.}(2008)\citenamefont{Rodr{\'i}guez,
  Egido, and Jungclaus}}]{rodriguez2008cadmium}
\bibinfo{author}{\bibfnamefont{T.}~\bibnamefont{Rodr{\'i}guez}},
  \bibinfo{author}{\bibfnamefont{J.}~\bibnamefont{Egido}}, \bibnamefont{and}
  \bibinfo{author}{\bibfnamefont{A.}~\bibnamefont{Jungclaus}},
  \bibinfo{journal}{Phys. Lett. B} \textbf{\bibinfo{volume}{668}},
  \bibinfo{pages}{410} (\bibinfo{year}{2008}).

\bibitem[{\citenamefont{Sakai and Rester}(1977)}]{SAKAI1977441}
\bibinfo{author}{\bibfnamefont{M.}~\bibnamefont{Sakai}} \bibnamefont{and}
  \bibinfo{author}{\bibfnamefont{A.~C.} \bibnamefont{Rester}},
  \bibinfo{journal}{Atomic Data and Nuclear Data Tables}
  \textbf{\bibinfo{volume}{20}}, \bibinfo{pages}{441} (\bibinfo{year}{1977}).

\bibitem[{\citenamefont{Garrett and Wood}(2010)}]{Garrett_2010}
\bibinfo{author}{\bibfnamefont{P.~E.} \bibnamefont{Garrett}} \bibnamefont{and}
  \bibinfo{author}{\bibfnamefont{J.~L.} \bibnamefont{Wood}},
  \textbf{\bibinfo{volume}{37}}, \bibinfo{pages}{069701}
  (\bibinfo{year}{2010}).

\bibitem[{\citenamefont{Siciliano
  et~al.}(2020{\natexlab{b}})\citenamefont{Siciliano, Zanon, Goasduff, John,
  Rodr\'iguez, P\'eru, Deloncle, Libert, Zieli\'{n}ska, Ashad
  et~al.}}]{siciliano2020coexistence}
\bibinfo{author}{\bibfnamefont{M.}~\bibnamefont{Siciliano}},
  \bibinfo{author}{\bibfnamefont{I.}~\bibnamefont{Zanon}},
  \bibinfo{author}{\bibfnamefont{A.}~\bibnamefont{Goasduff}},
  \bibinfo{author}{\bibfnamefont{P.~R.} \bibnamefont{John}},
  \bibinfo{author}{\bibfnamefont{T.~R.} \bibnamefont{Rodr\'iguez}},
  \bibinfo{author}{\bibfnamefont{S.}~\bibnamefont{P\'eru}},
  \bibinfo{author}{\bibfnamefont{I.}~\bibnamefont{Deloncle}},
  \bibinfo{author}{\bibfnamefont{J.}~\bibnamefont{Libert}},
  \bibinfo{author}{\bibfnamefont{M.}~\bibnamefont{Zieli\'{n}ska}},
  \bibinfo{author}{\bibfnamefont{D.}~\bibnamefont{Ashad}},
  \bibnamefont{et~al.}, \bibinfo{journal}{Phys. Rev. C}
  \textbf{\bibinfo{volume}{102}}, \bibinfo{pages}{014318}
  (\bibinfo{year}{2020}{\natexlab{b}}).

\bibitem[{\citenamefont{Poves et~al.}(2020)\citenamefont{Poves, Nowacki, and
  Alhassid}}]{poves2020shape}
\bibinfo{author}{\bibfnamefont{A.}~\bibnamefont{Poves}},
  \bibinfo{author}{\bibfnamefont{F.}~\bibnamefont{Nowacki}}, \bibnamefont{and}
  \bibinfo{author}{\bibfnamefont{Y.}~\bibnamefont{Alhassid}},
  \bibinfo{journal}{Phys. Rev. C} \textbf{\bibinfo{volume}{101}},
  \bibinfo{pages}{054307} (\bibinfo{year}{2020}).

\bibitem[{\citenamefont{Rocchini et~al.}(2021)\citenamefont{Rocchini,
  Hady\'{n}ska-Kl\c{e}k, Nannini, Goasduff, Zieli\'{n}ska, Testov,
  Rodr\'{i}guez, Gargano, Nowacki, De~Gregorio et~al.}}]{rocchini2021Zn66}
\bibinfo{author}{\bibfnamefont{M.}~\bibnamefont{Rocchini}},
  \bibinfo{author}{\bibfnamefont{K.}~\bibnamefont{Hady\'{n}ska-Kl\c{e}k}},
  \bibinfo{author}{\bibfnamefont{A.}~\bibnamefont{Nannini}},
  \bibinfo{author}{\bibfnamefont{A.}~\bibnamefont{Goasduff}},
  \bibinfo{author}{\bibfnamefont{M.}~\bibnamefont{Zieli\'{n}ska}},
  \bibinfo{author}{\bibfnamefont{D.}~\bibnamefont{Testov}},
  \bibinfo{author}{\bibfnamefont{T.~R.} \bibnamefont{Rodr\'{i}guez}},
  \bibinfo{author}{\bibfnamefont{A.}~\bibnamefont{Gargano}},
  \bibinfo{author}{\bibfnamefont{F.}~\bibnamefont{Nowacki}},
  \bibinfo{author}{\bibfnamefont{G.}~\bibnamefont{De~Gregorio}},
  \bibnamefont{et~al.}, \bibinfo{journal}{Phys. Rev. C}
  \textbf{\bibinfo{volume}{103}}, \bibinfo{pages}{014311}
  (\bibinfo{year}{2021}).

\bibitem[{\citenamefont{Wrzosek-Lipska
  et~al.}(2012)\citenamefont{Wrzosek-Lipska, Pr\'{o}chniak, Zieli\'{n}ska,
  Srebrny, Hady\'{n}ska-Kl\c{e}k, Iwanicki, Kisieli\'{n}ski, Kowalczyk,
  Napiorkowski, Pi\c{e}tak et~al.}}]{kasiaw2012Mo100}
\bibinfo{author}{\bibfnamefont{K.}~\bibnamefont{Wrzosek-Lipska}},
  \bibinfo{author}{\bibfnamefont{L.}~\bibnamefont{Pr\'{o}chniak}},
  \bibinfo{author}{\bibfnamefont{M.}~\bibnamefont{Zieli\'{n}ska}},
  \bibinfo{author}{\bibfnamefont{J.}~\bibnamefont{Srebrny}},
  \bibinfo{author}{\bibfnamefont{K.}~\bibnamefont{Hady\'{n}ska-Kl\c{e}k}},
  \bibinfo{author}{\bibfnamefont{J.}~\bibnamefont{Iwanicki}},
  \bibinfo{author}{\bibfnamefont{M.}~\bibnamefont{Kisieli\'{n}ski}},
  \bibinfo{author}{\bibfnamefont{M.}~\bibnamefont{Kowalczyk}},
  \bibinfo{author}{\bibfnamefont{P.~J.} \bibnamefont{Napiorkowski}},
  \bibinfo{author}{\bibfnamefont{D.}~\bibnamefont{Pi\c{e}tak}},
  \bibnamefont{et~al.}, \bibinfo{journal}{Phys. Rev. C}
  \textbf{\bibinfo{volume}{86}}, \bibinfo{pages}{064305}
  (\bibinfo{year}{2012}).

\bibitem[{\citenamefont{Hady\'{n}ska-Kl\c{e}k
  et~al.}(2018)\citenamefont{Hady\'{n}ska-Kl\c{e}k, Napiorkowski,
  Zieli\'{n}ska, Srebrny, Maj, Azaiez, Valiente~Dob\'on, Kici\'{n}ska-Habior,
  Nowacki, Na{\"i}dja et~al.}}]{kasiah2018Ca42}
\bibinfo{author}{\bibfnamefont{K.}~\bibnamefont{Hady\'{n}ska-Kl\c{e}k}},
  \bibinfo{author}{\bibfnamefont{P.~J.} \bibnamefont{Napiorkowski}},
  \bibinfo{author}{\bibfnamefont{M.}~\bibnamefont{Zieli\'{n}ska}},
  \bibinfo{author}{\bibfnamefont{J.}~\bibnamefont{Srebrny}},
  \bibinfo{author}{\bibfnamefont{A.}~\bibnamefont{Maj}},
  \bibinfo{author}{\bibfnamefont{F.}~\bibnamefont{Azaiez}},
  \bibinfo{author}{\bibfnamefont{J.~J.} \bibnamefont{Valiente~Dob\'on}},
  \bibinfo{author}{\bibfnamefont{M.}~\bibnamefont{Kici\'{n}ska-Habior}},
  \bibinfo{author}{\bibfnamefont{F.}~\bibnamefont{Nowacki}},
  \bibinfo{author}{\bibfnamefont{H.}~\bibnamefont{Na{\"i}dja}},
  \bibnamefont{et~al.}, \bibinfo{journal}{Phys. Rev. C}
  \textbf{\bibinfo{volume}{97}}, \bibinfo{pages}{024326}
  (\bibinfo{year}{2018}).

\bibitem[{\citenamefont{Morrison et~al.}(2020)\citenamefont{Morrison,
  Hady\'{n}ska-Kl\c{e}k, Podoly\'{a}k, Doherty, Gaffney, Kaya, Pr\'{o}chniak,
  Samorajczyk-Py\'{s}k, Srebrny, Berry et~al.}}]{morrison2020Xe130}
\bibinfo{author}{\bibfnamefont{L.}~\bibnamefont{Morrison}},
  \bibinfo{author}{\bibfnamefont{K.}~\bibnamefont{Hady\'{n}ska-Kl\c{e}k}},
  \bibinfo{author}{\bibfnamefont{Z.}~\bibnamefont{Podoly\'{a}k}},
  \bibinfo{author}{\bibfnamefont{D.~T.} \bibnamefont{Doherty}},
  \bibinfo{author}{\bibfnamefont{L.~P.} \bibnamefont{Gaffney}},
  \bibinfo{author}{\bibfnamefont{L.}~\bibnamefont{Kaya}},
  \bibinfo{author}{\bibfnamefont{L.}~\bibnamefont{Pr\'{o}chniak}},
  \bibinfo{author}{\bibfnamefont{J.}~\bibnamefont{Samorajczyk-Py\'{s}k}},
  \bibinfo{author}{\bibfnamefont{J.}~\bibnamefont{Srebrny}},
  \bibinfo{author}{\bibfnamefont{T.}~\bibnamefont{Berry}},
  \bibnamefont{et~al.}, \bibinfo{journal}{Phys. Rev. C}
  \textbf{\bibinfo{volume}{102}}, \bibinfo{pages}{054304}
  (\bibinfo{year}{2020}).

\bibitem[{\citenamefont{Yordanov et~al.}(2018)\citenamefont{Yordanov,
  Balabanski, Bissell, Blaum, Blazhev, Budin\v{c}evi{}\'{c}, Fr\"ommgen,
  Geppert, Grawe, Hammen et~al.}}]{PhysRevC.98.011303}
\bibinfo{author}{\bibfnamefont{D.~T.} \bibnamefont{Yordanov}},
  \bibinfo{author}{\bibfnamefont{D.~L.} \bibnamefont{Balabanski}},
  \bibinfo{author}{\bibfnamefont{M.~L.} \bibnamefont{Bissell}},
  \bibinfo{author}{\bibfnamefont{K.}~\bibnamefont{Blaum}},
  \bibinfo{author}{\bibfnamefont{A.}~\bibnamefont{Blazhev}},
  \bibinfo{author}{\bibfnamefont{I.}~\bibnamefont{Budin\v{c}evi{}\'{c}}},
  \bibinfo{author}{\bibfnamefont{N.}~\bibnamefont{Fr\"ommgen}},
  \bibinfo{author}{\bibfnamefont{C.}~\bibnamefont{Geppert}},
  \bibinfo{author}{\bibfnamefont{H.}~\bibnamefont{Grawe}},
  \bibinfo{author}{\bibfnamefont{M.}~\bibnamefont{Hammen}},
  \bibnamefont{et~al.}, \bibinfo{journal}{Phys. Rev. C}
  \textbf{\bibinfo{volume}{98}}, \bibinfo{pages}{011303(R)}
  (\bibinfo{year}{2018}).

\bibitem[{\citenamefont{Andreetto et~al.}(2019)\citenamefont{Andreetto,
  Chiarello, Costa, Crescente, Fantinel, Fanzago, Konomi, Mazzon, Menguzzato,
  Segatta et~al.}}]{CloudVeneto}
\bibinfo{author}{\bibfnamefont{P.}~\bibnamefont{Andreetto}},
  \bibinfo{author}{\bibfnamefont{F.}~\bibnamefont{Chiarello}},
  \bibinfo{author}{\bibfnamefont{F.}~\bibnamefont{Costa}},
  \bibinfo{author}{\bibfnamefont{A.}~\bibnamefont{Crescente}},
  \bibinfo{author}{\bibfnamefont{S.}~\bibnamefont{Fantinel}},
  \bibinfo{author}{\bibfnamefont{F.}~\bibnamefont{Fanzago}},
  \bibinfo{author}{\bibfnamefont{E.}~\bibnamefont{Konomi}},
  \bibinfo{author}{\bibfnamefont{P.~E.} \bibnamefont{Mazzon}},
  \bibinfo{author}{\bibfnamefont{M.}~\bibnamefont{Menguzzato}},
  \bibinfo{author}{\bibfnamefont{M.}~\bibnamefont{Segatta}},
  \bibnamefont{et~al.}, \bibinfo{journal}{Eur. Phys. J.: Web Conf.}
  \textbf{\bibinfo{volume}{214}}, \bibinfo{pages}{07010}
  (\bibinfo{year}{2019}).

\end{thebibliography}

\end{document}